\documentclass[10pt,titlepage]{article}

\usepackage{amsmath}
\usepackage{./here}
\usepackage{./authblk}
\usepackage{./comment}
\usepackage{color}
\usepackage{ulem}
\usepackage{graphicx}

\usepackage{amssymb}
\usepackage{comment}
\usepackage{cite}

\title{Improving the soil water module of the Decision Support System for Agrotechnology Transfer cropping system model for subsurface irrigation}
\author[a,*]{Dan Chen}
\author[a]{Yusuke Kikuchi}
\author[a]{Kenichiro Fujiyama}
\author[a]{Shunsuke Akimoto}
\author[a]{Shinji Oominato}
\author[b]{Toshihiro Hasegawa}

\affil[a]{Digital Platform Division, NEC, 1753, Shimonumabe, Nakahara-ku, Kawasaki, Kanagawa 211-8666, Japan}
\affil[b]{Tohoku Agricultural Research Center, NARO, 4 Akahira, Shimo-kuriyagawa, Morioka, Iwate 020-0198, Japan}
\affil[*]{Corresponding author. E-mail address: d-chen@cq.jp.nec.com (Dan Chen)}

\date{}

\begin{document}
	
	\maketitle
	
\section*{Abstract}
	\par Ensuring that crops use water and nutrients efficiently is an important strategy for increasing the profitability of farming and reducing the environmental load from agriculture.
	Subsurface irrigation can be an alternative to surface irrigation as a means of losing less irrigation water, but the application timing and amount are often difficult to determine.
	Well-defined soil and crop models are useful for assisting decision support, but most of the models developed to date have been for surface irrigation.
	The present study examines whether the Decision Support System for Agrotechnology Transfer (DSSAT, version 4.5) cropping system model is applicable for the production of processing tomatoes with subsurface irrigation, 
	and it revises the soil module to simulate irrigation schemes with subsurface irrigation.
	Five farmed fields in California, USA, are used to test the performance of the model.
	The original DSSAT model fails to produce fruit yield by overestimating the water deficiency.
	The soil water module is then revised by introducing the movement of soil moisture due to a vertical soil moisture gradient.
	Moreover, an external parameter optimization system is constructed to minimize the error between the simulation and observations.
	The revised module reduces the errors in the soil moisture profile at each field compared to those by the original DSSAT model.
	The average soil moisture error decreases from $0.065\,{\rm m}^3\,{\rm m}^{-3}$ to $0.029\,{\rm m}^3\,{\rm m}^{-3}$.
	The yields estimated by the modified model are in a reasonable range from 80 to 150$\,{\rm ton\,ha^{-1}}$, which is commonly observed under well-managed conditions.
	The present results show that although further testing is required for yield prediction, the present modification to the original DSSAT model improves the precision of the soil moisture profile under subsurface irrigation and can be used for decision support for efficient producting of processing tomatoes.
	\vspace{\baselineskip}
	\par \noindent {\it Keywords:} DSSAT, tomato, soil moisture, van Genuchten model
	
\clearpage

\section{Introduction}
	\par Reducing the amount of water and nutrients used for farming with no yield penalty would not only increase the profitability of farming but also reduce the negative environmental impacts of agriculture; the latter being a major source of pollution of the aquatic and terrestrial ecosystems \cite{Springmann}.
	Tomatoes are the most consumed vegetables in the world and require intensive management skills.
	In particular, high production efficiency (i.e., high yield with minimal resources) is needed when producing processing tomatoes because the farming cost and the yield are the main factors that determine the farmer's profit.
	\par Over the past half century, various types of micro-irrigation systems have been developed for efficient managing water and nutrients for greenhouse crops and field crops \cite{Ayars1999}.
	Of those systems, subsurface irrigation has been studied for more than 30 years and has been shown to be an efficient and productive system particularly in water-scarce environments such as California \cite{Ayars1999, Ayars2015, Kennedy2013}, which produces roughly $30\%$ of the world's processing tomatos \cite{WPTC}.
	The subsurface drip irrigation system is now commonly adopted in this region \cite{Hartz2008}.
	This involves burying irrigation pipes to a depth of roughly 30\,cm to minimize the loss of irrigation water through evaporation, and doing so improves water use efficiency considerably compared to using sprinkler irrigation systems on the surface.
	\par Despite the efficient system having been developed and introduced, the timing and amount of water and nutrients applied to the crops depend on the experience of the growers, and objective decision support information is much needed in the face of variable climatic conditions.
	Crop models are useful tools for simulating crop growth and then calculating the necessary and sufficient levels of irrigation and fertilizers (e.g., \cite{Jones2003}).
	Of those models, the Decision Support System for Agrotechnology Transfer (DSSAT) cropping system model is the one that is used most widely for agricultural technology transfer \cite{Jones2003, DSSAT1}.
	The DSSAT model is an open and versatile model for many crops, including soybean and maize, and it has been tested in locations all around the world.
	DSSAT provides time-series information about plant growth (e.g., leaf area index (LAI), root distribution, dry weight) using environment data (e.g., weather data) and initial parameters (e.g., nursery temperature, soil water conductivity constant) given at the beginning of the simulation.
	More importantly, although some tomato growth models have been developed around the world \cite{Heuvelink1999, Soto2014}, the DSSAT tomato model developed by Boote et al. (2012) is the only accessible open-source model that can be used for the production of processing tomatoes.
	\par Prior to this study, we tested the DSSAT tomato model (version 4.5) in an effort to understand the current efficiency of the tomato production system in California, USA.
	We found that the original model (i) is accurate for tomato growth in fields with surface irrigation but (ii) predicts much lower growth than that observed in fields with subsurface irrigation.
	We hypothesized that the original DSSAT model does not explain soil water movement properly for fields with subsurface irrigation, where water upflow is assumed to occur when the soil moisture content is above the saturation threshold at each layer.
	This soil water movement might be insufficient to simulate the vertical water movement in the soil when the water is supplied through subsurface irrigation.
	It is possible that the upward water movement takes place even where the soil moisture is not saturated in the lower soil layer, and we hypothesized that the original model requires a process that accounts for the unsaturated hydraulic movement to measure the efficiency of the subsurface drip irrigation system.
	Therefore, in the present work, we test a modified DSSAT model that incorporates the van Genuchten model \cite{vanGenuchten} to allow for the moisture movement due to subsurface irrigation.
	Furthermore, by introducing another parameter optimization system to adjust the physical soil parameters, we attempt to improve the accuracy of estimating soil physical properties that are closely related to soil moisture but that are difficult to measure.
	
\clearpage

\section{Materials and methods}
	
	\subsection{Modified model of soil water movement}
		\par DSSAT is a crop growth model that calculates fruit yield using data on weather, variety, and soil properties.
		Prior to running a simulation for the current season, ``spin-up'' simulations of multiple seasons are run to allow the soil organic matter and nutrients to reach a steady state.
		DSSAT calculates water and nutrition movement or balance in the soil plant system on a daily basis to simulate the growth, development, and yield of crop plants.
		\par
\begin{figure}[H]
			\begin{center}
			\includegraphics[scale=0.5, bb=0 0 840 383]{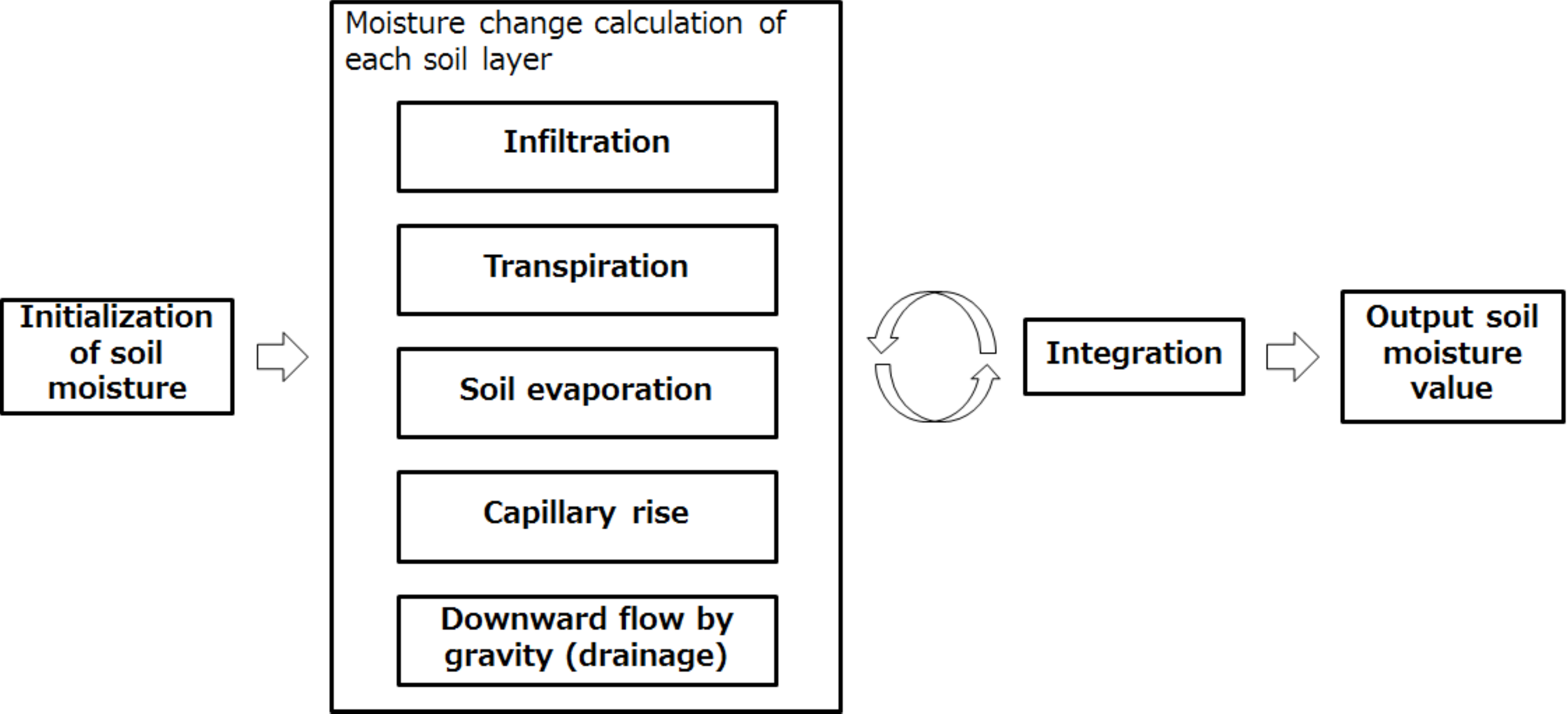}
			\caption{Soil water calculation in original DSSAT cropping system model. The values of the initial water content are read from either an input file or the results of previous seasonal calculations in the initialization process. The amounts of water movement in a day due to the five processes including downward flow by gravity (drainage) and transpiration are calculated in a moisture change calculation process for each soil layer. This calculation provides the changes in water content due to each process for each soil layer. The integration process integrates those water movements and fixes the soil moisture content at the end of the day. The moisture change calculation and integration are repeated until a user-given day for ending the simulation. Finally, the calculation results are output.}
			\label{calc_flow}
			\end{center}
		\end{figure}

 To calculate the movement of water in soil precisely and in a realistic calculation time, the soil is divided virtually into layers in DSSAT.
		The calculation of the water change in a soil layer in the original DSSAT is shown in Figure \ref{calc_flow} \cite{DSSATmanual}.
		In the first step, initial data on the soil moisture are read from an input file or from the results of previous seasonal calculations if those have been executed.
		Daily variation in soil moisture at each layer is then calculated based on five processes:
		(i) infiltration, wherein the water input stems from irrigation and precipitation;
		(ii) transpiration, wherein plants absorb the water;
		(iii) soil evaporation, wherein the water evaporates into the air from each soil layer;
		(iv) capillary rise, wherein water movement is caused by capillary flow;
		and (v) downward flow by gravity, wherein vertical water movement occurs between adjacent soil layers (i.e., drainage).
		This daily variation is integrated daily to update the value of the soil moisture in each layer until the designated day for ending the calculations, which can be the planned harvest day.
		The final output comprises the calculation results including all the daily results in the growth season.
		\par The original model is a ``bucket-type'' model \cite{bucket} in which the drainage (downward flow by gravity) is calculated from the top soil layer; if the amount of water in the next layer exceeds the upper limit, then excess water can move up to the upper layer.
		It also calculates capillary rise movement as an upward flow of water. However, the original model does not assume a water source in the middle of the soil layers, and it may be insufficient for reproducing the upward water movement in the subsurface drip irrigation system.
		\par In the present study, we add another water movement to the original DSSAT model to better account for vertical water movement from the subsurface irrigation based on the van Genuchten model \cite{vanGenuchten}.
		The van Genuchten model is a basic model that calculates the unsaturated hydraulic conductivity of soil from the volumetric water content and soil-specific parameters.
		With this, vertical soil water movement between adjacent soil layers can be calculated from the conductivity and gradient of the volumetric soil-moisture content between the layers.
		In our case, we use this logic to calculate the water movement between two vertically adjacent soil layers.
		Each layer is either a 10-cm-thick shallow layer ($<75\,{\rm cm}$ deep) or a 20- or 30-cm-thick deeper layer.
		\par The original DSSAT calculates soil moisture movement in a day in one step using the ``bucket-type'' model.
		Because this model was determined to be insufficiently is accurate for tomato growth in fields with subsurface irrigation in our previous test in California, USA, we divide the soil moisture movement calculation of one day in the original DSSAT into small time step calculations to achieve higher accuracy.
		The duration time of one step is indicated by $\Delta t$ bellow, and the volumetric water content change in $\Delta t$ is indicated by $\Delta \theta$.
		We use $\Delta t = 10.0$\,s in our calculations to ensure a sufficiently low discretization error.
		This means that the calculations of the soil moisture movements between soil layers in a day are divided into 8,640 short-time calculations.
		The trend of discretization errors depending on $\Delta t$ is estimated in Appendix \ref{dt_trend}, which shows that 10.0\, s is enough short time to suppress the discretization errors.
		In a time as short as 10\,s, the water movement between a pair of adjacent soil layers is much less than soil moisture content in the layers.
		This means that a water movement calculation between a pair of adjacent layers has an insignificant effect on the calculations between the other pairs.
		For that reason, we can calculate independently the water movement between each pair of adjacent layers.
		Therefore, we calculate the water movement from the pair of 10\,cm and 20\,cm layers to lower layers.
		Moreover, although the water that enters the soil from rain and irrigation is contiguous during times of rainfall and irrigation, in our model irrigation and rain are input whole at midnight because we could not get the start and end time of them but only the total amount.  
		The saturation water flow is then calculated using the original DSSAT logic.
		Finally, the water movement via the moisture gradient $\Delta \theta$ is calculated step by step using the time step $\Delta t$ and is given by 
		\begin{equation}
			\Delta \theta (L) = \frac{ V_{in}(L) - V_{out}(L)}{V_{layer}(L)}, \label{Dtheta}
		\end{equation}
		where $V_{in}(L)$ is the amount of water that moves from the upper layer into layer $L$, $V_{out}(L)$ is the amount of water that moves to the lower layer, and $V_{layer}(L)$ is the volume of layer L.
		If the water moves from lower layer $L+1$ to layer $L$ by capillarity, then $V_{out}(L)$ is negative.
		The top-most layer receives precipitation, which is included in $V_{in}$.
		At the irrigation layer, irrigation water is included in $V_{in}$.
		At the other layers, we have that $V_{out}(L) = V_{in}(L+1)$.
		To simplify the model, we assume that all daily irrigation is applied at the first calculation step on the day.
		According to the van Genuchten model, $V_{out}(L)$ can be determined using the following equation \cite{CDJones}:
		\begin{equation}
			V_{out}\left( L \right) = \left[ D_{GM}\frac{\theta\left( L \right) - \theta\left( L+1 \right)}{\frac{\Delta z(L) + \Delta z(L+1)}{2}} + K_{GM} \right]\Delta t, \label{deltatheta}
		\end{equation}
		where $\Delta z(L)$ is the thickness of layer $L$, $D_{GM}$ is the geometric mean of the hydraulic diffusivity of layers $L$ and $L+1$, and $K_{GM}$ is the geometric mean of the hydraulic conductivity:
		\begin{eqnarray}
			D_{GM} &=& \sqrt{D(L)D(L+1)}, \label{DGM} \\
			K_{GM} &=& \sqrt{K(L)K(L+1)}, \label{KGM} \\
			D &=& K \left| \frac{\mathrm{d} \Psi}{\mathrm{d} \theta} \right|, \label{Dsiki}
		\end{eqnarray}
		where $\Psi$ is the soil suction head.
		According to the van Genuchten model, $K(\theta)$ is given by 
		\begin{eqnarray}
			K(\theta) &=& K_sS_e^{l}(\theta)\left[ 1-\left( 1- S_e^{\frac{1}{m}}(\theta) \right) \right]^2, \label{Ksiki} \\
			S_e &=& \frac{\theta - \theta_r}{\theta_s - \theta_r}, \label{Sesiki} \\
			m &=& 1-\frac{1}{n}, \label{ndsiki}
		\end{eqnarray}
		where $K_s$ is the saturated hydraulic conductivity, $l$ is the pore-connectivity coefficient, and $n, \theta_r,$ and $\theta_s$ are the van Genuchten parameters. 
		According to the work of Mualem et. (1976) and Kosugi (2007), we used 0.5 for the value of $l$, and $\Psi$ is given by 
		\begin{equation}
			\Psi = \frac{1}{\alpha}\left[ \left( \frac{\theta_s - \theta_r}{\theta - \theta_r} \right)^{\frac{1}{m}} - 1 \right]^{\frac{1}{n}}, \label{Psisiki}
		\end{equation}
		where $\alpha$ is another parameter in the van Genuchten model.
		
	\subsection{Information about test fields and observation data}
		\par
\begin{figure}[H]
			\begin{center}
				\includegraphics[scale=0.3, bb=0 0 1293 600]{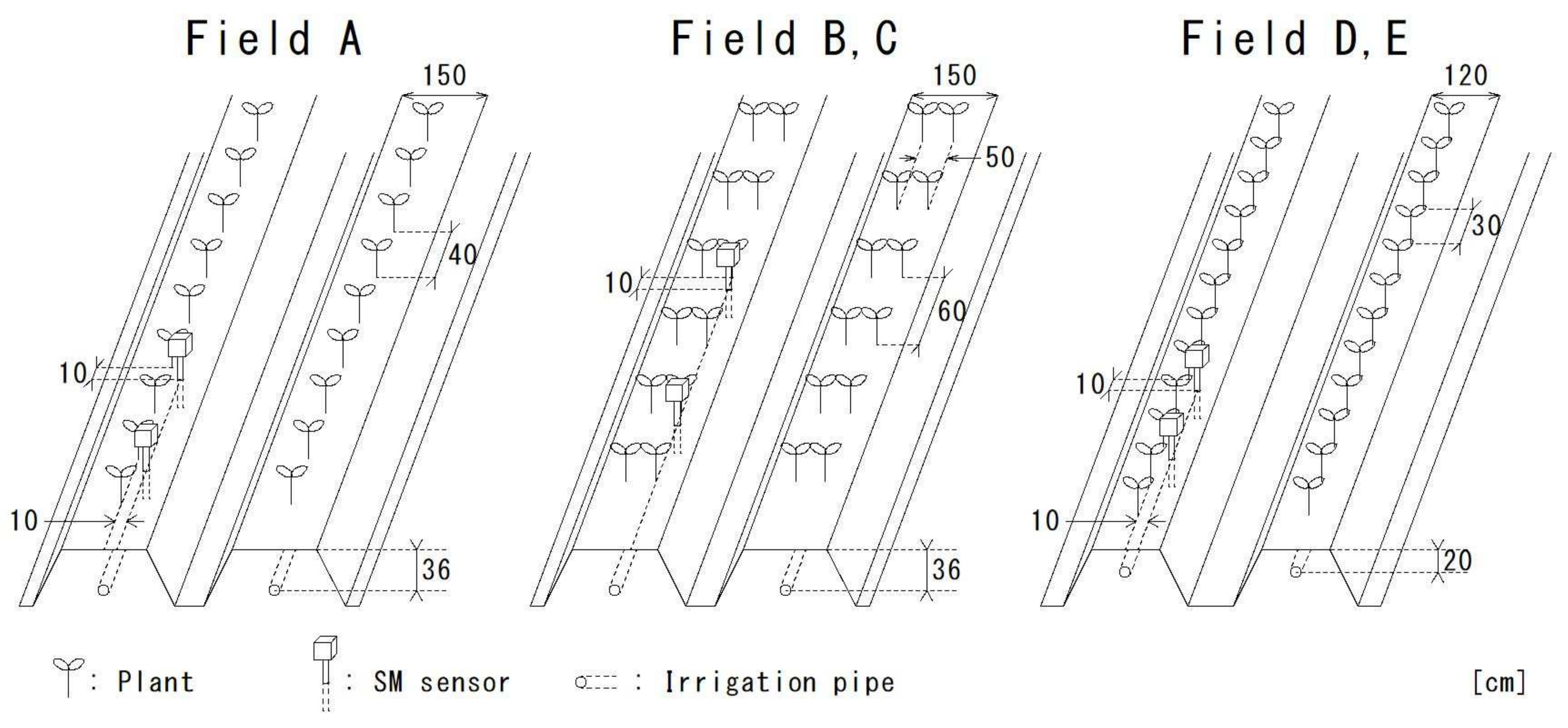}
				\caption{Positions of plants, irrigation pipes, and sensors at each field. Fields A, D, and E have one subsurface irrigation tube for one plant row, while fields B and C have one tube for two plant rows. The soil moisture sensors are installed 10\,cm from the plants and the irrigation tube to avoid damaging them.}
				\label{local_position}
			\end{center}
		\end{figure}

 To test the modified model, we use five fields in California, USA, in which the subsurface drip irrigation systems are used.
		The planting pattern and depth of the irrigation pipes vary from field to field (Figure \ref{local_position}), but ridge bed cultivation is practiced in all the fields, with the ridge bed width being 120 or 150\,cm.
		The seedlings are planted in one row per bed in fields A, D, and E and in two rows per bed in fields B and C.
		The depth of the irrigation pipes ranged from 20 to 36\,cm below the top of the ridge bed; this depth varies according to the field (as shown in Figure \ref{local_position}) along with the positions of the plants and the soil moisture sensors.
		All the crop management including the irrigation and fertilization is done by the farmers, and the farming data provided by the farmers are among the input data for the simulation.
		The weather data (i.e., air temperature, humidity, solar radiation, precipitation, and wind speed) for the simulation are collected by weather stations \cite{DACOM_W} provided by DACOM \cite{DACOM} or weather sensors \cite{CIMIS_W} provided by CIMIS \cite{CIMIS} depending on the distance between the field and weather stations.
		The measured LAI data are converted from the observation data of the Sentinel-2 satellite \cite{sentinel2}, which are from band 4 (red: $0.665\,{\rm \mu m}$) and band 8 (near-infrared: $0.842\,{\rm \mu m}$).
		The conversion equation \cite{Clevers, JGPW} is 
		\begin{eqnarray}
			LAI = -\frac{1}{a}\ln\left( 1-\frac{WDVI}{WDVI_\infty} \right), \label{LAIeq}
		\end{eqnarray}
		where $a$ is the combined value of the extinction and scattering coefficients, and $WDVI_\infty$ is the asymptotic value of $WDVI$, the weighted difference vegetation index, which is calculated as
		\begin{eqnarray}
			WDVI = r_{ir} - C_{soil}r_r, \label{WDVIeq}
		\end{eqnarray}
		where $r_{ir}$ is the measured near-infrared reflectance, $r_r$ is the measured red reflectance and
		\begin{eqnarray}
			C_{soil} = \frac{r_{s,ir}}{r_{s,r}}, \label{WDVI_C}
		\end{eqnarray}
		where $r_{s,ir}$ and $r_{s,r}$ are the near-infrared and red reflectance of the soil respectively.
		We estimate $C_{soil}$ from the satellite data obtained just before planting at each field.
		We use the TerraSen soil moisture sensor \cite{DACOM} to measure the soil moisture at depths of 10, 20, 30, 40, and 50\,cm.
		The measurements made at two points in one field could differ because of the errors and installation point of each sensor; we treat the data from each sensor as true values expect for the clearly anomalous values at L1 as discussed later.
		We run the parameter optimization system described in Section \ref{param_opt_exp} and the original/modified DSSAT for each sensor.
		\par We use the soil analysis data from the beginning of the season for the initial conditions (organic carbon, total nitrogen, pH, cation exchange capacity) and soil type (ratio of clay, silt, and sand) in DSSAT.
		To obtain the information in a satellite picture element, we take samples for soil analysis from five points in a $10 \times 10\,{\rm m^2}$ square centered on the position of the soil moisture sensors.
		We sample two different soil layers (5-15 and 25-40\,cm) at each sampling point, and we analyze a mixed sample from the five points for each layer.
		\par
\begin{figure}[H]
		\begin{center}
		\includegraphics[scale=0.6, bb=0 0 582 296]{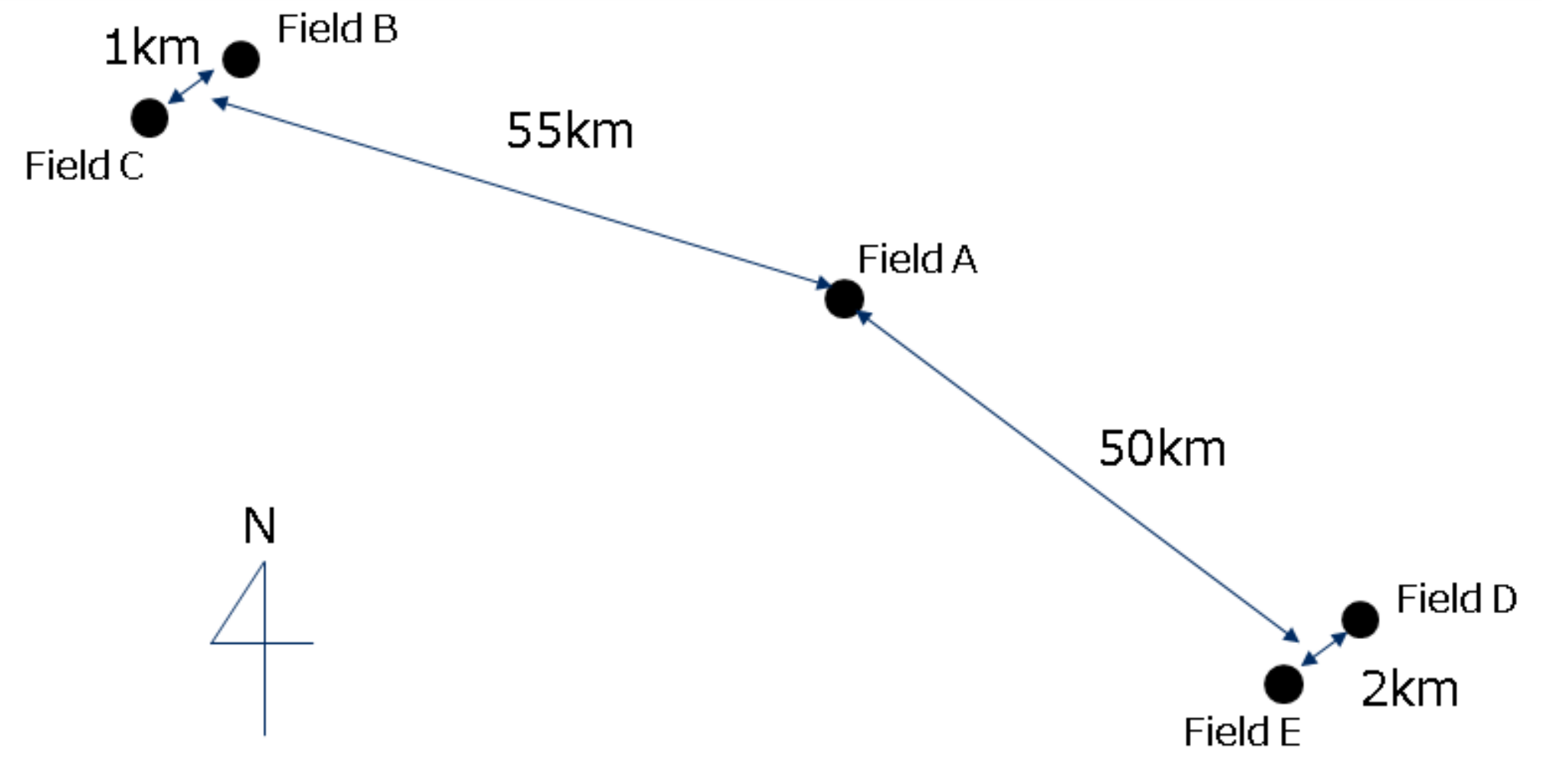}
		\caption{Relative field positions in California. Because the fields are spread across three areas, the present model can be tested using three sets of weather conditions and soil types.}
		\label{field_position}
		\end{center}
		\end{figure}

 Detailed information regarding the fields is given in Table \ref{filed_info_1} and \ref{filed_info_2}, and the relative field positions in California are shown in Figure \ref{field_position}.
		
		\begin{table}[H]
		\caption{Field information 1. Five soil samples for each depth are collected for analysis. These five fields differ somewhat in texture.}
		\begin{tabular}{p{14mm}p{17.5mm}p{17.5mm}ccccccccc} \hline
			& & & \multicolumn{9}{c}{Soil sampling result (texture) [\%]} \\ \cline{4-12}
			Field name & Irrigation depth [cm] & Soil sampling date & \multicolumn{3}{c}{10\,cm} & \multicolumn{3}{c}{30\,cm} & \multicolumn{3}{c}{Average} \\ \cline{4-12}
			& & & Clay & Silt & Sand & Clay & Silt & Sand & Clay & Silt & Sand \\ \hline
			Field A & 36 & 19-Apr-18 & 17 & 27 & 56 & 17 & 31 & 52 & 17 & 29 & 54 \\ \hline
			Field B & 36 & 20-Mar-18 & 27 & 24 & 49 & 26 & 27 & 47 & 27 & 26 & 48 \\ \hline
			Field C & 36 & 20-Mar-18 & 32 & 31 & 37 & 29 & 36 & 35 & 31 & 34 & 36 \\ \hline
			Field D & 20 & 20-Mar-18 & 14 & 22 & 64 & 11 & 24 & 65 & 13 & 23 & 65 \\ \hline
			Field E & 20 & 20-Mar-18 & 10 & 23 & 67 & 13 & 16 & 71 & 12 & 20 & 69 \\ \hline
		\end{tabular}
		\label{filed_info_1}
		\end{table}
		
		\begin{table}[H]
		\caption{Field information 2. $SRAD$ is solar radiation, $T_{max}$ is the maximum temperature that day, and $T_{min}$ is the minimum temperature that day.}
		\begin{tabular}{ccp{24.5mm}p{24.5mm}p{24.5mm}p{24.5mm}} \hline
			& & \multicolumn{4}{c}{Observed weather information for four months from planting date} \\ \cline{3-6}
			Field name & Planting date & Ave $SRAD$ [${\rm MJ}/{\rm m}^2 {\rm day}$] & Ave $T_{max}$ [$\ {}^\circ\mathrm{C}$]  & Ave $T_{min}$ [$\ {}^\circ\mathrm{C}$] & Total rain [mm] \\ \hline
			Field A & 15-Apr-18 & \hfil 27.0 \hfil & \hfil 32.3 \hfil & \hfil 15.1 \hfil & \hfil 2.0 \hfil \\ \hline
			Field B & 5-Apr-18  & \hfil 26.4 \hfil & \hfil 33.1 \hfil & \hfil 16.1 \hfil & \hfil 4.2 \hfil \\ \hline
			Field C & 5-Apr-18  & \hfil 26.4 \hfil & \hfil 33.1 \hfil & \hfil 16.1 \hfil & \hfil 4.6 \hfil \\ \hline
			Field D & 27-Mar-18 & \hfil 24.7 \hfil & \hfil 28.8 \hfil & \hfil 14.0 \hfil & \hfil 11.7 \hfil  \\ \hline
			Field E & 28-Mar-18 & \hfil 24.7 \hfil & \hfil 28.9 \hfil & \hfil 14.1 \hfil & \hfil 11.1 \hfil  \\ \hline
		\end{tabular}
		\label{filed_info_2}
		\end{table}
		
	\subsection{Parameter optimization}\label{param_opt_exp}
		\par The parameters necessary for the modified DSSAT listed in Table \ref{tune_para} were estimated through optimization using the measurement data.
		\begin{table}[H]
		\caption{The parameters in original and modified DSSAT. These parameters were selected as being either influential ones or parameters that are difficult to measure and were instead estimated using particle swarm optimazation (PSO).}
		\begin{tabular}{p{31.5mm}cp{63mm}} \hline
			& Parameter & Definition \\ \hline
			Plant growth parameters in original DSSAT & $PENV [{\rm ^oC}]$ & Average air temperature in the nursery \\ \cline{2-3}
			& $PAGE [{\rm day}]$ & Age of seedlings at transplanting \\ \cline{2-3}
			& $PH2T5$ [photothermal day] & Time from end of juvenile phase to first flowering under optimal conditions. \\ \hline
			Soil parameters in original DSSAT & $SLDR [{\rm day^{-1}}]$ & Soil water conductivity constant. \\ \cline{2-3}
			& $LL(L) [{\rm m^3\,m^{-3}}]$ & Volumetric soil moisture content in soil layer L at lower limit. \\ \cline{2-3}
			& drainage upper limit $(DUL(L)) [{\rm m^3\,m^{-3}}]$ & Volumetric soil moisture content at Drained Upper Limit in soil layer L. \\ \cline{2-3}
			& $SAT(L) [{\rm m^3\,m^{-3}}]$ & Volumetric soil moisture content in layer L at saturation. \\ \hline
			Soil parameters in additional model & $K_s(L) [{\rm cm\,day^{-1}}]$ & Saturated hydraulic conductivity in layer L. \\ \cline{2-3}
			& $\alpha [{\rm cm^{-1}}]$ & van Genuchten model parameter \\ \cline{2-3}
			& $\theta_s [{\rm m^3\,m^{-3}}]$ & van Genuchten model parameter \\ \cline{2-3}
			& $\theta_r [{\rm m^3\,m^{-3}}]$ & van Genuchten model parameter \\ \cline{2-3}
			& $n [{\rm -}]$ & van Genuchten model parameter \\ \hline
		\end{tabular}
		\label{tune_para}
		\end{table}
		\par In the original DSSAT system, the user must in principle specify the values of all parameters before simulation.
		However, despite having considerable influence on the simulation results, several of the parameters are difficult to specify or measure.
		In the present study, we introduce an additional parameter optimization system that estimates some of the key parameters and initial values using observed data obtained prior to beginning the calculations; these include initial soil moisture data and satellite images.
		Moreover, the accuracy can be improved during the season by running the parameter optimization system every day using all the data from before that day.
		In other words, the parameters are updated with all newly observed data, including soil moisture data and satellite data.
		We choose three days in a season as ``analysis days'' for which to show the calculation results in Section \ref{Calc_result} and the Appendices.
		Simulation on an analysis day (e.g., May 10) means a simulation using data obtained prior to that date to calibrate the parameters using the parameter optimization system.
		\par
\begin{figure}[H]
			\begin{center}
				\includegraphics[scale=0.6, bb=0 0 554 136]{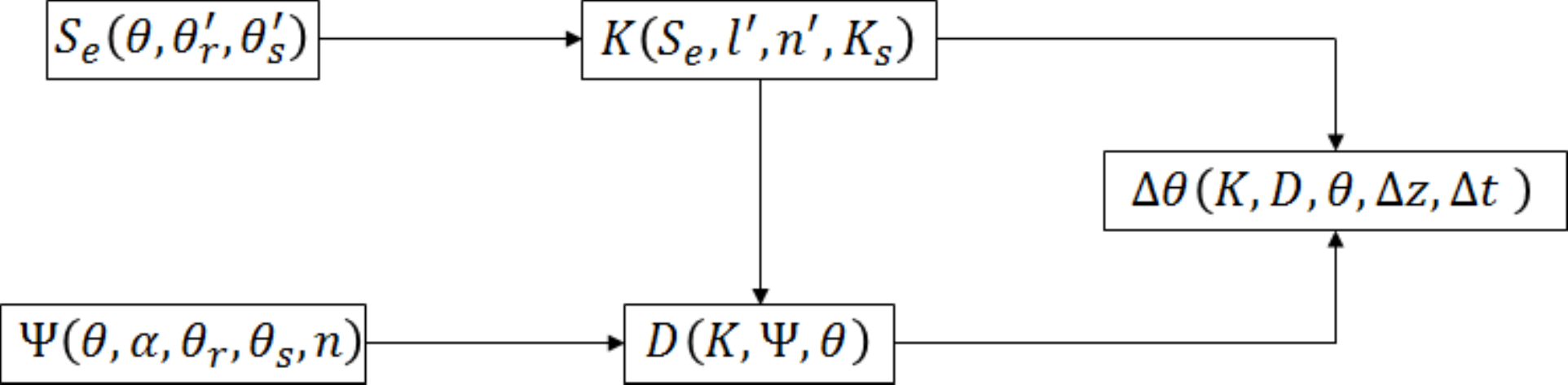}
				\caption{Flow of parameter use. A water movement $\Delta \theta$ is calculated using the hydraulic diffusivity $D$ and the hydraulic conductivity $K$. 
				$K$ builds up with the van Genuchten model parameters $l', n', \theta_r',$ and $\theta_S'$, the saturated hydraulic conductivity $K_s$, and the soil moisture $\theta$.
				Furthermore, $D$ is made up of $\alpha, \theta_r, \theta_S,$ and $n$, which are the van Genuchten model parameters $K$ and $\theta$.
				To simplify the parameter optimization, the van Genuchten model parameters used in $K$ are fixed at literature values.}
				\label{param_use_flow}
			\end{center}
		\end{figure}

 The parameter optimization system in the revised system comprises the original DSSAT parameter optimization system (OPOS) and an additional parameter optimization system (APOS).
		OPOS calculates the parameters of the original DSSAT as given in Table \ref{tune_para} using PSO, whose evaluation function comprises (i) the difference between the simulated LAI value and the estimated from satellite images from Sentinel-2 \cite{sentinel2} and (ii) the difference between simulated soil moisture value and that observed from the soil moisture sensor of five layers.
		We used the satellite data to create band WDVI data that we then converted into LAI values as observed LAI data. 
		APOS estimates the parameters of the additional model as given in Table \ref{tune_para} again using PSO with an evaluation function that uses the difference between observed data and simulated value of five layers soil moisture.
		To maintain a realistic number of estimated parameters for one PSO, we separate the parameters into OPOS and APOS.
		In addition, we execute OPOS once before APOS and once after to avoid convergence at local minima.
		In the APOS calculation, Ks and the parameters for the pF curve ($\Psi$) are estimated.
		Meanwhile, we use literature values \cite{vanGenuchten, SAKAITORIDE} for all parameters except $K_s$ for the $K(\theta)$ calculation to simplify the parameter optimization.
		Figure \ref{param_use_flow} shows a diagram of the sequence of parameter estimation.
		Literature values are used for $\theta_r', \theta_s', l', $ and $n'$ for the calculation of $K_{GM}$, whereas $\alpha, \theta_r, \theta_s,$ and $n$ for the calculation of $\Psi$ are estimated via APOS.
		The same literature values are used for the initial value in APOS.
		Regarding the initial values of $K_s$, because the literature value of $K_s$ have large dispersion, we integrated the information from six sources \cite{SAKAITORIDE, USDA_HP, SBDovey, Acknowledgements, SomnathT, BlobalSecurity} to obtain the reference values listed in Table \ref{Ks_value}.
		\begin{table}[H]
		\caption{$K_s$ value for each soil type. Six references were used to make this table. Because the values have a wide range even for a given soil type, we use the middle of the range as the initial value for parameter optimization.}
		\begin{center}
		\begin{tabular}{cc} \hline
			Soil type 		& $K_s[{\rm cm/day}]$ \\ \hline \hline
			Sand 			& $500 	- 1500$ \\ \hline
			Loamy sands 	& $350 	- 500$ \\ \hline
			Sandy loam 		& $120 	- 350$ \\ \hline
			Loam 			& $30 	- 120$ \\ \hline
			Silty loam		& $30 	- 120$ \\ \hline
			Clay loam		& $10 	- 35$ \\ \hline
			Sandy clay		& $3 	- 15$ \\ \hline
			Silty clay		& $0.1 	- 15$ \\ \hline
			Clay			& $0.1 	- 15$ \\ \hline
			Silt			& $0.5 	- 8000$ \\ \hline
		\end{tabular}
		\end{center}
		\label{Ks_value}
		\end{table}
		\par We run the parameter optimization systems (APOS and OPOS) and original/modified DSSAT simulation on three analysis days (May 10, May 30, and June 19) in the season for each of the two sensors in the five fields.
		In the analysis on each analysis day, only the data before the day of the analysis are used of the calculation processes.
		\par To assess the effects of the modified DSSAT, we also show the results given by the original DSSAT with slight modification to change the irrigation input layer.
		For fair comparison, this original DSSAT calculation also uses the same OPOS to estimate the parameters in the original DSSAT.
		\par
\begin{figure}[H]
		\begin{center}
		\includegraphics[scale=0.5, bb=0 0 994 338]{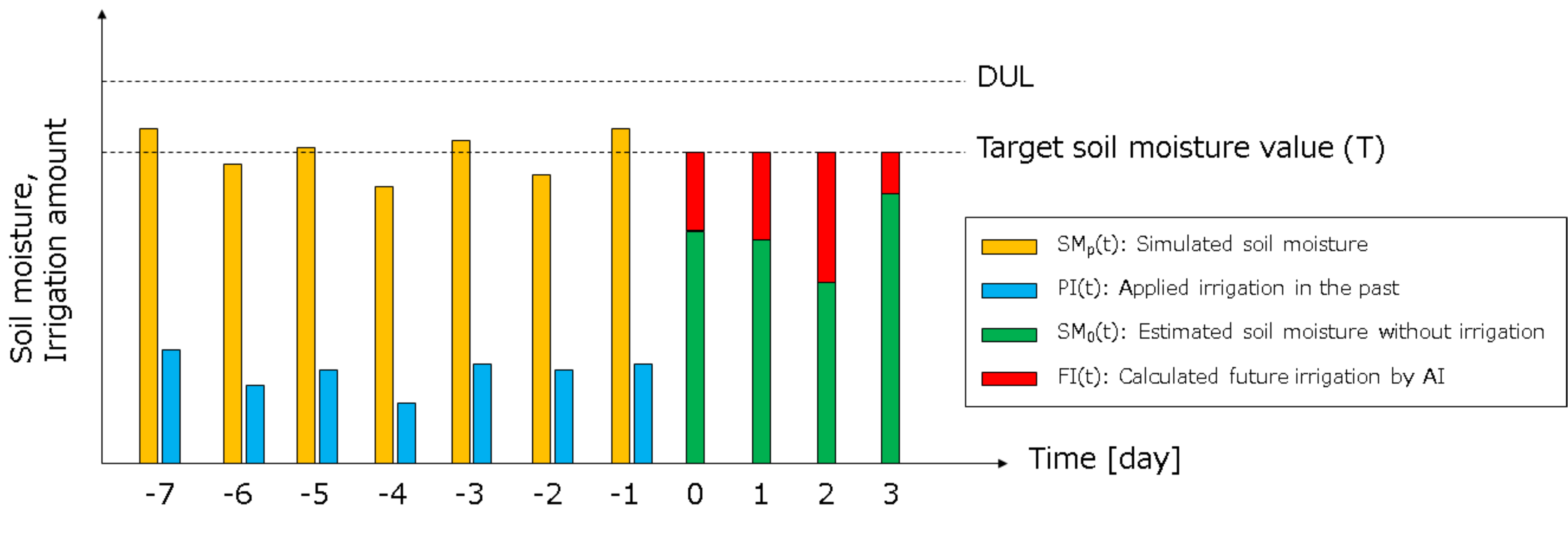}
		\caption{Schematic of method for calculating future irrigation. On the vertical axis, soil moisture and irrigation amount are expressed in the same units (e.g., cm) to allow comparison. DUL is the volumetric soil moisture content at the drained upper limit. $SM_0(0)$ is the soil moisture with irrigation on the analysis day (time $t=0$) as calculated by DSSAT. The target soil moisture $T$ is set at the sum of $SM_0(0)$ and the average of the irrigation over the past seven days ($PI(-7) \sim PI(-1)$). The calculated future irrigation $FI(t)$ is the difference between $T$ and $SM_0(t)$.}
		\label{AI_calc}
		\end{center}
		\end{figure}

 To determine the amount of irrigation subsequent to the analysis day, we use the ``automatic irrigation'' (AI) function as implemented originally in DSSAT.
		The AI function calculates the future irrigation from the differences between the simulated soil moisture values at all layers without irrigation on a future day and the target soil moisture value.
		The target value is often set as the DUL to keep enough water in the soil.
		In our case, we tune the target value by using irrigation values applied in the past.
		Figure \ref{AI_calc} shows how the target value is set and how the output future irrigation values from the analysis day (i.e., at time$t=0$) are calculated.
		First, DSSAT calculates $SM_0(0)$, which is the estimated soil moisture value at $t=0$ with no irrigation. 
		The target soil moisture value $T$ is then calculated as 
		\begin{eqnarray}
			T = SM_0(0) + \sum_{t=-7}^{-1}\frac{PI(t)}{7},
		\end{eqnarray}
		where $PI(t)$ is the irrigation amount (also including precipitation) applied in the past.
		Finally, the future irrigation $FI(t)$ is calculated as 
		\begin{eqnarray}
			FI(t) = T - SM_0(t) - R(t),
		\end{eqnarray}
		where $R(t)$ is the precipitation forecast on day $t$.
		This method of calculating the irrigation can prevent large differences from the levels of irrigation used in the recent past and can incorporate the views of the farmer if he/she determined those previous levels of irrigation.
		Because the AI function and method to tune the target value method calculate future irrigation, it does not affect the calculation results from before the analysis day that are used to compare with the observed values.
		
		\begin{figure}[H]
			\begin{center}
			\includegraphics[scale=0.5, bb=0 0 780 372]{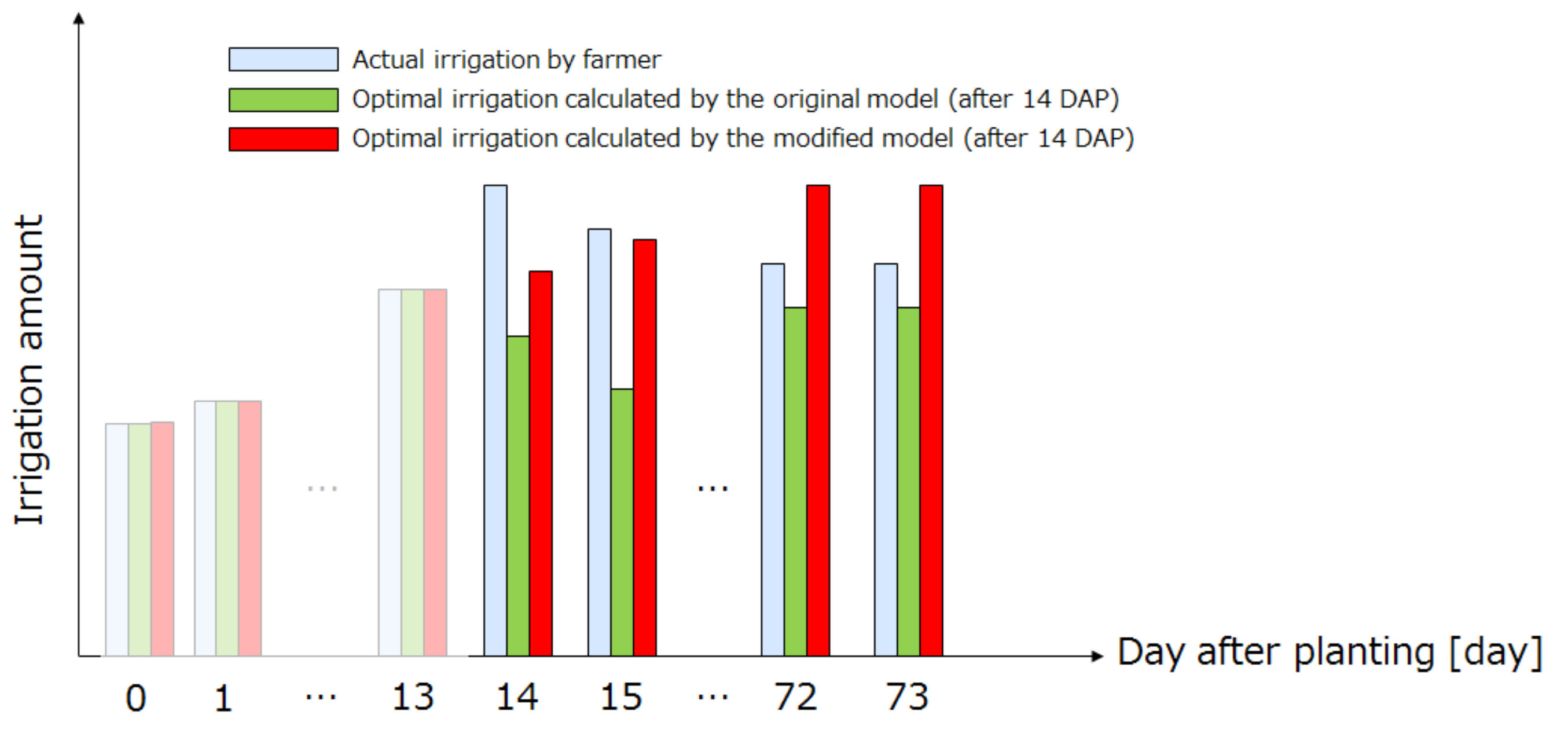}
			\caption{Conceptual figure of irrigation amount comparison calculation. Irrigation calculations by the original model and modified model are executed from the 14th day after the planting. Irrigation amount between the planting day and the 13th day is assumed same in all calculation. At the end, sum of the irrigation value of each model and actual farming between 14th day and 73th day is calculated as the total irrigation amount.}
			\label{irrigation_calc}
			\end{center}
		\end{figure}
		\par To compare the irrigation amounts calculated by the original model and the modified model, we use the method described below.
		A conceptual example is shown in Figure \ref{irrigation_calc}.
		Because irrigation control at the start of the season depends not only on plant growth but also on other farm operation such as weeding and agrochemical use, even if the original or modified DSSAT provides the farming schedule on the base of the basic growth of the plant, it cannot be performed.
		Therefore, to practically compare the models, we assume that the same irrigation is applied in the first two weeks.
		For this purpose, for all the calculations we use the actual irrigation given for 14 days from the transplant day.
		We then use each model to calculate the irrigation for 60 days after the 14th day.
		The parameters in these calculations are estimated from a simulation whose analysis day is June 19.
		In Section \ref{improved_model_result}, comparison result of the total irrigation over the 60 days given by the two models is shown.
		
	\subsection{Method for calculating errors in soil moisture simulation}
		\par In this section, we describe the method for estimating the errors due to the original DSSAT and the modified DSSAT, and we compare the two models quantitatively.
		The raw calculation results on which the quantitative comparison is based are given in the Appendices.
		\par As above, there are two sensors in a field, and we run the parameter optimization system and the models for three analysis days for each sensor.
		Accordingly, we have six sets of results for one field.
		We calculate the root mean square error (RMSE) $\epsilon$ between the simulation results and the observed data for the original DSSAT and the modified DSSAT by
		\begin{eqnarray}
			\epsilon = \sqrt{\frac{\Sigma_d\left( O_{d,l} - S_{d,l} \right)^2}{N_d}}, \label{RMSE_l}
		\end{eqnarray}
		where $\epsilon$ is the RMSE, $d$ is an index for the day, $l$ is an index for the soil layer, $O_{d,l}$ and $S_{d,l}$ are the observed and simulated values, respectively, for the given day and layer, and $N_d$ is the total number of observed days.
		In the RMSE calculation, days with no observed soil moisture data are ignored.
		The values of $\epsilon$ depend on the model ($m\!:$ ori or mod), field ($f\!: A,\cdots,E$), sensor ($S\!:$ 1 or 2), soil layer($l\!: 1,\cdots,5$), and analysis day ($A_d\!:$ May 10, May 30, or June 19).
		All of the calculated values of $\epsilon$ are given in the Appendices.
		To evaluate the improvement of the modified DSSAT compared to the original DSSAT, the following error $E$ is calculated:
		\begin{eqnarray}
			E_{m,f} = \frac{\Sigma_S \Sigma_{A_d} \Sigma_l \epsilon_{m, f, S, l, A_d}}{N_l \cdot N_{A_d} \cdot N_S}, \label{RMSE_total}
		\end{eqnarray}
		where $N_l$ is the total number of layers used in the calculation, $N_{A_d}$ is the total number of analysis days (i.e., three) and $N_S$ is the total number of sensors (i.e., two).
		
\clearpage

\section{Results}\label{Calc_result}
	\subsection{Simulations using original DSSAT model}\label{calc_by_ori}
		\par
\begin{center}
		\begin{figure}[H]
			\raisebox{5em}{
			\begin{minipage}{0.3\hsize}
			\begin{center}
				\includegraphics[width=50mm, bb=0 0 216 151]{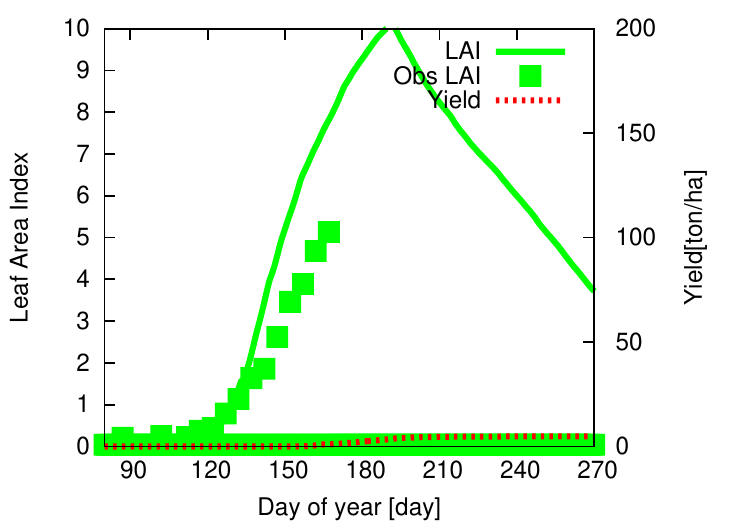}
			\end{center}
			\end{minipage}
			\begin{minipage}{0.3\hsize}
			\begin{center}
				\includegraphics[width=50mm, bb=0 0 216 151]{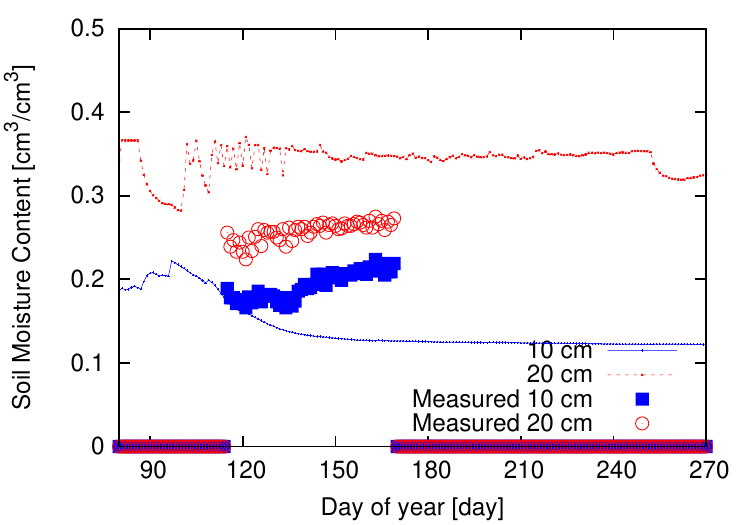}
			\end{center}
			\end{minipage}
			\begin{minipage}{0.3\hsize}
			\begin{center}
				\includegraphics[width=50mm, bb=0 0 216 151]{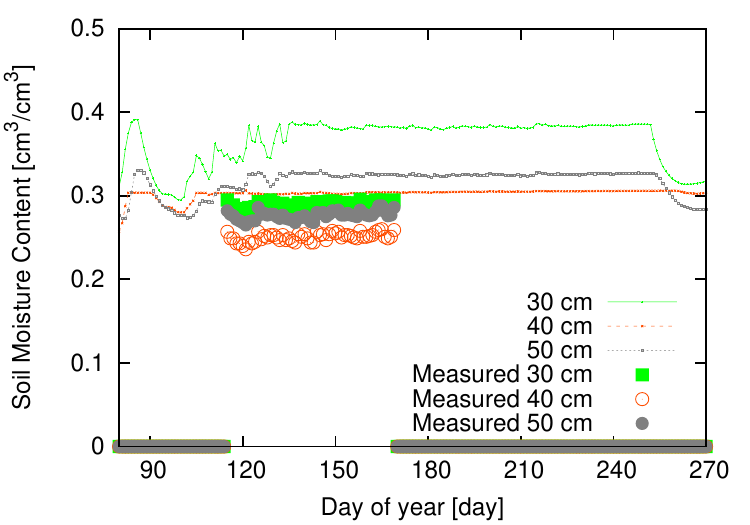}
			\end{center}
			\end{minipage}
			}
			\caption{Simulation results obtained using original DSSAT (field E, June 19). 
					Left: observed and simulated LAI and simulated yield (total weight of fruit).
					Middle: observed and simulated soil moisture in $10\,{\rm cm}$ and $20\,{\rm cm}$ layers.
					Right: observed and simulated soil moisture in $30\,{\rm cm}$, $40\,{\rm cm}$, and $50\,{\rm cm}$ layers.
					There are large difference between the simulation and observed data, and the calculated yield is very low.
					These results indicate that the model could not simulate the plant and soil behavior.}
			\label{Field_E_old_model}
		\end{figure}
		\end{center}

 An example of the simulation results obtained using the original DSSAT is shown in Figure \ref{Field_E_old_model}.
		These results are for field E on June 19 as the analysis day.
		There are clearly large differences between the observed and simulated soil moisture in all the layers: the average difference among the five layers is $0.068\,{\rm m^3\,m^{-3}}$.
		In particular, the simulated soil moisture in the 10\,cm layer decreases at the beginning of the season, whereas the observed value remained almost constant.
		This implies that the original DSSAT model does not reproduce the upward movement of water from the 20-cm-deep irrigation pipe well.
		Consequently, the original model underestimates leaf area and total crop weight.
		The actual yield of field E exceeds $100\,{\rm ton\,ha^{-1}}$, but the simulation fails to produce any fruit.
		
	\subsection{Improved soil moisture simulation using modified DSSAT}\label{improved_model_result}
		\par
\begin{figure}[H]
			\begin{center}
			\raisebox{5em}{
			\begin{minipage}{0.3\hsize}
			\begin{center}
				\includegraphics[width=50mm, bb=0 0 216 151]{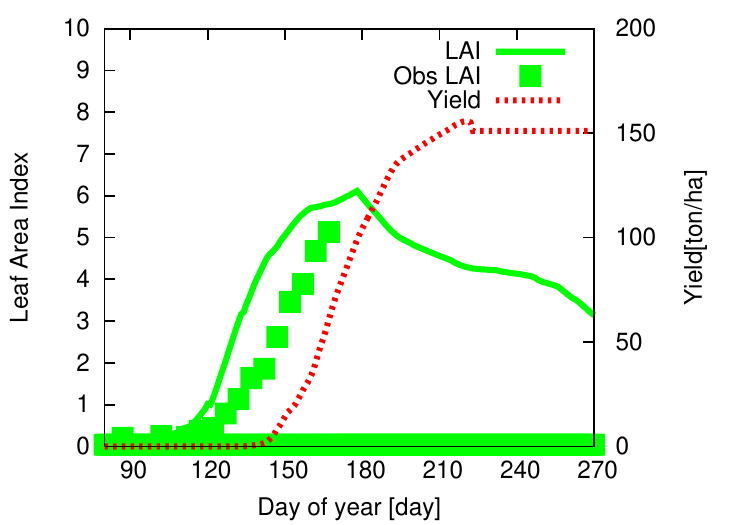}
			\end{center}
			\end{minipage}
			\begin{minipage}{0.3\hsize}
			\begin{center}
				\includegraphics[width=50mm, bb=0 0 216 151]{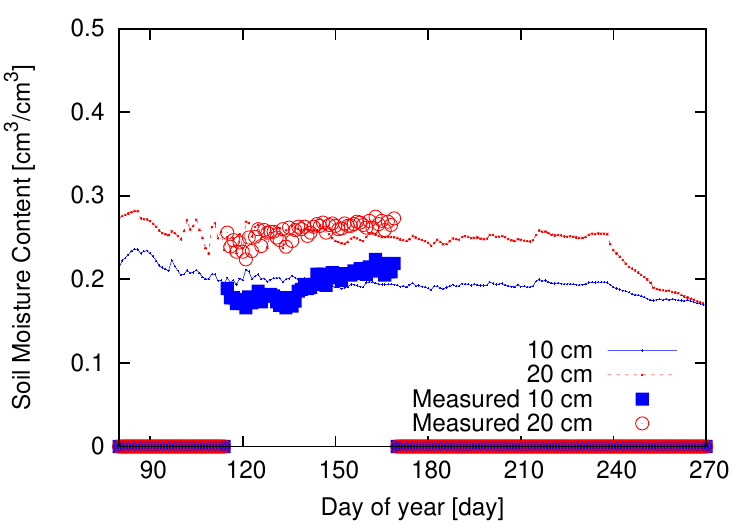}
			\end{center}
			\end{minipage}
			\begin{minipage}{0.3\hsize}
			\begin{center}
				\includegraphics[width=50mm, bb=0 0 216 151]{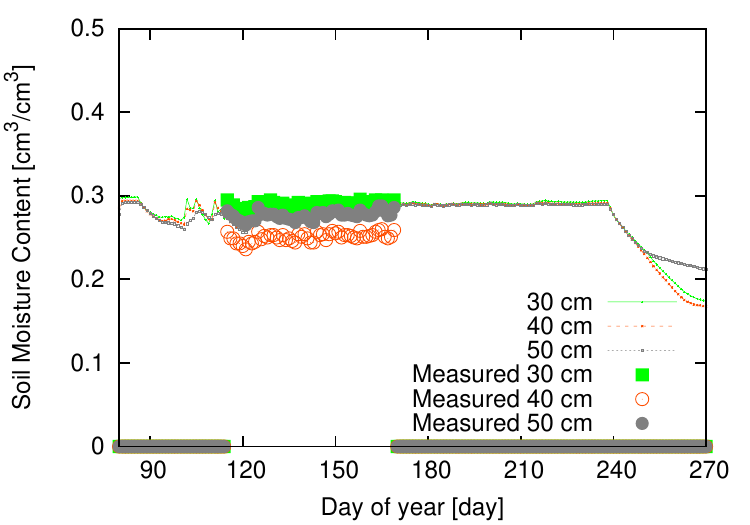}
			\end{center}
			\end{minipage}
			}
			\caption{Simulation results obtained using modified DSSAT (field E, June 19). 
					Left: observed and simulated LAI and simulated yield (total weight of fruit).
					Middle: observed and simulated soil moisture in 10\,cm and 20\,cm layers.
					Right: observed and simulated soil moisture in 30\,cm, 40\,cm, and 50\,cm layers.
					The errors between the simulation and observed data are much smaller than those for the original DSSAT.
					Furthermore, the calculated yield of $150\,{\rm ton\,ha^{-1}}$ is a reasonable value for the California area.}
			\label{Field_E_new_model}
			\end{center}
		\end{figure}

\begin{figure}[H]
			\centering
			\includegraphics[scale=1.0, bb=0 0 360 176]{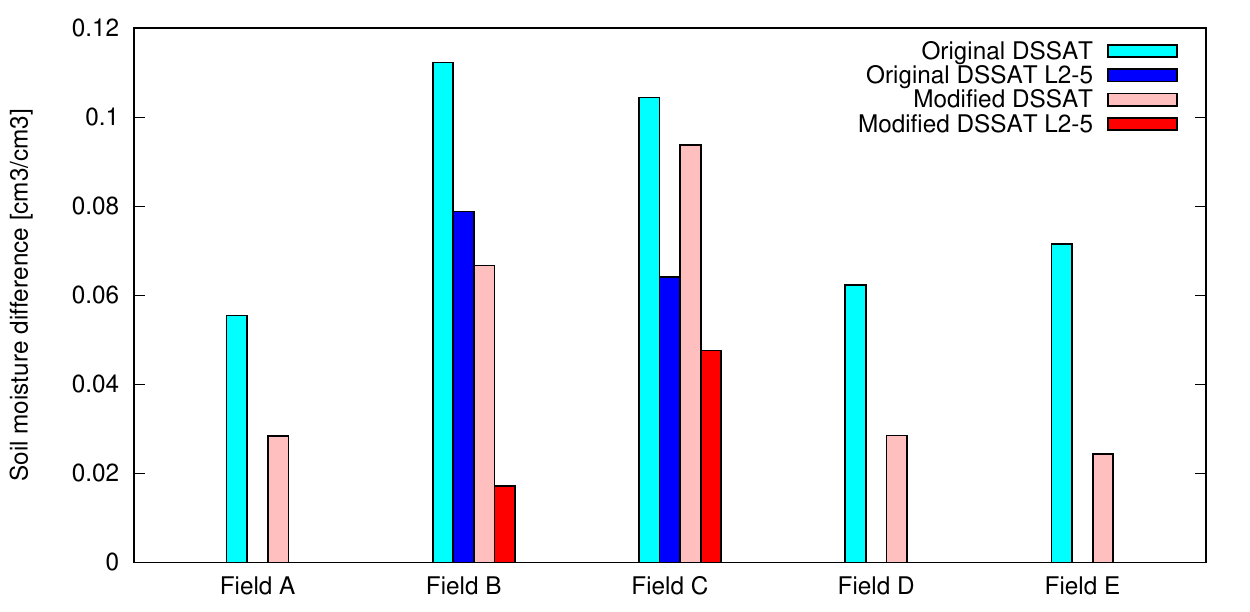}
			\caption{Soil moisture differences between original model and modified model: $E_{m,f}$.
					``Original DSSAT'' and ``Modified DSSAT'' are the results obtained by simulation using all the observed soil moisture data. ``Original DSSAT L2-5'' and ``Modified DSSAT L2-5'' are the results without data at 10\,cm in fields B and C. The ignored data for these fields contain large errors because of the high clay content of the soil.
					The results with the modified DSSAT are better than those with the original DSSAT in each case.}
			\label{accu_all}
		\end{figure}

\begin{figure}[H]
			\begin{center}
			\includegraphics[scale=1.0, bb=0 0 360 176]{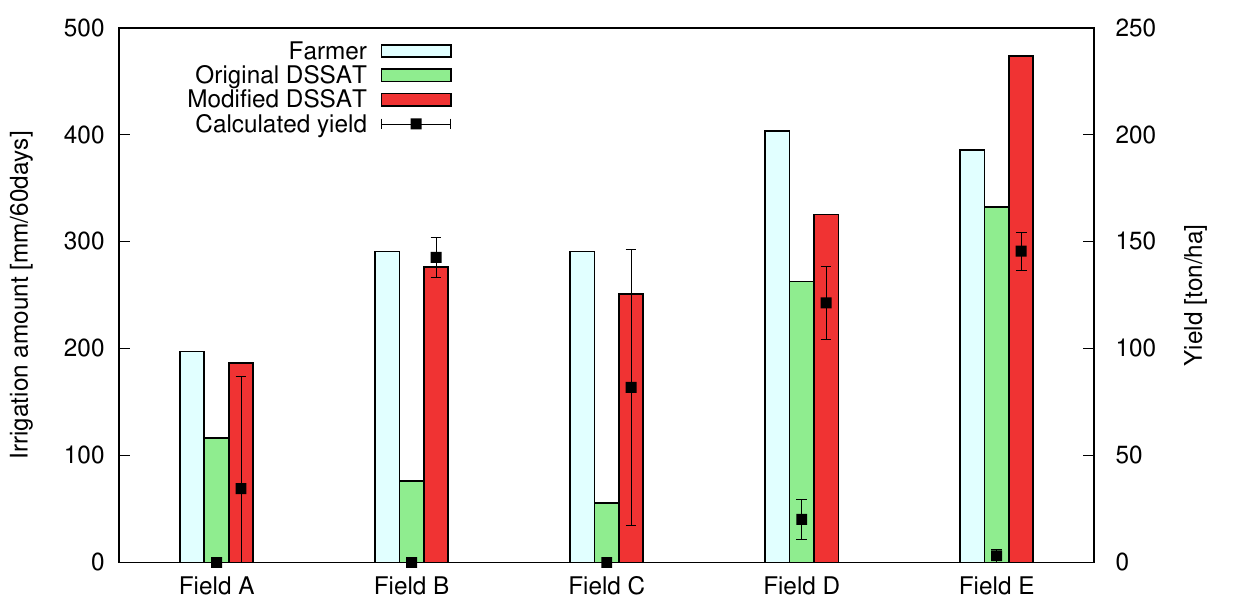}
			\caption{Comparison of irrigation amounts for 60 days in a season among farmer allocation, original DSSAT, and modified DSSAT with calculated average yield.
			Whereas the irrigation with the original DSSAT is much lower than that allocated by the farmer, the modified DSSAT calculates an amount that is comparable with that allocated by the farmer.
			In addition, the yields calculated by the modified DSSAT are reasonable.}
			\label{irri_comp_1}
			\end{center}
		\end{figure}

 The modified DSSAT model with the same input data as in Section \ref{calc_by_ori} results in very different simulation results (Figure \ref{Field_E_new_model}) from those obtained using the original DSSAT model (Figure \ref{Field_E_old_model}).
		The difference in soil moisture between the measured data and the simulation results is $0.018\,{\rm m^3\,m^{-3}}$, which is much smaller than the difference of $0.068\,{\rm m^3\,m^{-3}}$ in the case of the original DSSAT model as shown above.
		Note that the simulation yield increases to $150\,{\rm ton}\,{\rm ha}^{-1}$, which is close to the area's average yield $120\,{\rm ton}\,{\rm ha}^{-1}$ \cite{tomato_yield}.
		The quantitative comparison results using all the soil layers ($N_l=5, l = 1,\cdots,5$) are shown in Figure \ref{accu_all} as ``Original DSSAT'' and ``Modified DSSAT''.
		According to these results, the errors in the soil moisture simulation are suppressed for all the fields.
		The average error $\Sigma_f E_{m,f}$ for all the fields decreases from $0.081\,{\rm m}^3\,{\rm m}^{-3}$ to $0.048\,{\rm m}^3\,{\rm m}^{-3}$.
		The errors for fields B and C, which are high clay fields, are larger than those for the other fields regardless of the model used.
		We found that this is due to some problems with the soil moisture sensors because the measured soil moisture data for the 10\,cm layer are very low compared to those of 20\,cm layer.
		For this reason, we redo the calculations for those two fields without using the data for the 10\,cm layer.
		These results are also shown in Figure \ref{accu_all}, and they show the advantage of the modified DSSAT over the original DSSAT for all the fields.
		The all-field average error $\Sigma_f E_{m,f}$ decreases from $0.065\,{\rm m}^3\,{\rm m}^{-3}$ to $0.029\,{\rm m}^3\,{\rm m}^{-3}$.
		The results for the calculated irrigation are compared in Figure \ref{irri_comp_1}.
		The amount of irrigation calculated by the original DSSAT is very much lower than the actual irrigation because the plants in the model could not grow well under the poor irrigation input caused by unrealistic soil moisture.
		By contrast, the irrigation amount calculated by the modified DSSAT is similar to the actual value.
		Furthermore, four of the five fields have lower calculated irrigation amounts relative to the actual irrigation.
		This indicates the possibility of reducing the amount of irrigation by using this model.
		\par The calculated yield is also shown in Figure \ref{irri_comp_1}.
		The average tomato yield in this area is $-120\,{\rm ton\,ha^{-1}}$ \cite{tomato_yield}, whereas the original DSSAT calculates yields that do not exceed $50\,{\rm ton\,ha^{-1}}$.
		This indicates that there is a large discrepancy between actual fields and the model.
		By contrast, 24 of 30 analyses (= 80\,\%) using the modified DSSAT results in more than $50\,{\rm ton\,ha^{-1}}$ (see Appendix \ref{all_result_of_new}).
		In particular, all of the calculated yields for June 19 exceed $50\,{\rm ton\,ha^{-1}}$, which means that the accuracy of the simulation analysis increases with an increasing amount of observation data.
		\par As shown in Figure \ref{accu_all}, the soil moisture values simulated by the modified DSSAT are closer to the measured data than are those the original DSSAT in each case.
		Moreover, the calculated yield, which was almost zero with the original DSSAT, is realistic with the modified DSSAT.
		Consequently, we have improved the water balance calculation in DSSAT, especially for fields with subsurface irrigation.

\clearpage

\section{Discussion}
	\par The original DSSAT model is not developed for the irrigation tubes located underground and it performs poorly in the subsurface irrigation system.
	However, subsurface irrigation systems are used widely around the world for tomatoes to save water.
	Therefore, we modified the model by not only changing the irrigation position but also adding a soil water movement model driven by the soil moisture gradient to improve the calculation of water movement in the soil of fields with subsurface irrigation.
	We collected basic field data, satellite data, soil moisture data, weather data, and farming data for five tomato fields in California.
	The parameters used in the model were estimated by PSO to suppress the differences between the simulation and observed data; the modified DSSAT halved the errors in the soil moisture compared to the original model.
	Consequently, the simulated crop yield, which was almost zero with the original model, became realistic and close to the average yield for the studied area, namely $120\,{\rm ton\,ha^{-1}}$ \cite{tomato_yield}.
	\par The present study led to the following improvements for simulations with much higher accuracy.
	\par 1. Improved coherence of soil parameters.
	To simplify the parameter optimization and reduce the calculation cost, we used literature values for $\theta'_r, \theta'_S,$ and $n'$ to optimize $K_{GM}$, which were not matching with the $\theta_r, \theta_S,$ and $n$ values used in the pF curve ($\Psi$).
	Even the effect can be ignored for this first trial modified model because the initial values for $\Phi$ optimization are the same as those used in the $K_{GM}$ optimization.
	If we can estimate $K_{GM}$ and $\Phi$ at the same time using the same parameters, then the parameters can be closer to the true values.
	\par 2. Improved measurement of soil moisture in the 10\,cm layer in high clay fields.
	We could not use the soil moisture data for the 10\,cm layer for parameter optimization in the present high clay fields because we found that the measured soil moisture data were not realistic for those fields.
	Sensor calibration or improved sensor installation would resolve this issue and improve the simulation accuracy.
	\par 3. Improved accuracy of water input timing.
	Irrigation and precipitation were assumed to be input at midnight in the calculations.
	However, if we could use timed information about irrigation and precipitation, then we could input that information into the simulation at the same times as in reality.
	Precise flow meters and rain gauges are needed for this improvement.
	\par In the present study, we compared the irrigation amounts calculated by the original and modified models with the actual amounts in five fields.
	The most effective irrigation reduction was that in the calculation for field D, whose irrigation amount calculated by the modified model was 19\% less than that allocated by the farmer.
	Although one of the total irrigation amounts calculated by the modified model was more than that allocated by tha farmer, for four of the five fields the modified model calculated a lower amount of irrigation.
	This suggests the possibility of saving water by using the model to control the irrigation amount.
	Enhancing the simulation accuracy would be a realistic approach to saving water in agriculture.
	\par Our aim in the present study was to make a practical system by improving the accuracy while using minimal data.
	From the results, our new system achieved sufficiently high accuracy to be applied in practice to actual fields with subsurface irrigation because of the new soil moisture movement logic and the calibration with observed data.
	In future work, we will develop an even newer system by using multi-seasonal data besides the improvements herein to improve the accuracy of simulations and irrigation calculations.

\clearpage

\appendix
\section{All analysis results from original DSSAT}
\par All of the analysis results obtained using the original DSSAT are presented in this section.
There were two soil moisture sensors (S1 and S2) in each field, and we chose three analysis days for each sensor.
In the each analysis, only the data collected before the day of the analysis are used for the calculation process to assess the application potency of the model in a particular season.
In each group of three graphs, the one on the left shows the LAI and the yield (total fruit weight), the one in the middle shows the soil moisture in the 10\,cm and 20\,cm layers, and the one on the right shows the soil moisture in the 30\,cm, 40\,cm, and 50\,cm layers.
Lines show the results calculated by the model, and points show the observed data.
Most of the results show unnatural growth (either very high/low LAI or very low yield), behavior that is inconsistent with the observed data, or both.
	\begin{table}[H]
		\caption{Field A}
		\begin{tabular}{ccc}
			\hline & S1 & S2 \\
			\hline 10th May  & \begin{minipage}{70mm}
									\centering
									\scalebox{0.29}{\includegraphics[scale=1.0, bb=0 0 216 151]{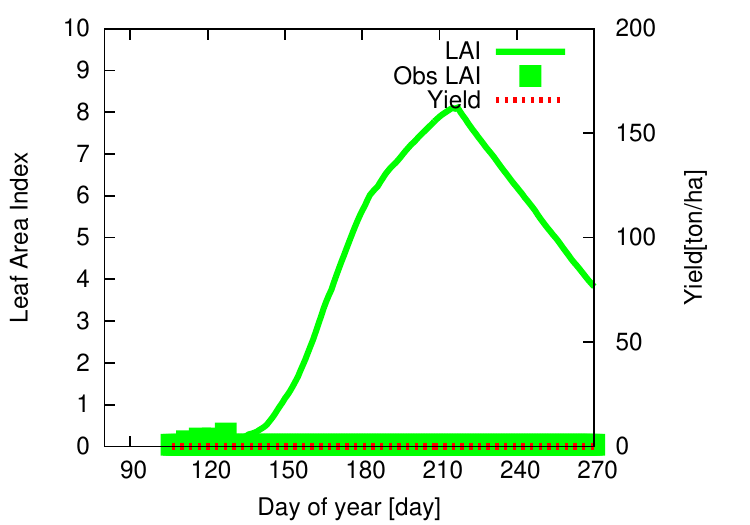}}
									\scalebox{0.29}{\includegraphics[scale=1.0, bb=0 0 216 151]{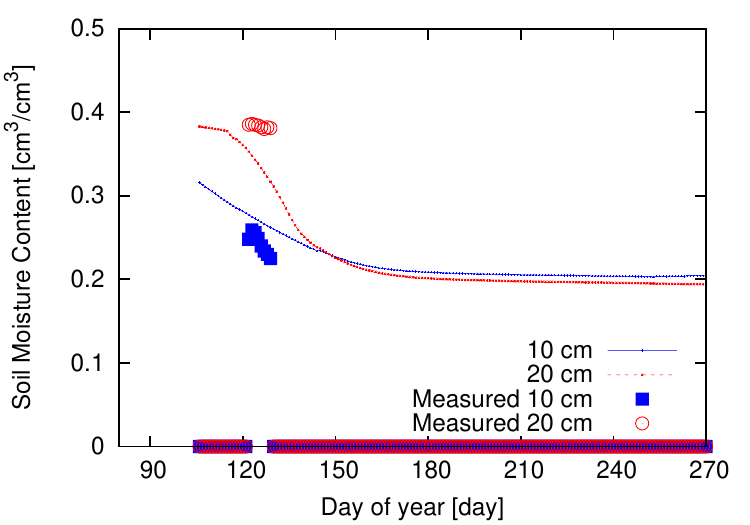}}
									\scalebox{0.29}{\includegraphics[scale=1.0, bb=0 0 216 151]{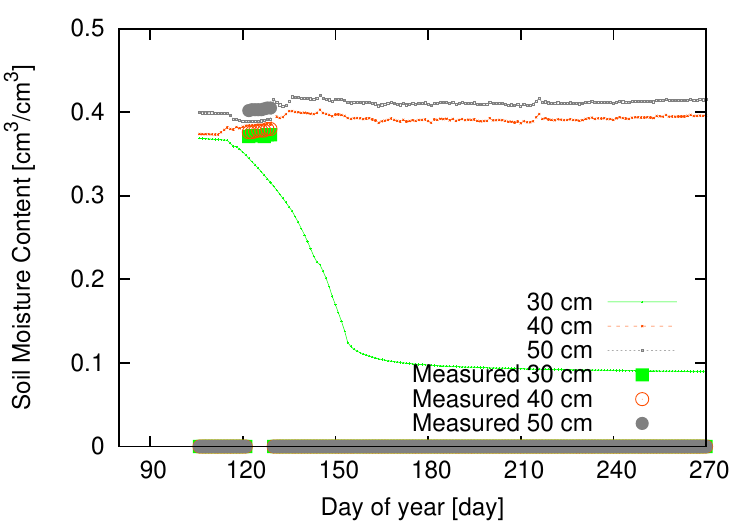}}
								\end{minipage} &
								\begin{minipage}{70mm}
									\centering
									\scalebox{0.29}{\includegraphics[scale=1.0, bb=0 0 216 151]{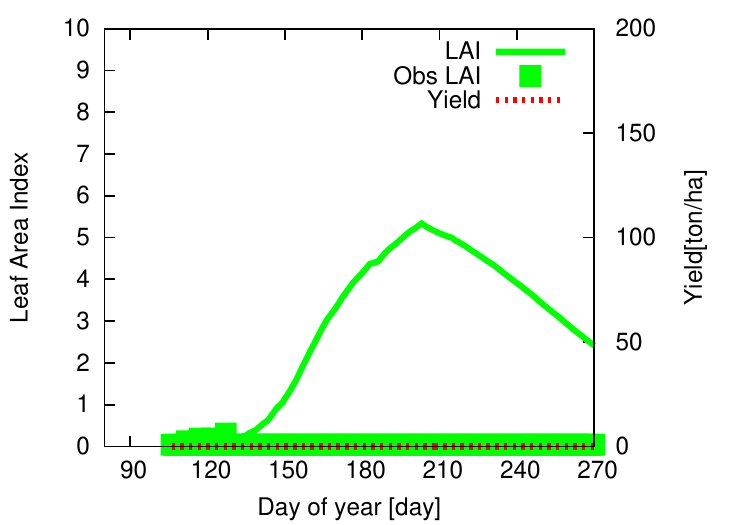}}
									\scalebox{0.29}{\includegraphics[scale=1.0, bb=0 0 216 151]{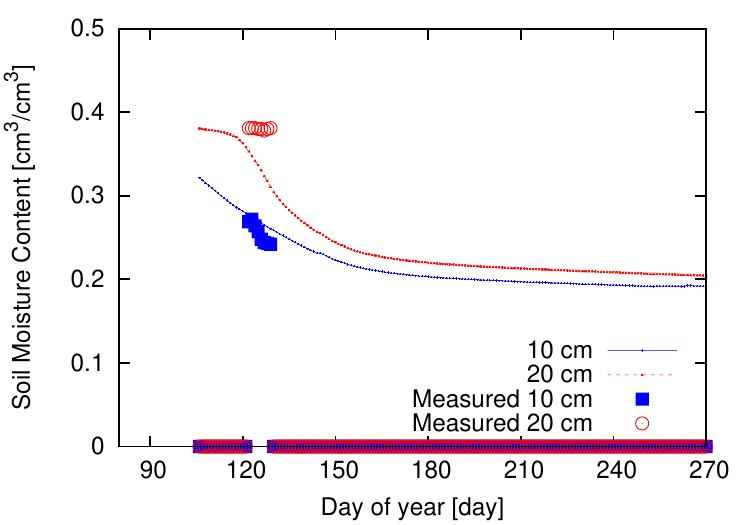}}
									\scalebox{0.29}{\includegraphics[scale=1.0, bb=0 0 216 151]{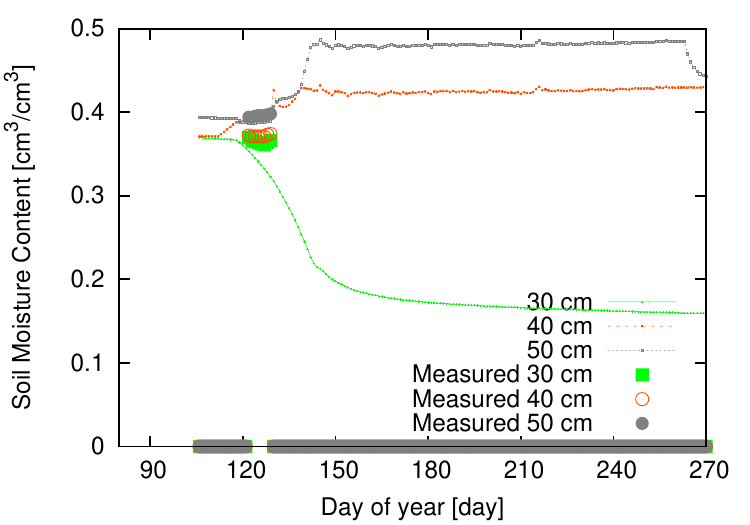}}
								\end{minipage} \\
			\hline 30th May  & \begin{minipage}{70mm}
									\centering
									\scalebox{0.29}{\includegraphics[scale=1.0, bb=0 0 216 151]{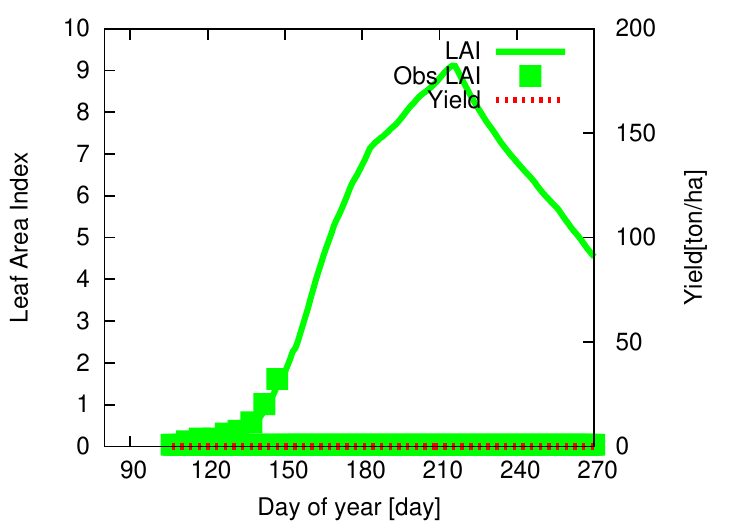}}
									\scalebox{0.29}{\includegraphics[scale=1.0, bb=0 0 216 151]{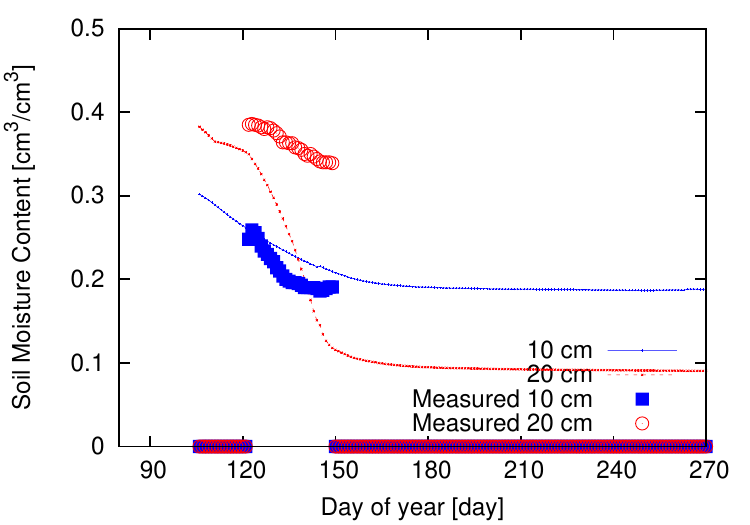}}
									\scalebox{0.29}{\includegraphics[scale=1.0, bb=0 0 216 151]{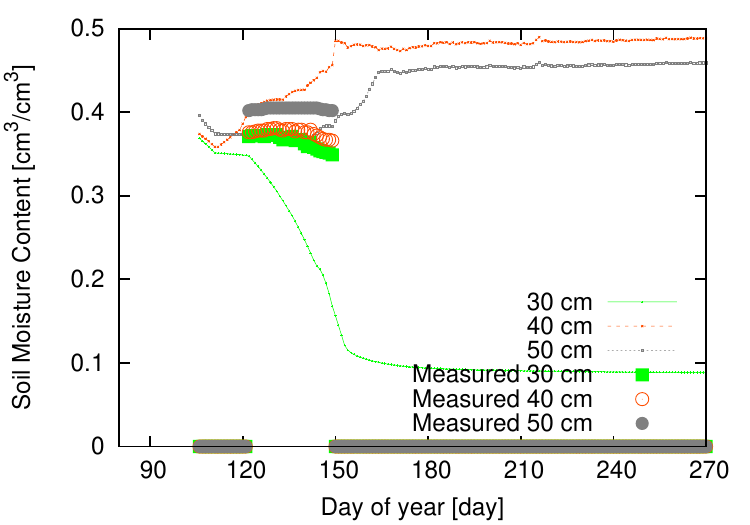}}
								\end{minipage} &
								\begin{minipage}{70mm}
									\centering
									\scalebox{0.29}{\includegraphics[scale=1.0, bb=0 0 216 151]{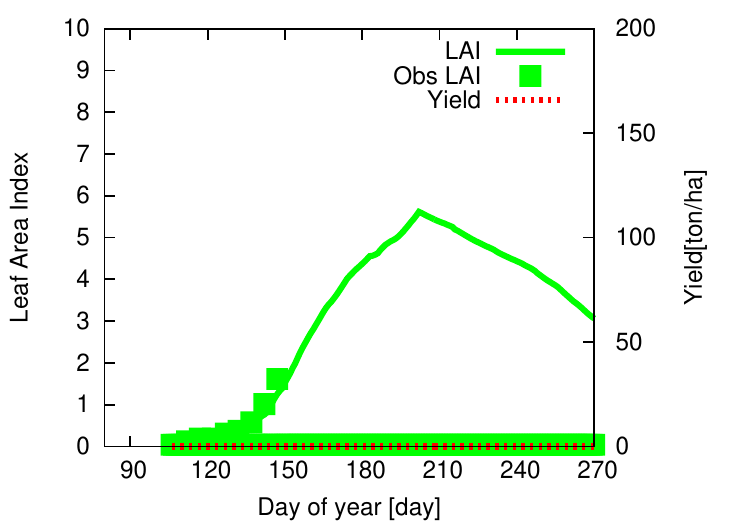}}
									\scalebox{0.29}{\includegraphics[scale=1.0, bb=0 0 216 151]{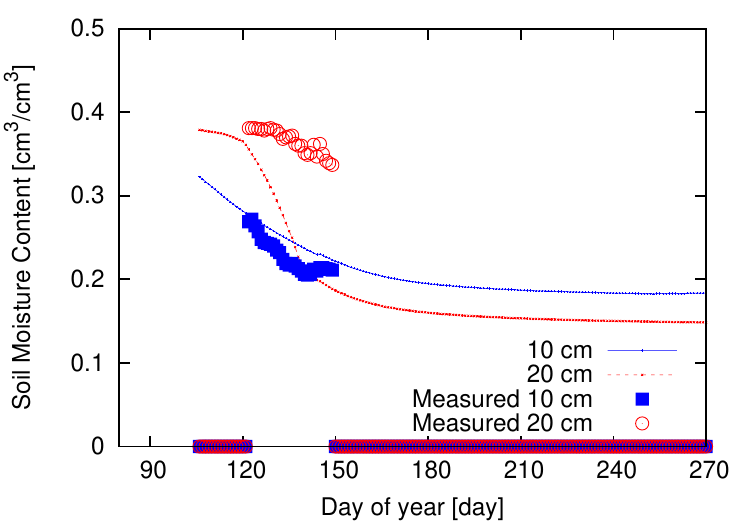}}
									\scalebox{0.29}{\includegraphics[scale=1.0, bb=0 0 216 151]{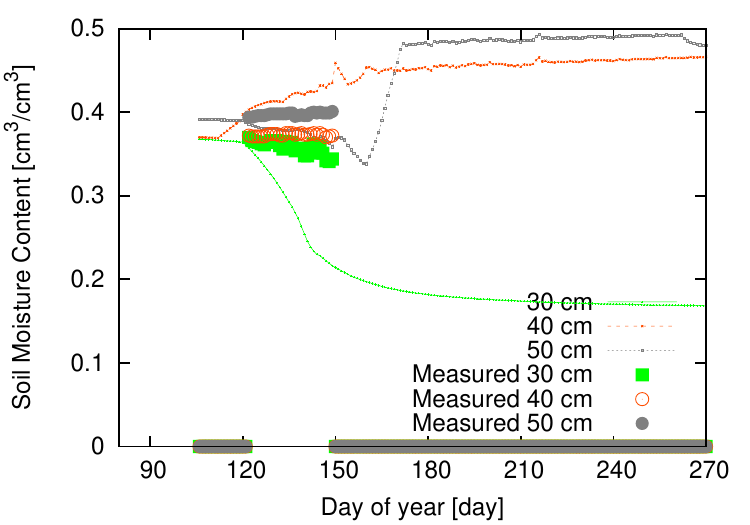}}
								\end{minipage} \\
			\hline 19th June & \begin{minipage}{70mm}
									\centering
									\scalebox{0.29}{\includegraphics[scale=1.0, bb=0 0 216 151]{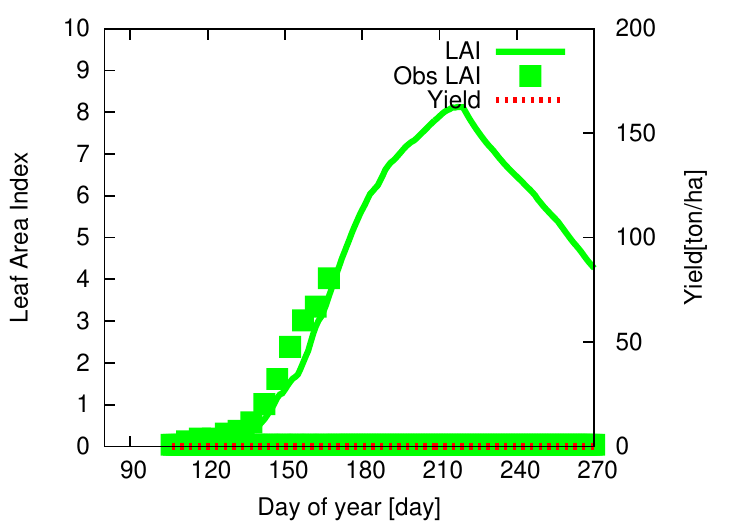}}
									\scalebox{0.29}{\includegraphics[scale=1.0, bb=0 0 216 151]{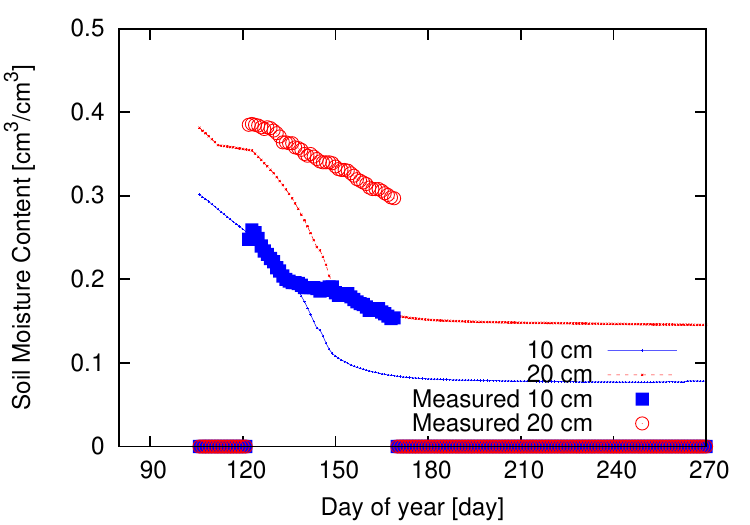}}
									\scalebox{0.29}{\includegraphics[scale=1.0, bb=0 0 216 151]{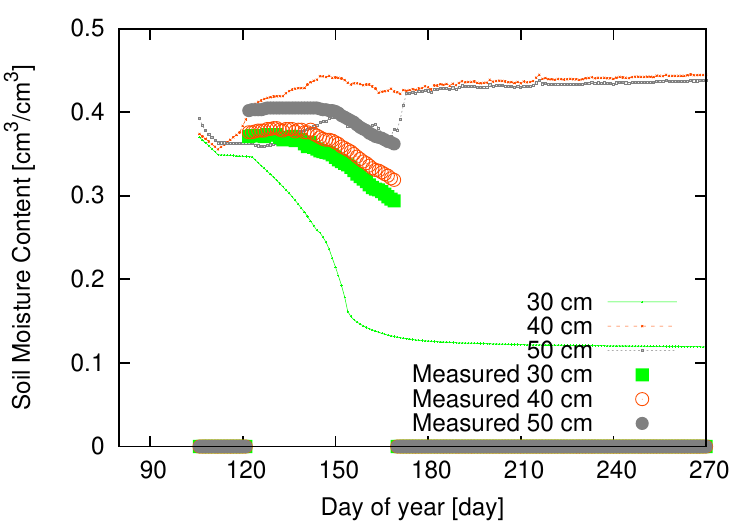}}
								\end{minipage} &
								\begin{minipage}{70mm}
									\centering
									\scalebox{0.29}{\includegraphics[scale=1.0, bb=0 0 216 151]{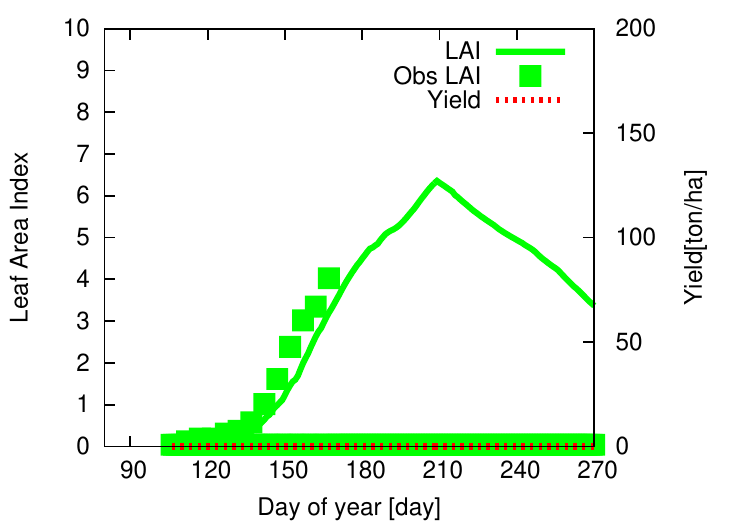}}
									\scalebox{0.29}{\includegraphics[scale=1.0, bb=0 0 216 151]{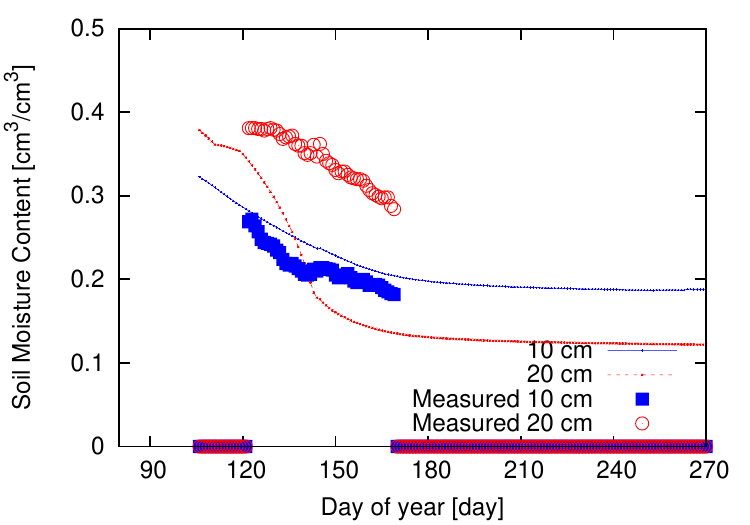}}
									\scalebox{0.29}{\includegraphics[scale=1.0, bb=0 0 216 151]{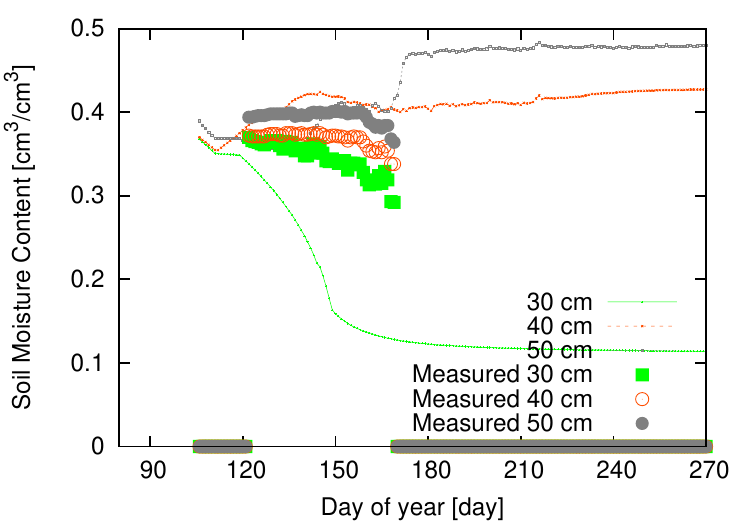}}
								\end{minipage} \\ \hline
		\end{tabular}
	\end{table}
	
	\begin{table}[H]
		\caption{Field B}
		\begin{tabular}{ccc}
			\hline & S1 & S2 \\
			\hline 10th May  & \begin{minipage}{70mm}
									\centering
									\scalebox{0.29}{\includegraphics[scale=1.0, bb=0 0 216 151]{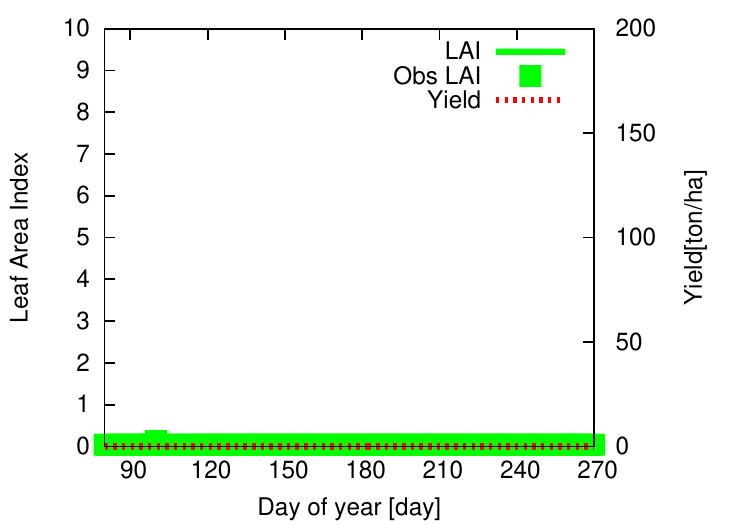}}
									\scalebox{0.29}{\includegraphics[scale=1.0, bb=0 0 216 151]{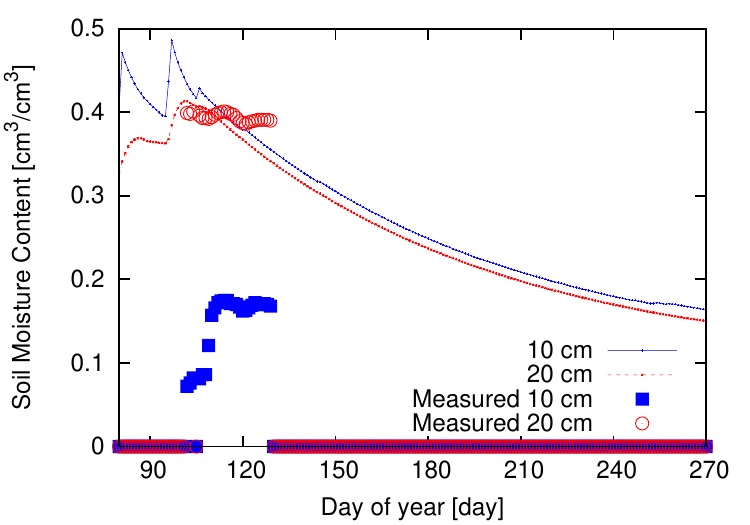}}
									\scalebox{0.29}{\includegraphics[scale=1.0, bb=0 0 216 151]{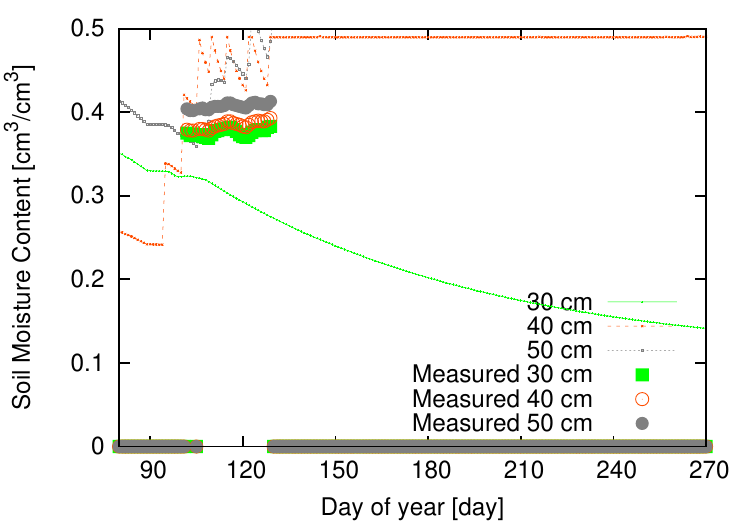}}
								\end{minipage} &
								\begin{minipage}{70mm}
									\centering
									\scalebox{0.29}{\includegraphics[scale=1.0, bb=0 0 216 151]{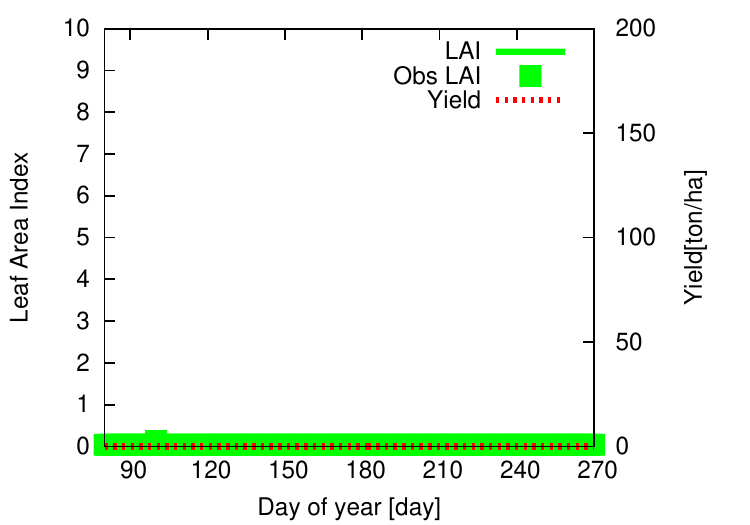}}
									\scalebox{0.29}{\includegraphics[scale=1.0, bb=0 0 216 151]{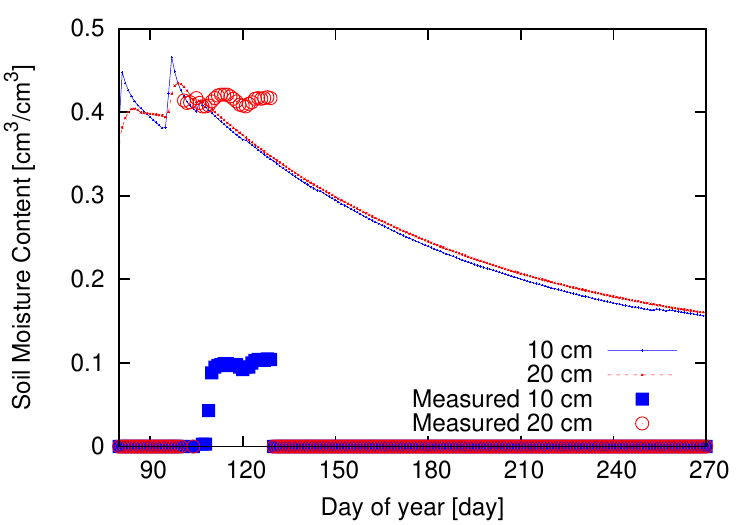}}
									\scalebox{0.29}{\includegraphics[scale=1.0, bb=0 0 216 151]{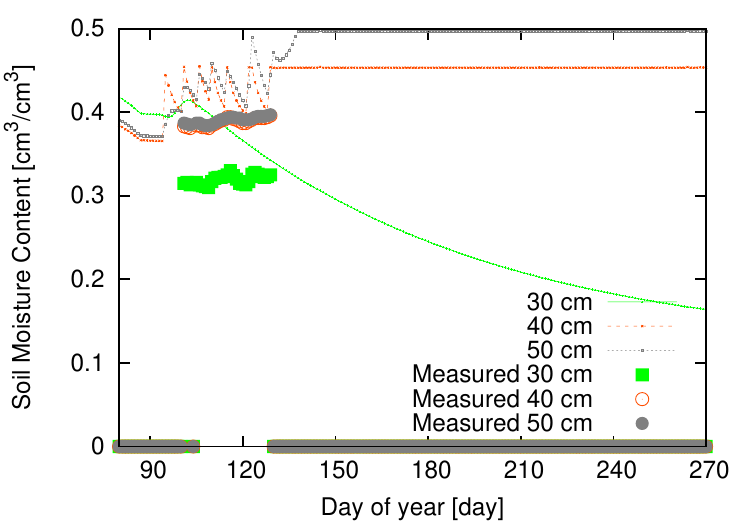}}
								\end{minipage} \\
			\hline 30th May  & \begin{minipage}{70mm}
									\centering
									\scalebox{0.29}{\includegraphics[scale=1.0, bb=0 0 216 151]{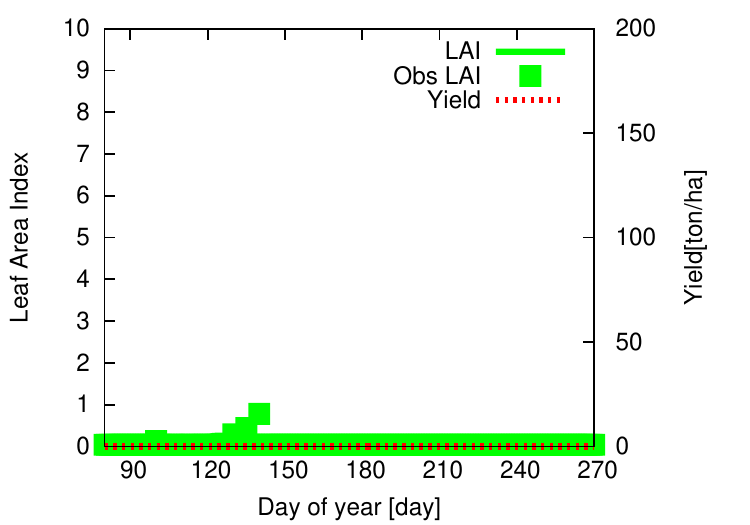}}
									\scalebox{0.29}{\includegraphics[scale=1.0, bb=0 0 216 151]{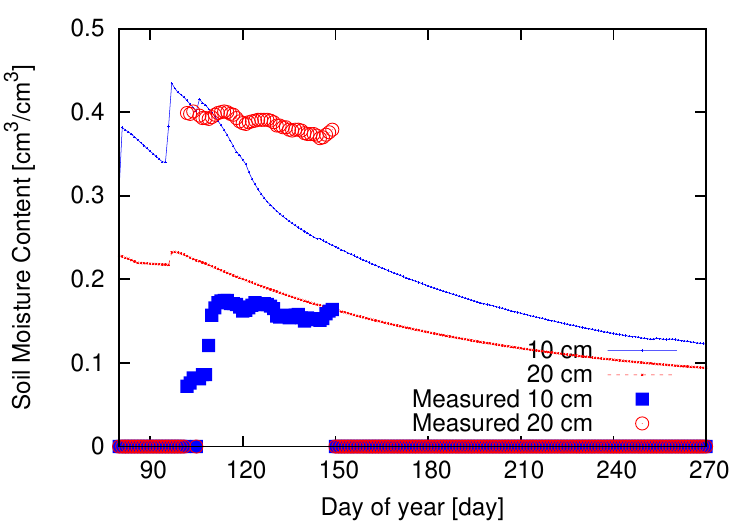}}
									\scalebox{0.29}{\includegraphics[scale=1.0, bb=0 0 216 151]{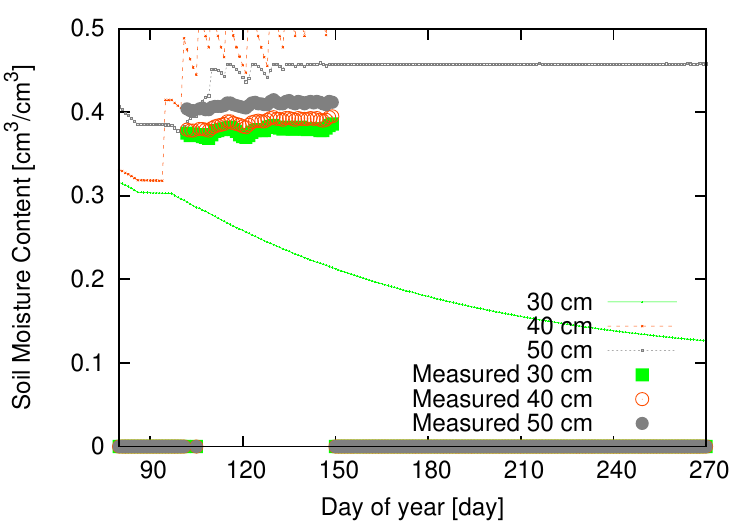}}
								\end{minipage} &
								\begin{minipage}{70mm}
									\centering
									\scalebox{0.29}{\includegraphics[scale=1.0, bb=0 0 216 151]{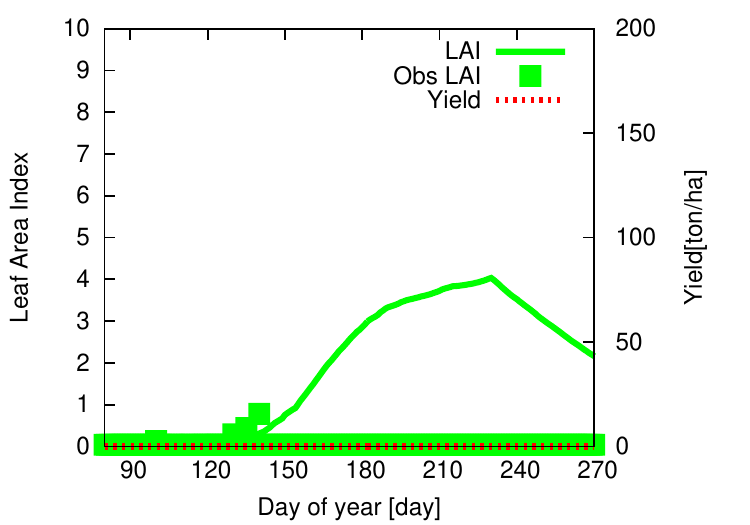}}
									\scalebox{0.29}{\includegraphics[scale=1.0, bb=0 0 216 151]{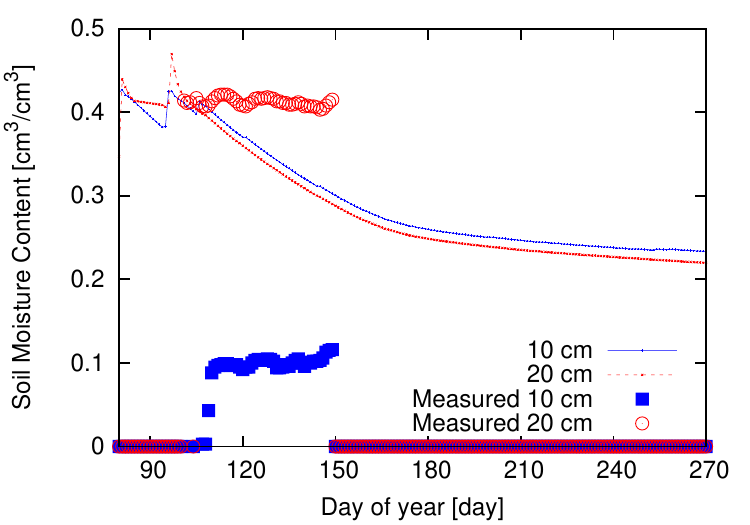}}
									\scalebox{0.29}{\includegraphics[scale=1.0, bb=0 0 216 151]{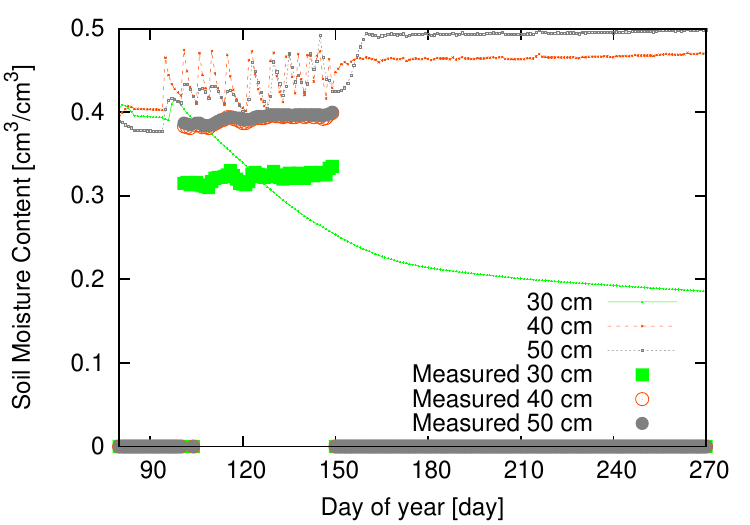}}
								\end{minipage} \\
			\hline 19th June & \begin{minipage}{70mm}
									\centering
									\scalebox{0.29}{\includegraphics[scale=1.0, bb=0 0 216 151]{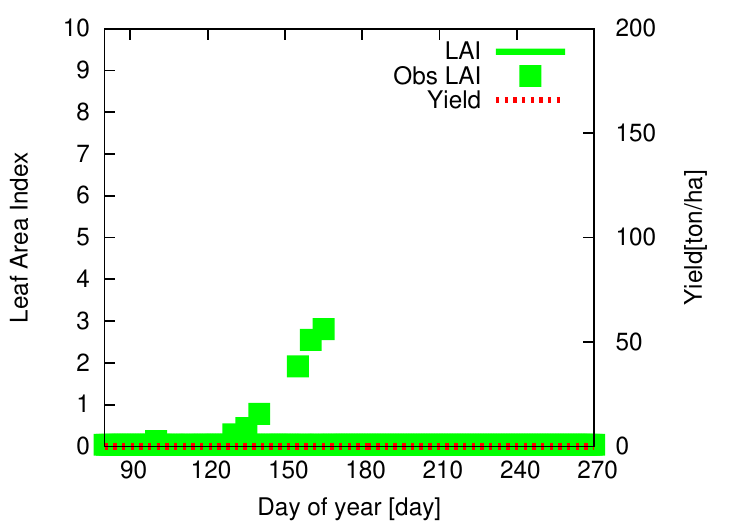}}
									\scalebox{0.29}{\includegraphics[scale=1.0, bb=0 0 216 151]{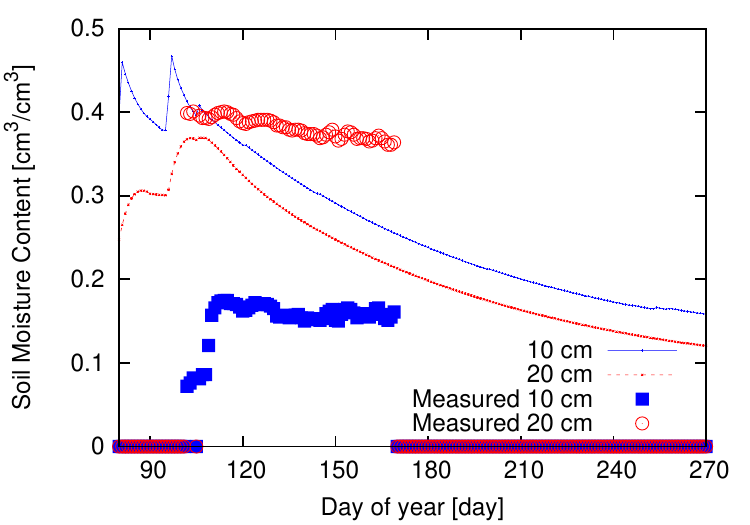}}
									\scalebox{0.29}{\includegraphics[scale=1.0, bb=0 0 216 151]{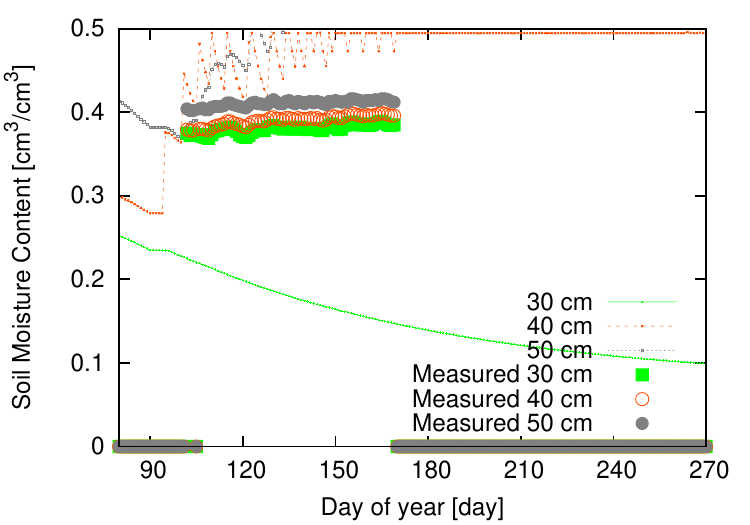}}
								\end{minipage} &
								\begin{minipage}{70mm}
									\centering
									\scalebox{0.29}{\includegraphics[scale=1.0, bb=0 0 216 151]{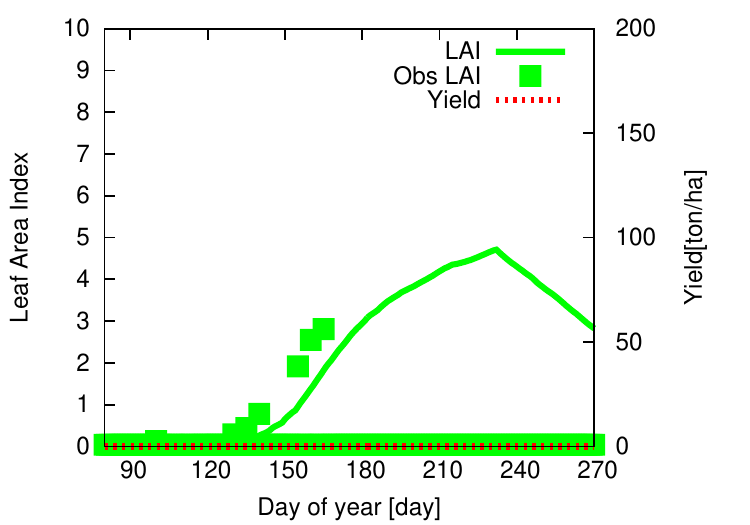}}
									\scalebox{0.29}{\includegraphics[scale=1.0, bb=0 0 216 151]{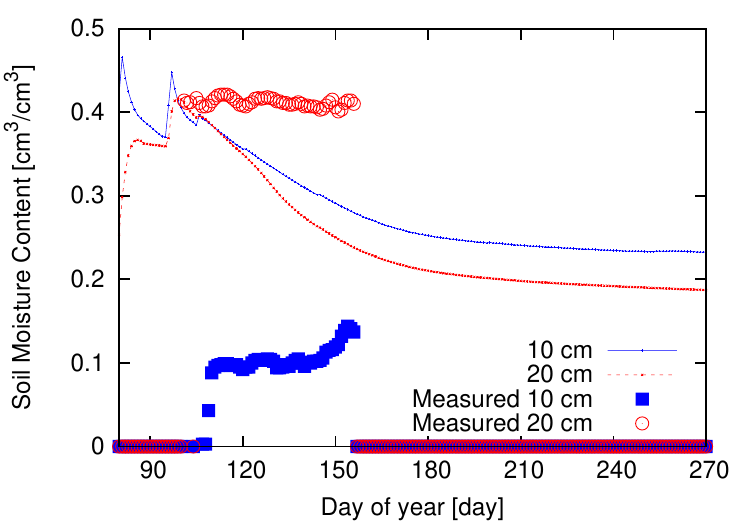}}
									\scalebox{0.29}{\includegraphics[scale=1.0, bb=0 0 216 151]{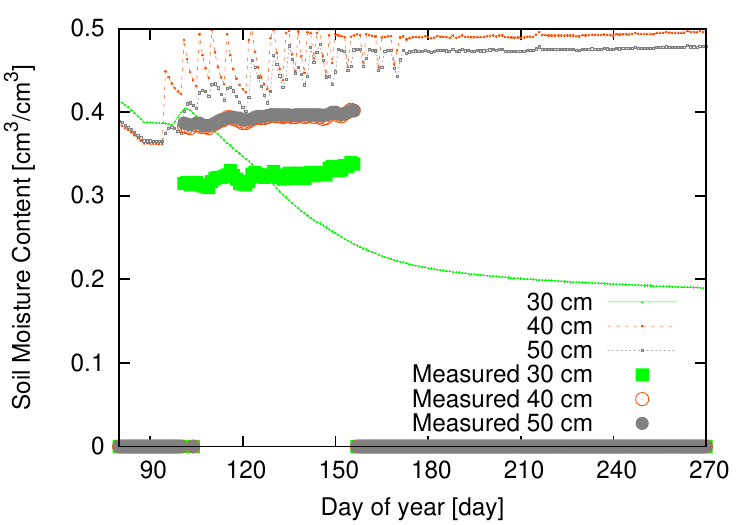}}
								\end{minipage} \\ \hline
		\end{tabular}
	\end{table}
	
	\begin{table}[H]
		\caption{Field C}
		\begin{tabular}{ccc}
			\hline & S1 & S2 \\
			\hline 10th May  & \begin{minipage}{70mm}
									\centering
									\scalebox{0.29}{\includegraphics[scale=1.0, bb=0 0 216 151]{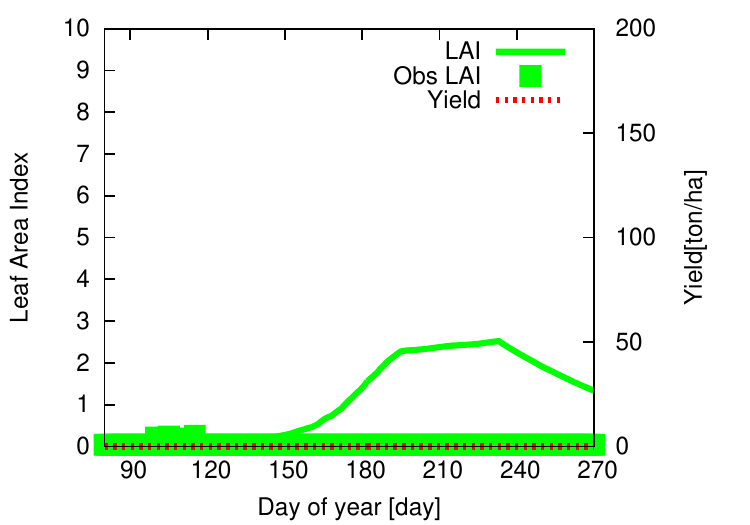}}
									\scalebox{0.29}{\includegraphics[scale=1.0, bb=0 0 216 151]{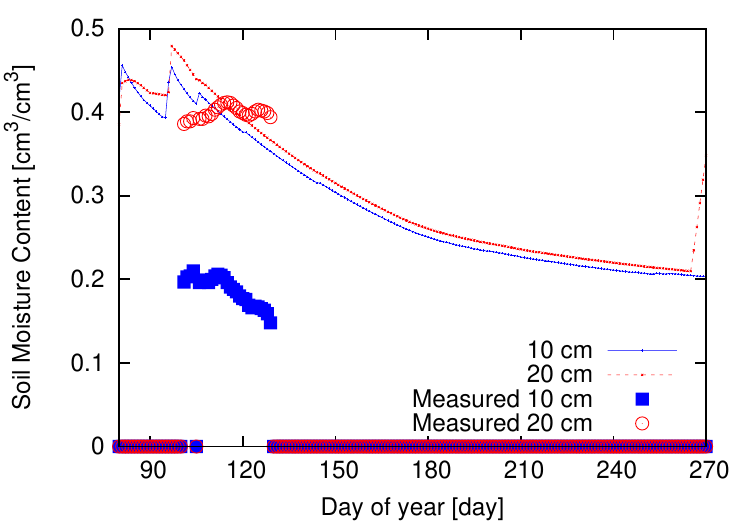}}
									\scalebox{0.29}{\includegraphics[scale=1.0, bb=0 0 216 151]{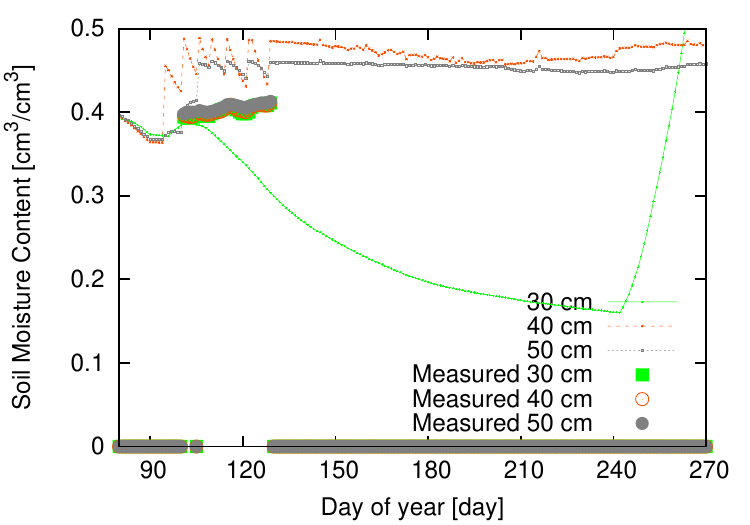}}
								\end{minipage} &
								\begin{minipage}{70mm}
									\centering
									\scalebox{0.29}{\includegraphics[scale=1.0, bb=0 0 216 151]{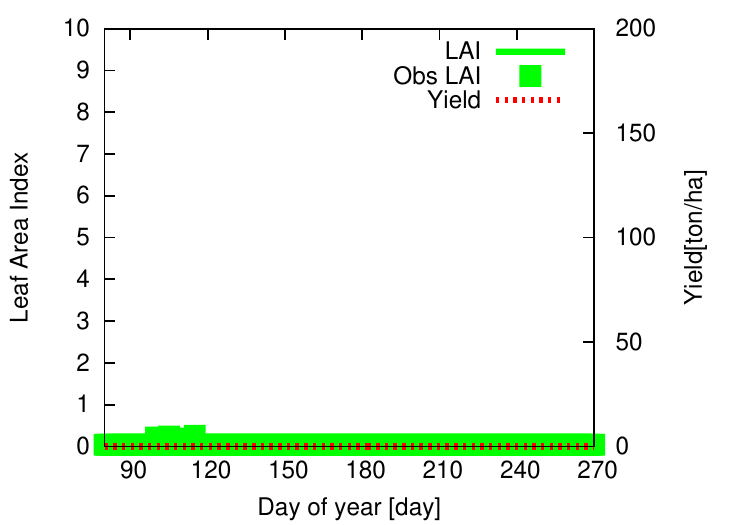}}
									\scalebox{0.29}{\includegraphics[scale=1.0, bb=0 0 216 151]{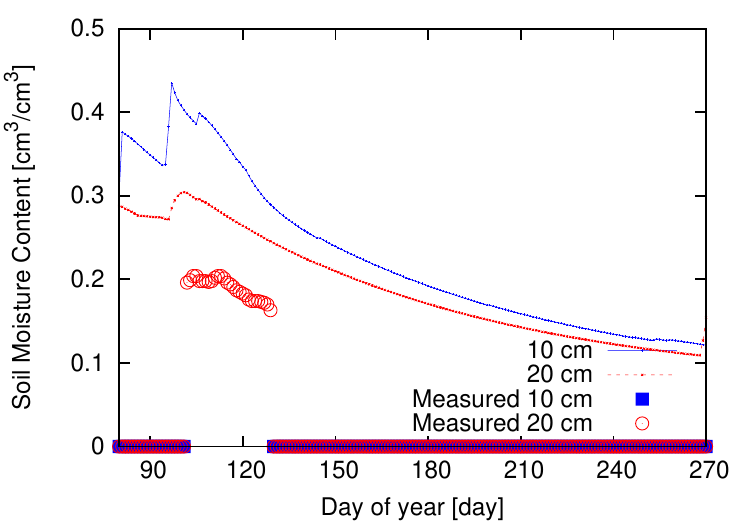}}
									\scalebox{0.29}{\includegraphics[scale=1.0, bb=0 0 216 151]{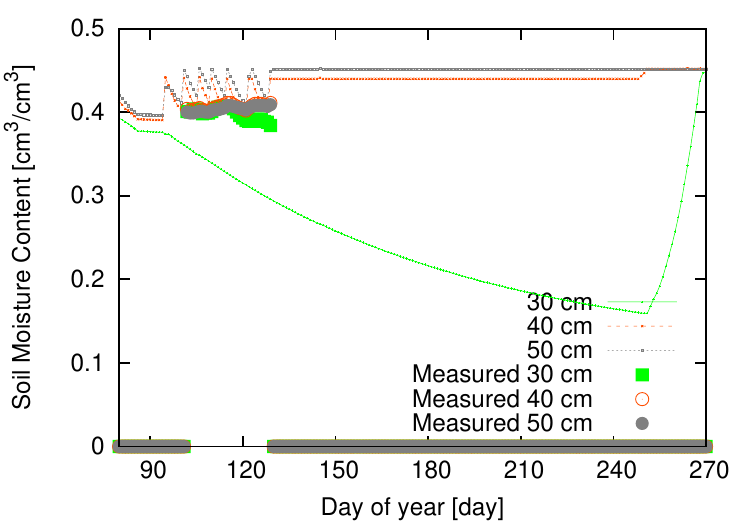}}
								\end{minipage} \\
			\hline 30th May  & \begin{minipage}{70mm}
									\centering
									\scalebox{0.29}{\includegraphics[scale=1.0, bb=0 0 216 151]{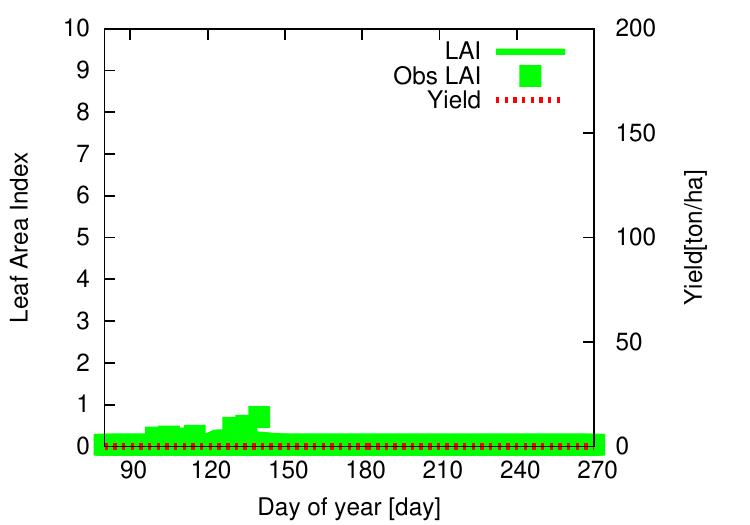}}
									\scalebox{0.29}{\includegraphics[scale=1.0, bb=0 0 216 151]{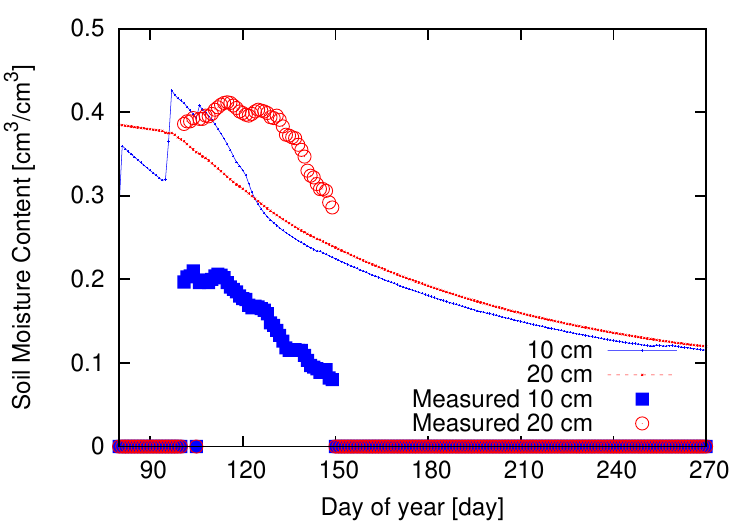}}
									\scalebox{0.29}{\includegraphics[scale=1.0, bb=0 0 216 151]{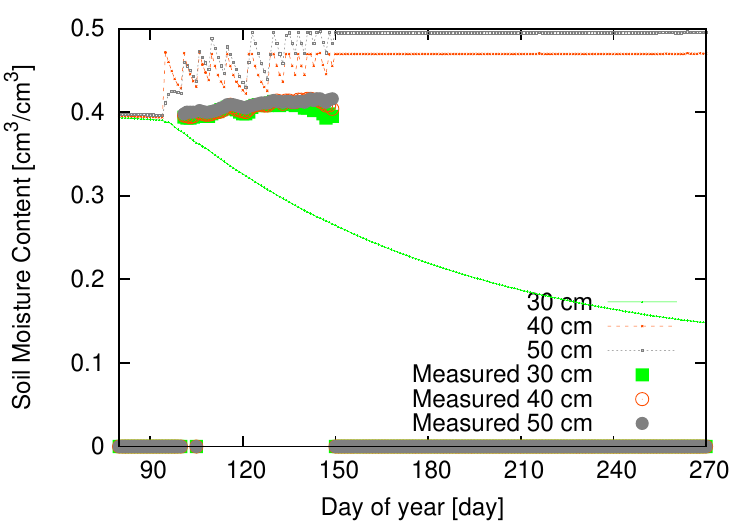}}
								\end{minipage} &
								\begin{minipage}{70mm}
									\centering
									\scalebox{0.29}{\includegraphics[scale=1.0, bb=0 0 216 151]{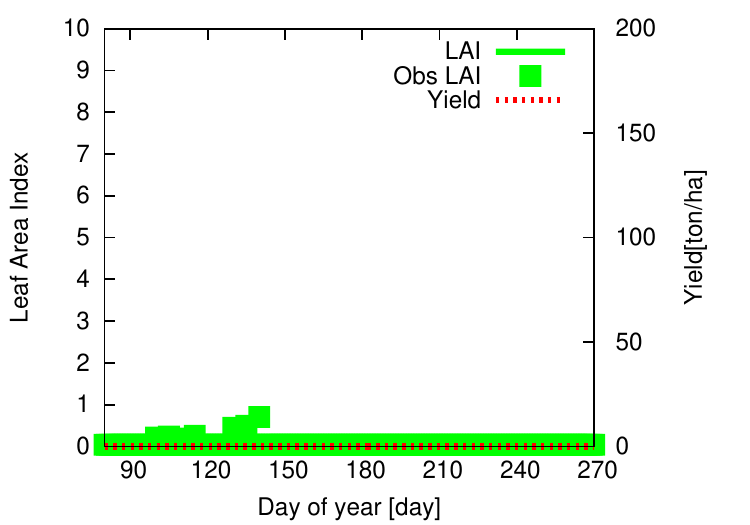}}
									\scalebox{0.29}{\includegraphics[scale=1.0, bb=0 0 216 151]{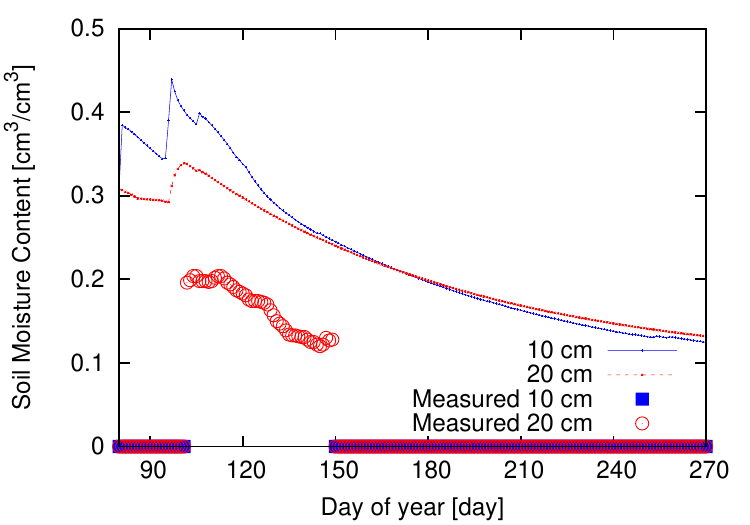}}
									\scalebox{0.29}{\includegraphics[scale=1.0, bb=0 0 216 151]{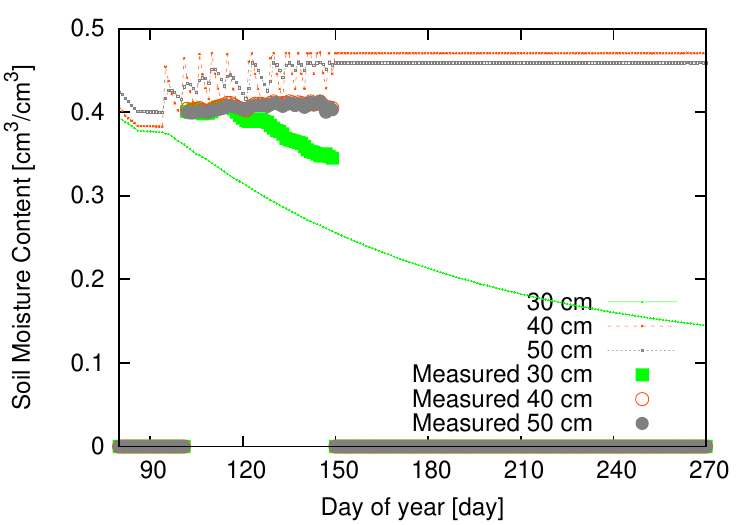}}
								\end{minipage} \\
			\hline 19th June & \begin{minipage}{70mm}
									\centering
									\scalebox{0.29}{\includegraphics[scale=1.0, bb=0 0 216 151]{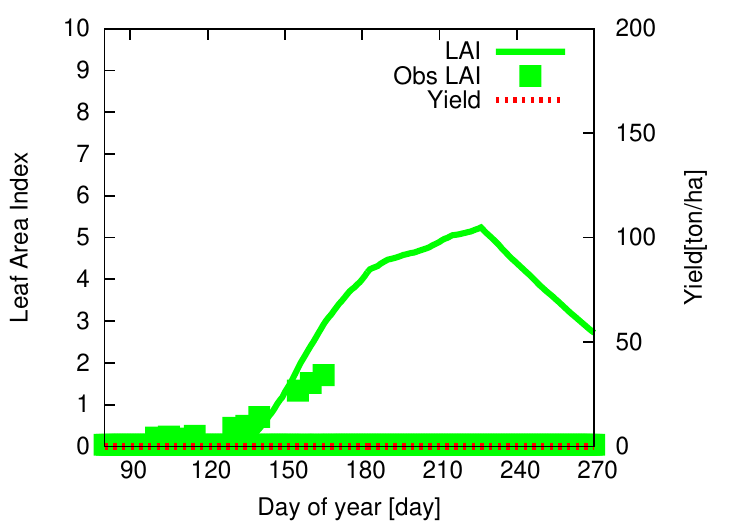}}
									\scalebox{0.29}{\includegraphics[scale=1.0, bb=0 0 216 151]{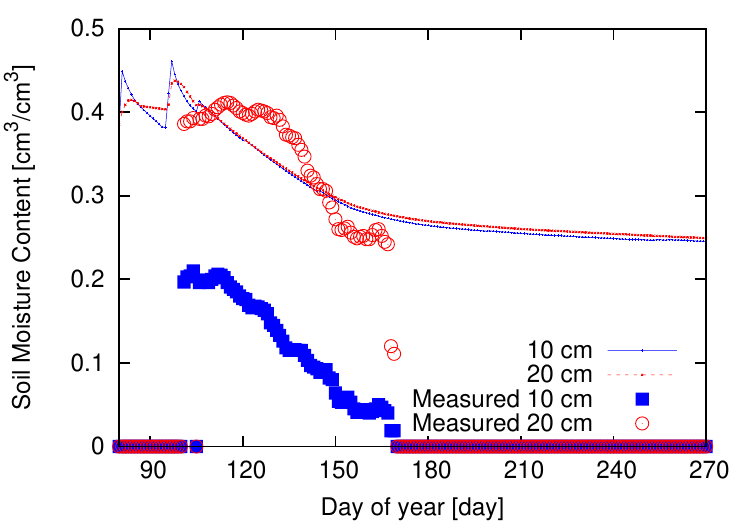}}
									\scalebox{0.29}{\includegraphics[scale=1.0, bb=0 0 216 151]{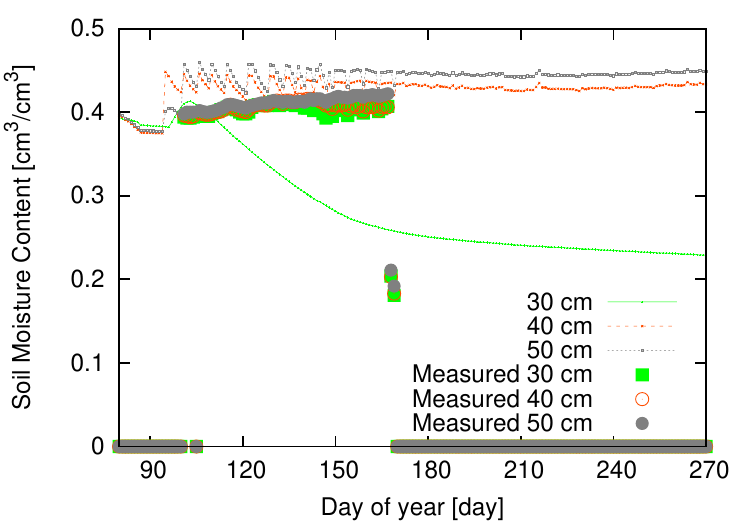}}
								\end{minipage} &
								\begin{minipage}{70mm}
									\centering
									\scalebox{0.29}{\includegraphics[scale=1.0, bb=0 0 216 151]{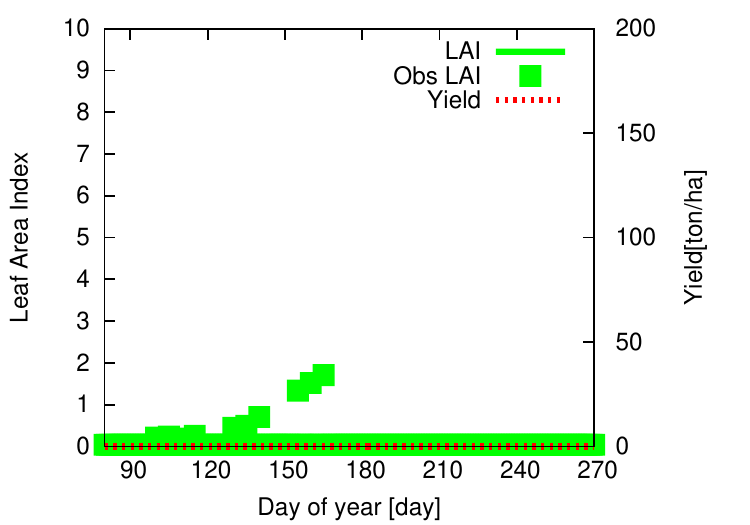}}
									\scalebox{0.29}{\includegraphics[scale=1.0, bb=0 0 216 151]{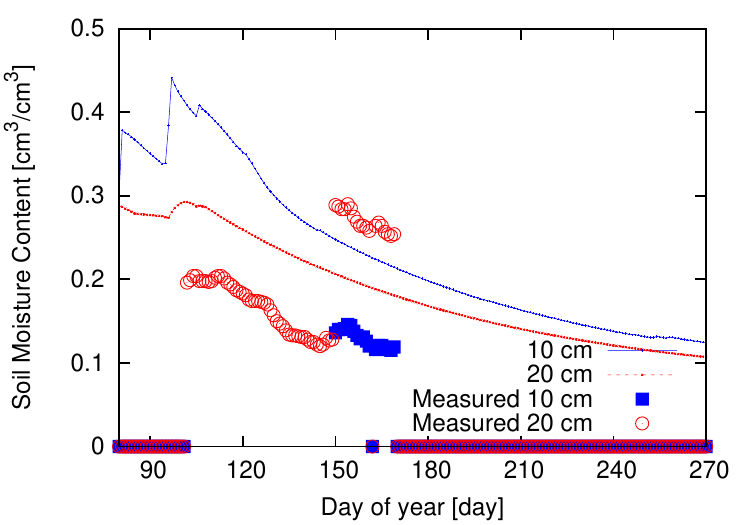}}
									\scalebox{0.29}{\includegraphics[scale=1.0, bb=0 0 216 151]{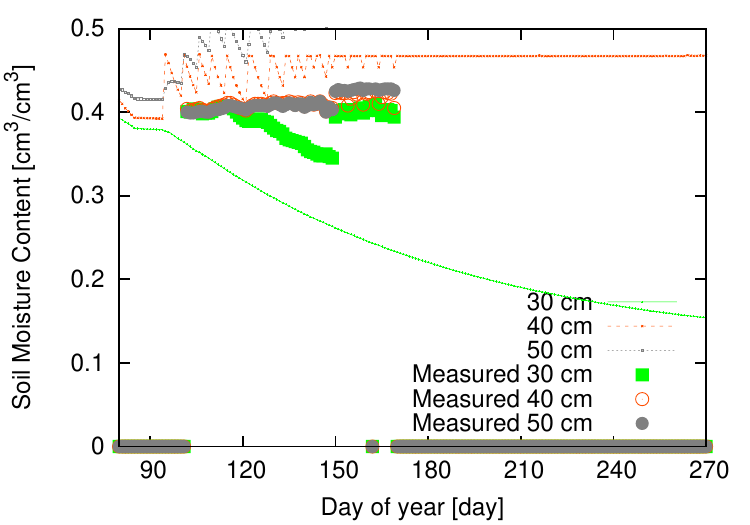}}
								\end{minipage} \\ \hline
		\end{tabular}
	\end{table}
	
	\begin{table}[H]
		\caption{Field D}
		\begin{tabular}{ccc}
			\hline & S1 & S2 \\
			\hline 10th May  & \begin{minipage}{70mm}
									\centering
									\scalebox{0.29}{\includegraphics[scale=1.0, bb=0 0 216 151]{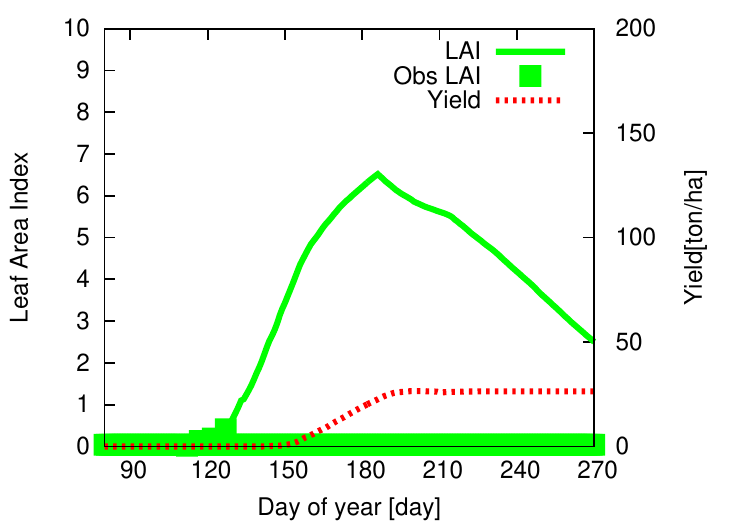}}
									\scalebox{0.29}{\includegraphics[scale=1.0, bb=0 0 216 151]{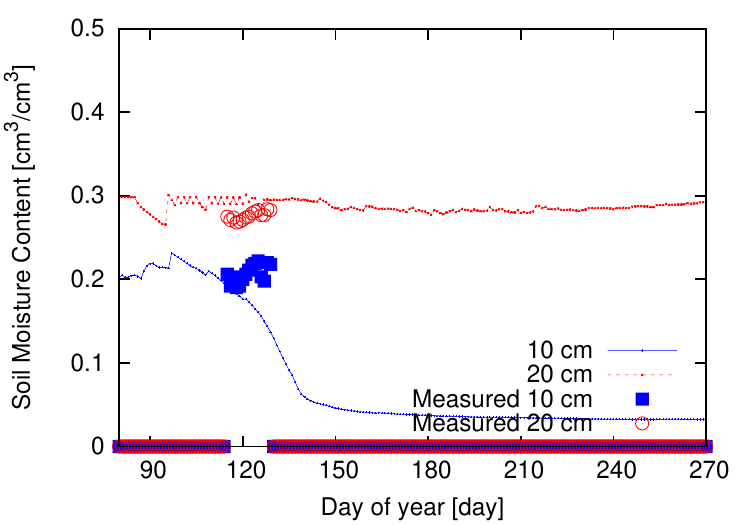}}
									\scalebox{0.29}{\includegraphics[scale=1.0, bb=0 0 216 151]{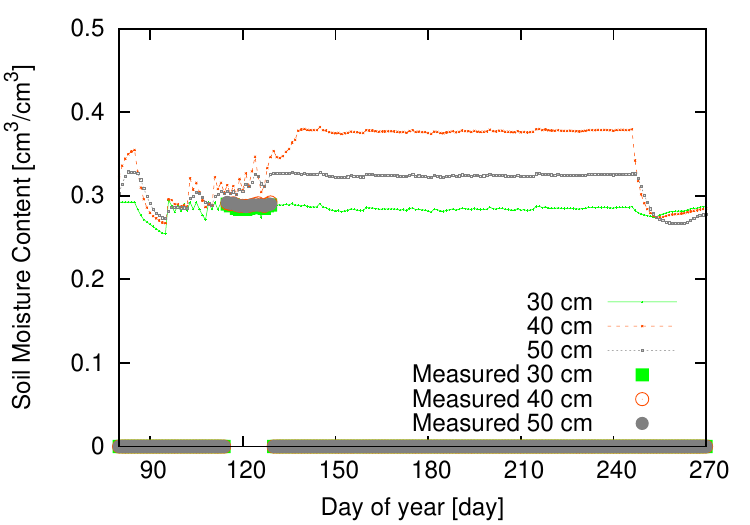}}
								\end{minipage} &
								\begin{minipage}{70mm}
									\centering
									\scalebox{0.29}{\includegraphics[scale=1.0, bb=0 0 216 151]{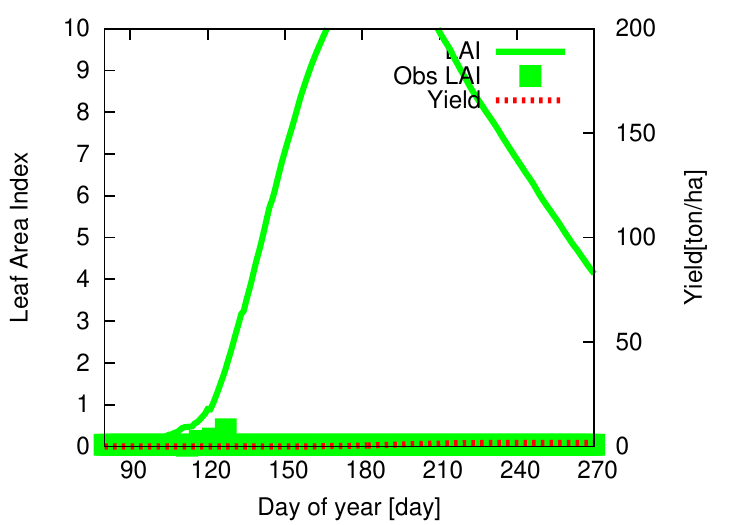}}
									\scalebox{0.29}{\includegraphics[scale=1.0, bb=0 0 216 151]{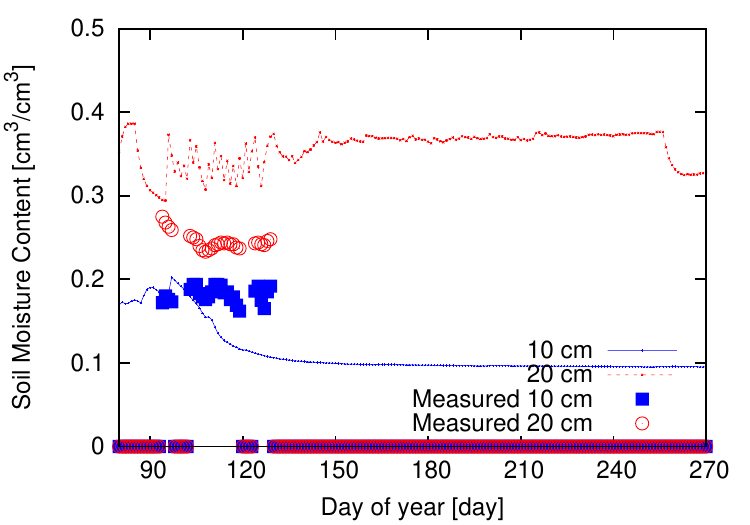}}
									\scalebox{0.29}{\includegraphics[scale=1.0, bb=0 0 216 151]{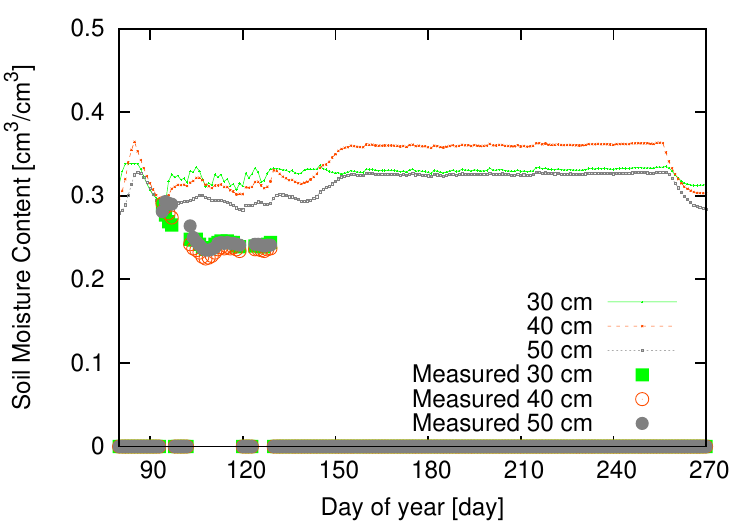}}
								\end{minipage} \\
			\hline 30th May  & \begin{minipage}{70mm}
									\centering
									\scalebox{0.29}{\includegraphics[scale=1.0, bb=0 0 216 151]{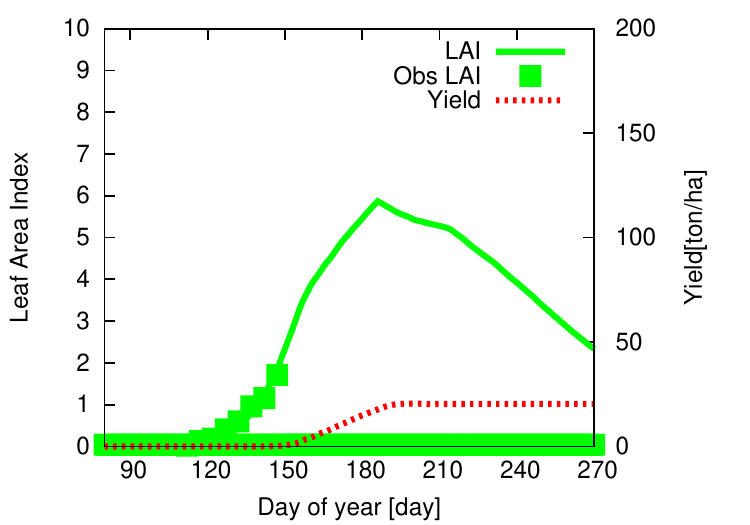}}
									\scalebox{0.29}{\includegraphics[scale=1.0, bb=0 0 216 151]{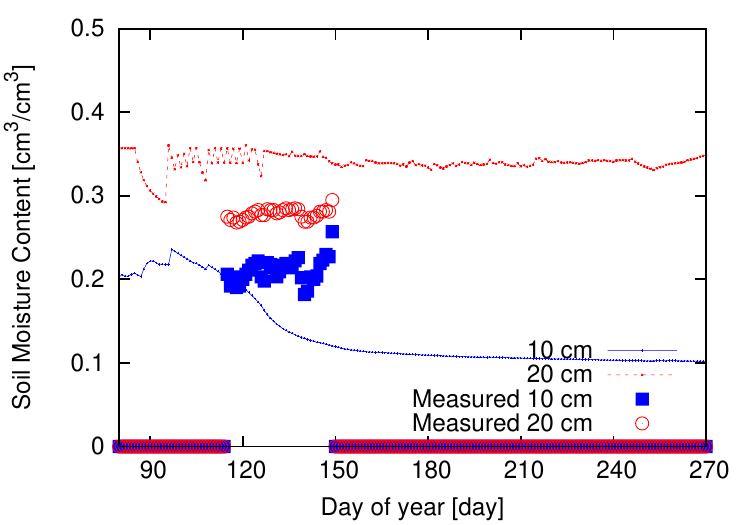}}
									\scalebox{0.29}{\includegraphics[scale=1.0, bb=0 0 216 151]{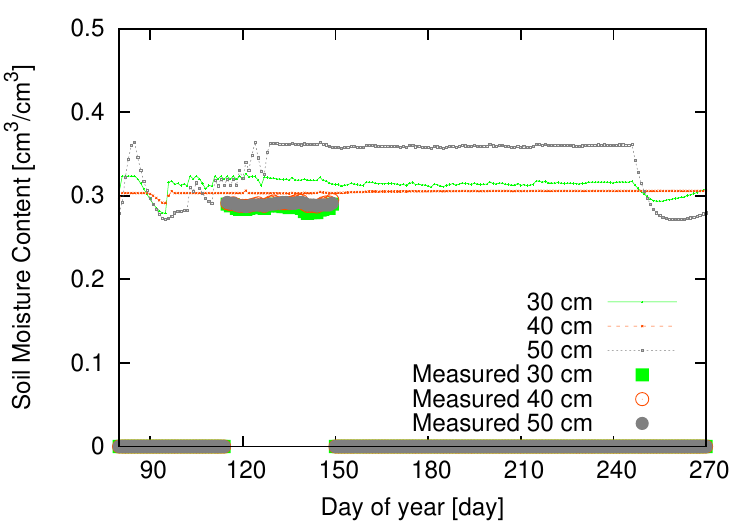}}
								\end{minipage} &
								\begin{minipage}{70mm}
									\centering
									\scalebox{0.29}{\includegraphics[scale=1.0, bb=0 0 216 151]{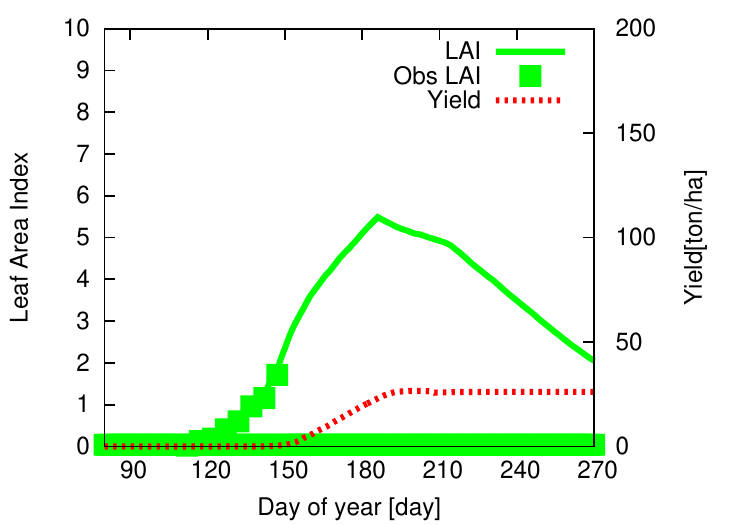}}
									\scalebox{0.29}{\includegraphics[scale=1.0, bb=0 0 216 151]{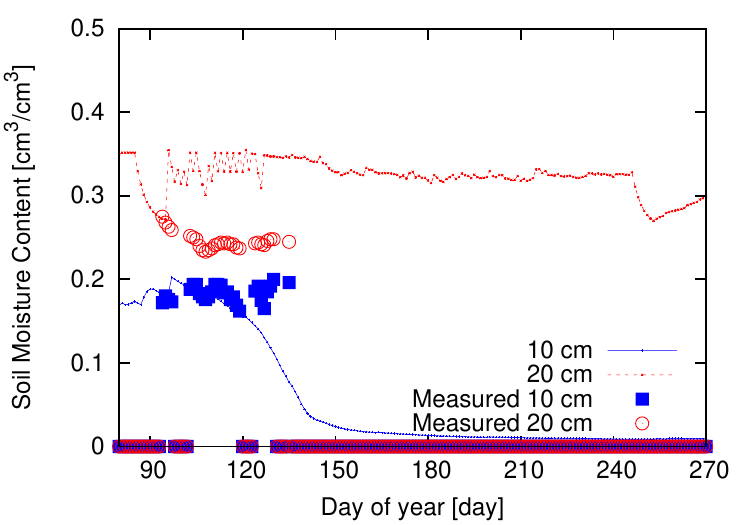}}
									\scalebox{0.29}{\includegraphics[scale=1.0, bb=0 0 216 151]{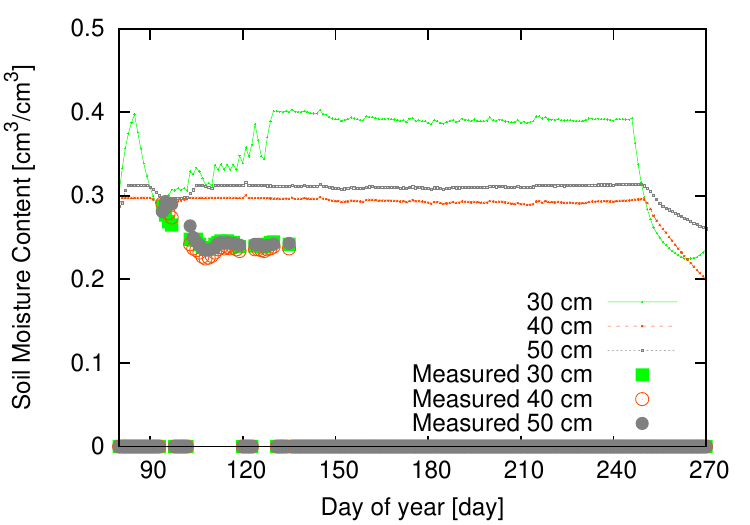}}
								\end{minipage} \\
			\hline 19th June & \begin{minipage}{70mm}
									\centering
									\scalebox{0.29}{\includegraphics[scale=1.0, bb=0 0 216 151]{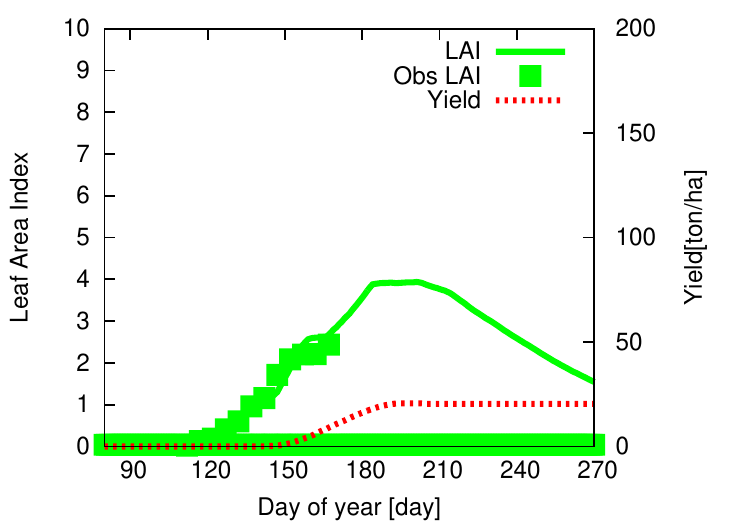}}
									\scalebox{0.29}{\includegraphics[scale=1.0, bb=0 0 216 151]{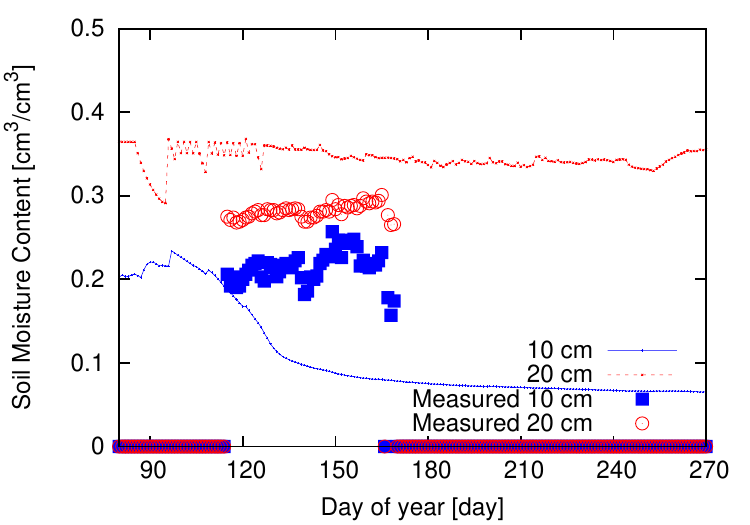}}
									\scalebox{0.29}{\includegraphics[scale=1.0, bb=0 0 216 151]{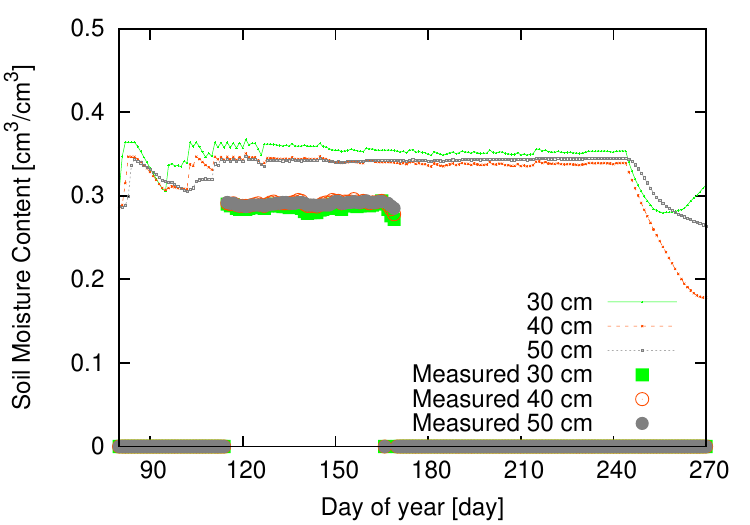}}
								\end{minipage} &
								\begin{minipage}{70mm}
									\centering
									\scalebox{0.29}{\includegraphics[scale=1.0, bb=0 0 216 151]{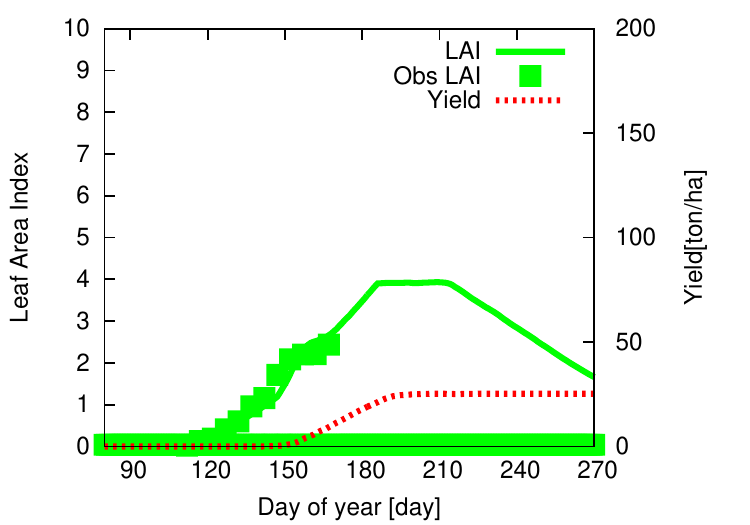}}
									\scalebox{0.29}{\includegraphics[scale=1.0, bb=0 0 216 151]{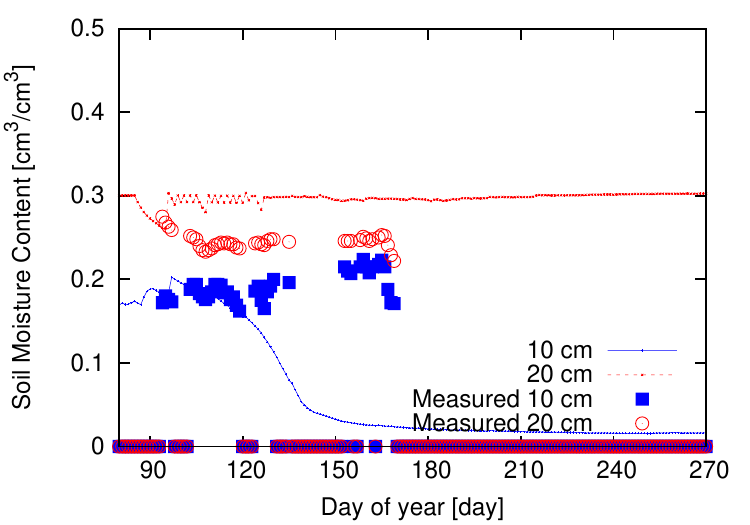}}
									\scalebox{0.29}{\includegraphics[scale=1.0, bb=0 0 216 151]{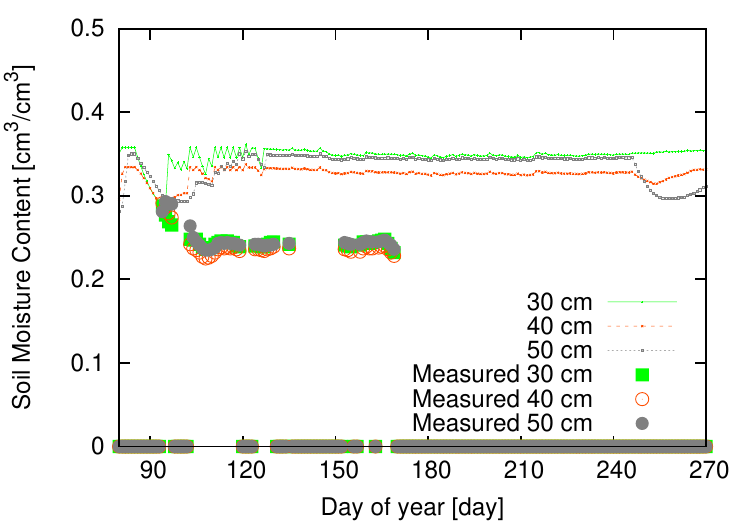}}
								\end{minipage} \\ \hline
		\end{tabular}
	\end{table}
	
	\begin{table}[H]
		\caption{Field E}
		\begin{tabular}{ccc}
			\hline & S1 & S2 \\
			\hline 10th May  & \begin{minipage}{70mm}
									\centering
									\scalebox{0.29}{\includegraphics[scale=1.0, bb=0 0 216 151]{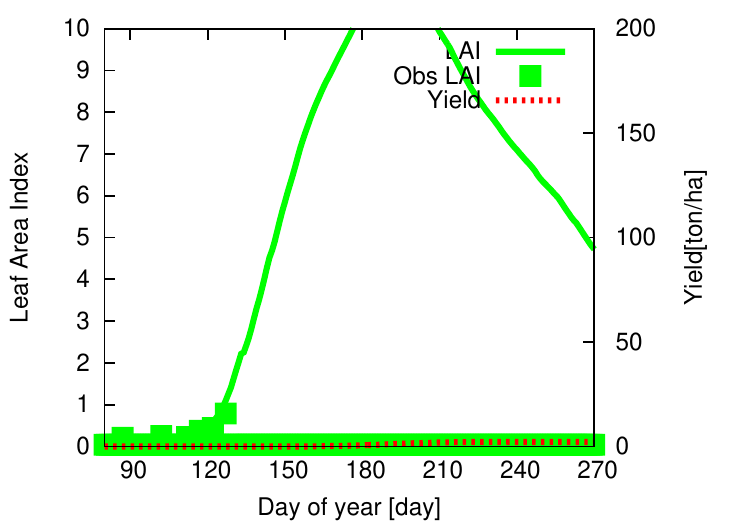}}
									\scalebox{0.29}{\includegraphics[scale=1.0, bb=0 0 216 151]{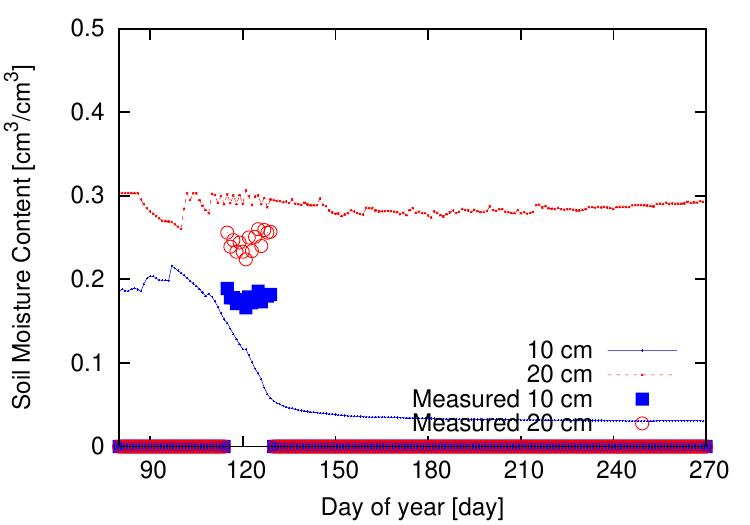}}
									\scalebox{0.29}{\includegraphics[scale=1.0, bb=0 0 216 151]{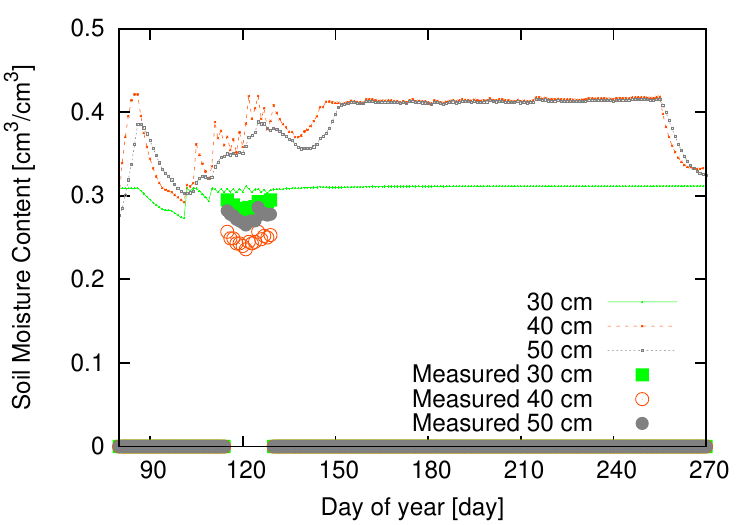}}
								\end{minipage} &
								\begin{minipage}{70mm}
									\centering
									\scalebox{0.29}{\includegraphics[scale=1.0, bb=0 0 216 151]{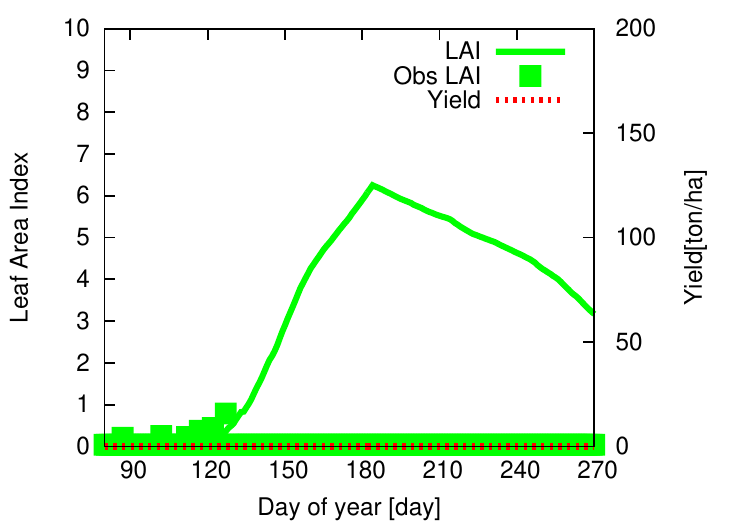}}
									\scalebox{0.29}{\includegraphics[scale=1.0, bb=0 0 216 151]{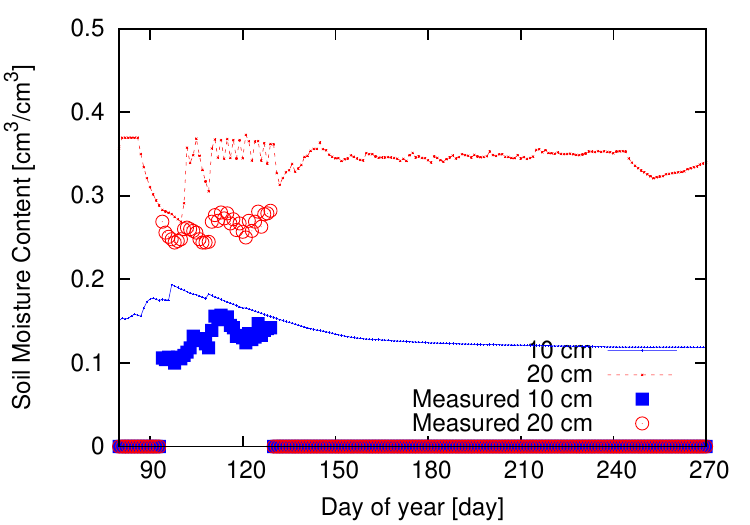}}
									\scalebox{0.29}{\includegraphics[scale=1.0, bb=0 0 216 151]{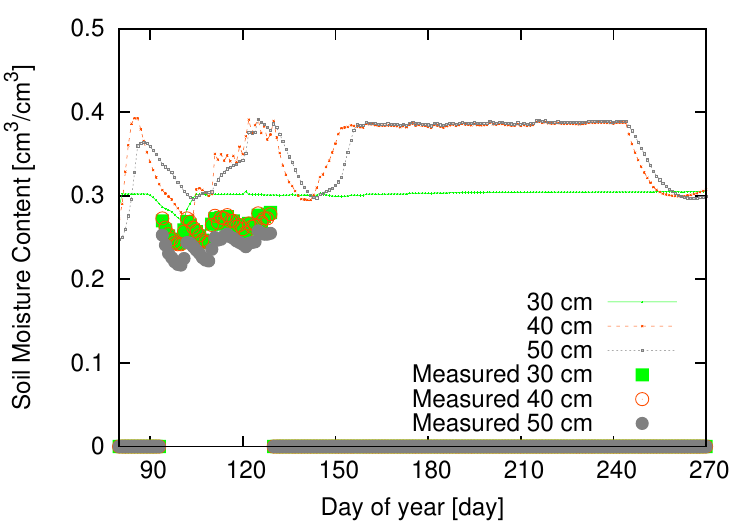}}
								\end{minipage} \\
			\hline 30th May  & \begin{minipage}{70mm}
									\centering
									\scalebox{0.29}{\includegraphics[scale=1.0, bb=0 0 216 151]{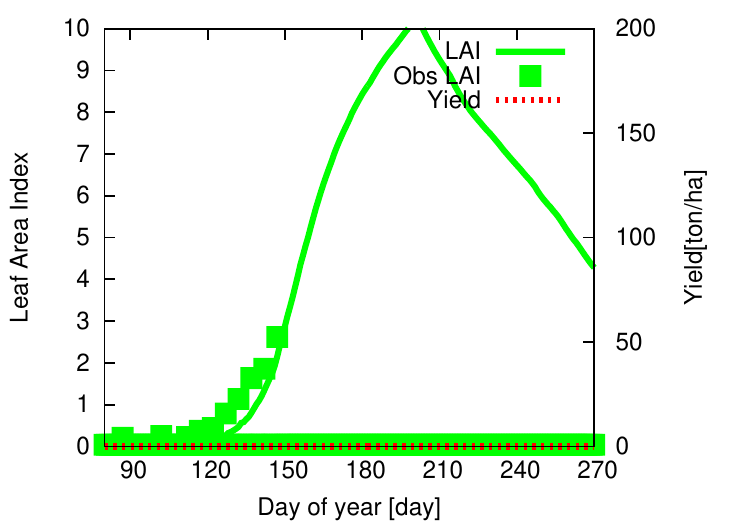}}
									\scalebox{0.29}{\includegraphics[scale=1.0, bb=0 0 216 151]{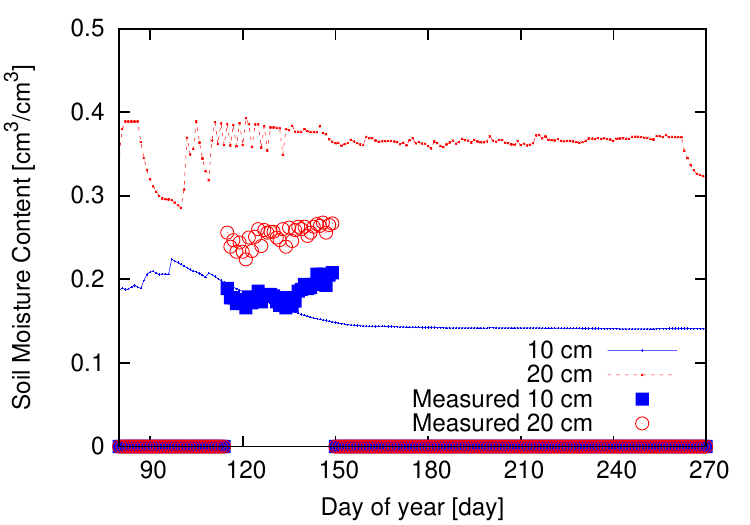}}
									\scalebox{0.29}{\includegraphics[scale=1.0, bb=0 0 216 151]{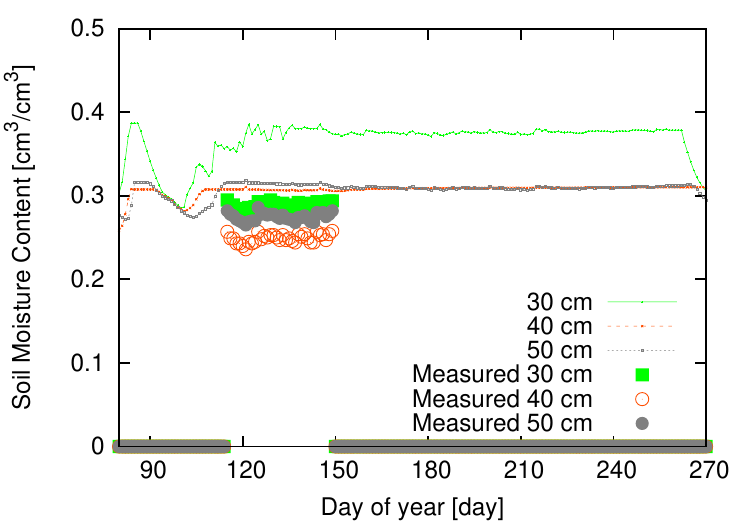}}
								\end{minipage} &
								\begin{minipage}{70mm}
									\centering
									\scalebox{0.29}{\includegraphics[scale=1.0, bb=0 0 216 151]{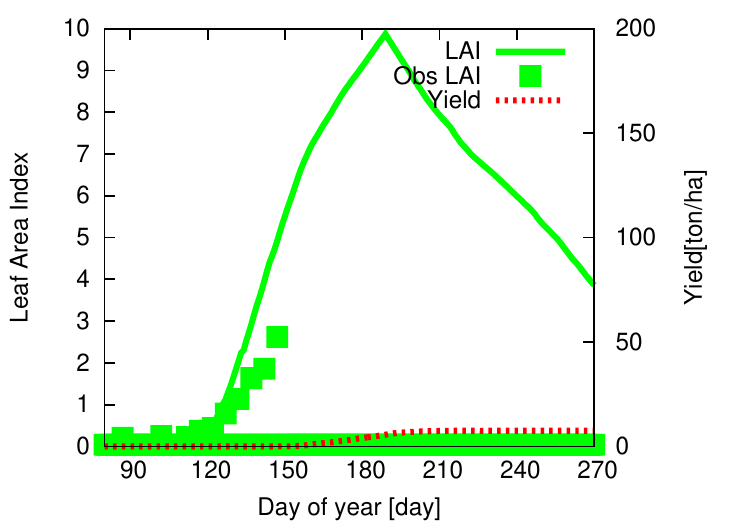}}
									\scalebox{0.29}{\includegraphics[scale=1.0, bb=0 0 216 151]{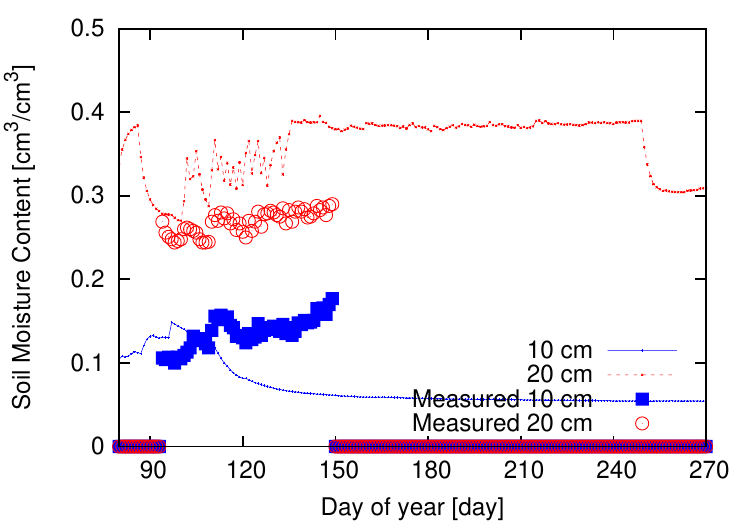}}
									\scalebox{0.29}{\includegraphics[scale=1.0, bb=0 0 216 151]{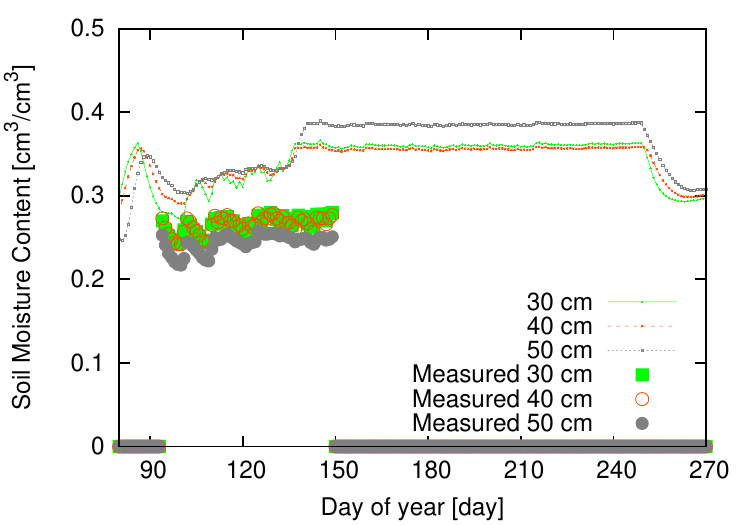}}
								\end{minipage} \\
			\hline 19th June & \begin{minipage}{70mm}
									\centering
									\scalebox{0.29}{\includegraphics[scale=1.0, bb=0 0 216 151]{./O_p_g_LAI_Y_us_FieldE_0619_190306002_S3.pdf}}
									\scalebox{0.29}{\includegraphics[scale=1.0, bb=0 0 216 151]{./O_p_g_SW_shallow_us_FieldE_0619_190306002_S3.pdf}}
									\scalebox{0.29}{\includegraphics[scale=1.0, bb=0 0 216 151]{./O_p_g_SW_deep_us_FieldE_0619_190306002_S3.pdf}}
								\end{minipage} &
								\begin{minipage}{70mm}
									\centering
									\scalebox{0.29}{\includegraphics[scale=1.0, bb=0 0 216 151]{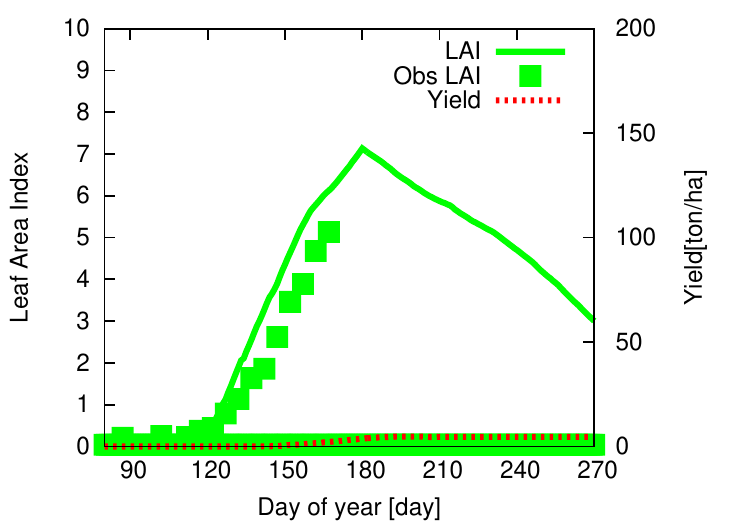}}
									\scalebox{0.29}{\includegraphics[scale=1.0, bb=0 0 216 151]{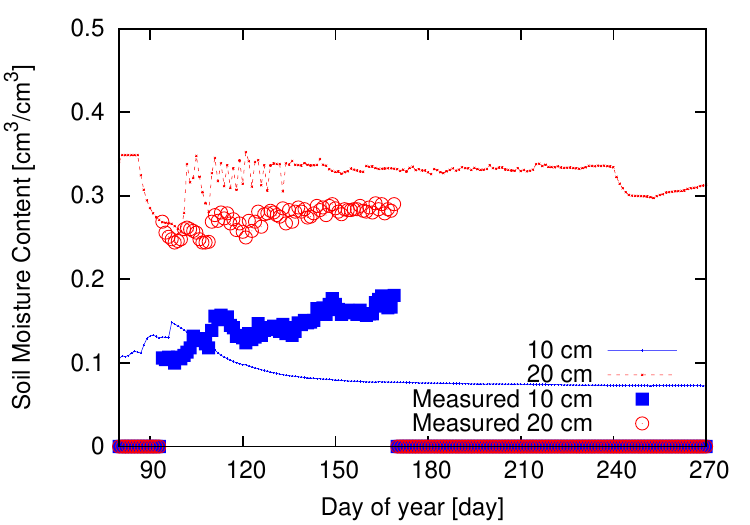}}
									\scalebox{0.29}{\includegraphics[scale=1.0, bb=0 0 216 151]{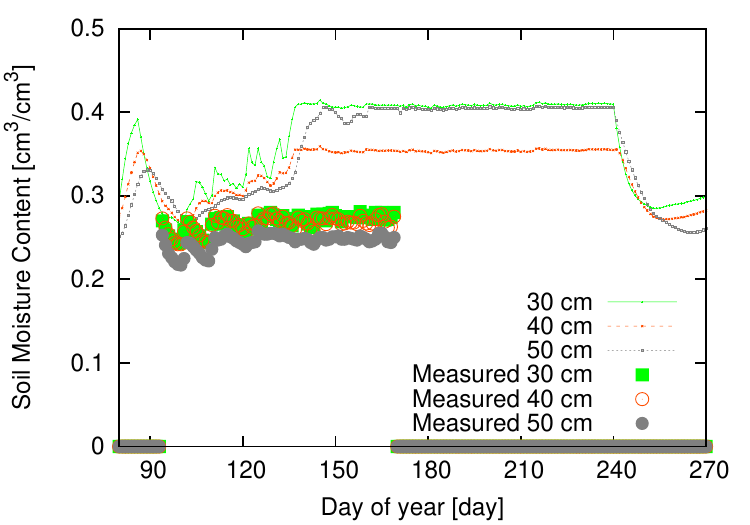}}
								\end{minipage} \\ \hline
		\end{tabular}
	\end{table}

\section{All analysis results from modified DSSAT}\label{all_result_of_new}
\par All of the analysis result by the modified DSSAT is shown in this section.
There were two soil moisture sensors (S1 and S2) in each field, and we chose three analysis days for each sensor.
In the each analysis, only the data collected before the day of the analysis are used for the calculation process to assess the application potency of the model in a particular season.
In each group of three graphs, the one on the left shows the LAI and the yield (total fruit weight), the one in the middle shows the soil moisture in the 10\,cm and 20\,cm layers, and the one on the right shows the soil moisture in the 30, 40, and 50\,cm layers.
Lines show the results calculated by the model, and points show the observed data.
Most of the results are consistent with the observed data and a reasonable yield is calculated, which means that the model has been improved.
	\begin{table}[H]
		\caption{Field A}
		\begin{tabular}{ccc}
			\hline & S1 & S2 \\
			\hline 10th May  & \begin{minipage}{70mm}
									\centering
									\scalebox{0.29}{\includegraphics[scale=1.0, bb=0 0 216 151]{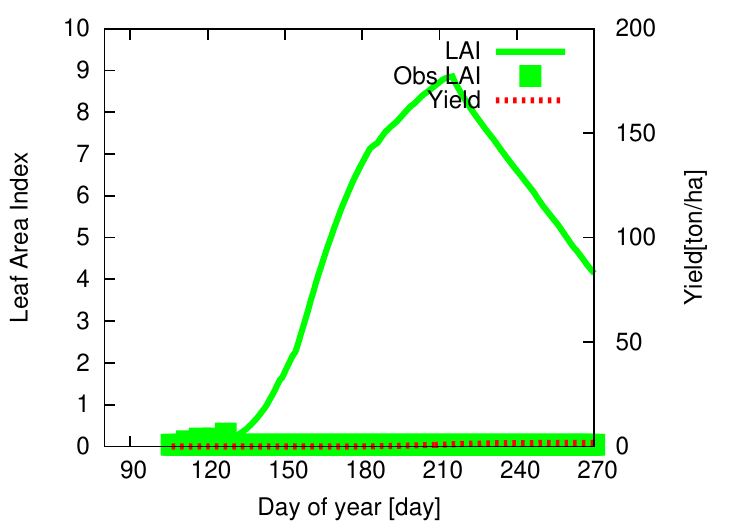}}
									\scalebox{0.29}{\includegraphics[scale=1.0, bb=0 0 216 151]{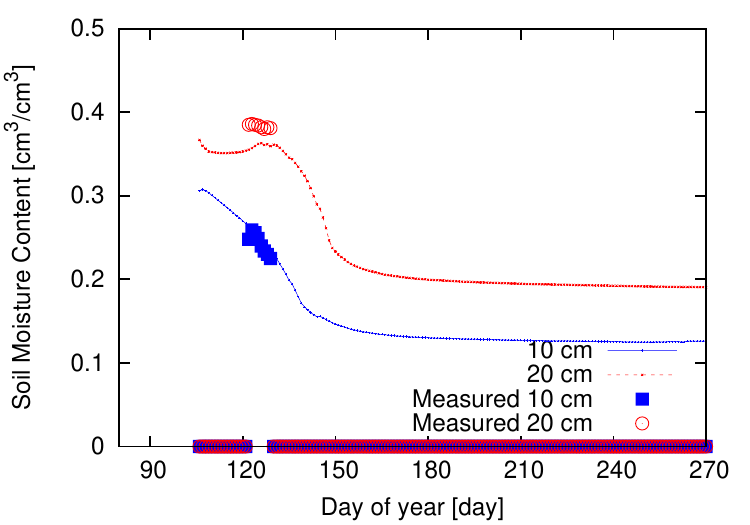}}
									\scalebox{0.29}{\includegraphics[scale=1.0, bb=0 0 216 151]{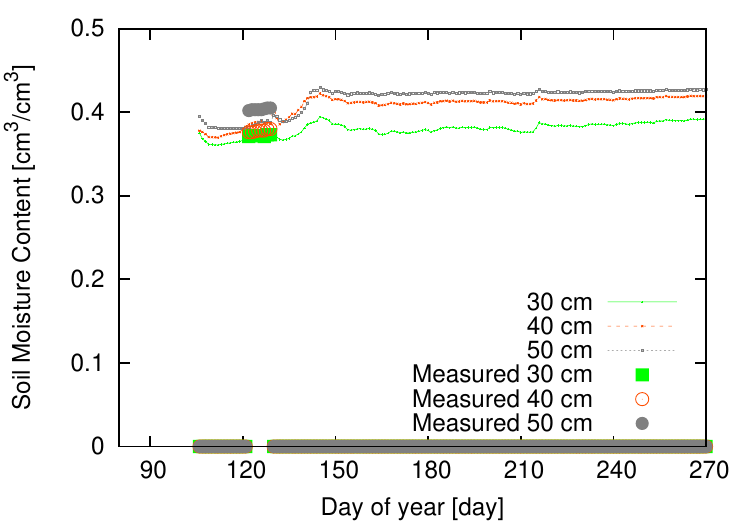}}
								\end{minipage} &
								\begin{minipage}{70mm}
									\centering
									\scalebox{0.29}{\includegraphics[scale=1.0, bb=0 0 216 151]{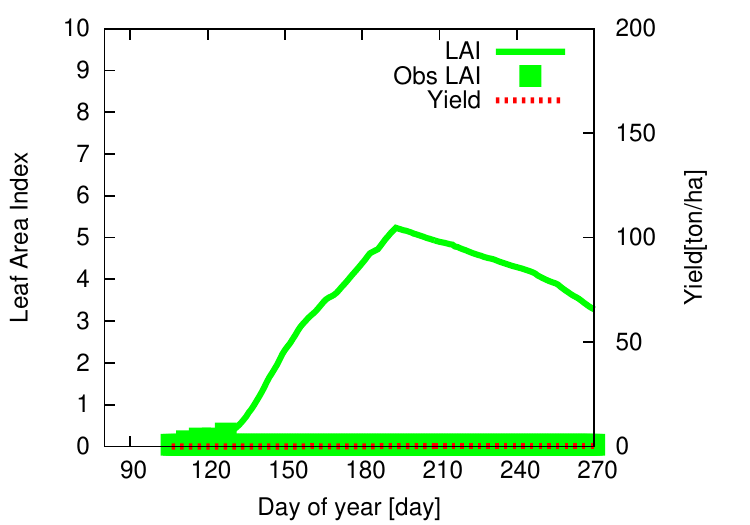}}
									\scalebox{0.29}{\includegraphics[scale=1.0, bb=0 0 216 151]{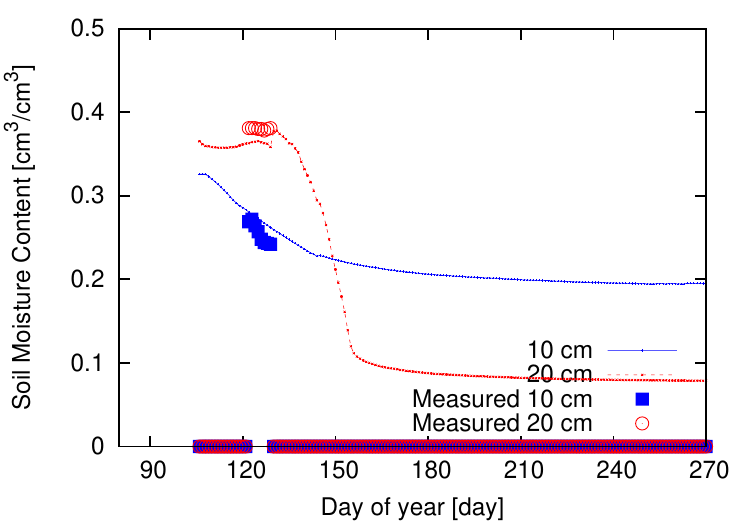}}
									\scalebox{0.29}{\includegraphics[scale=1.0, bb=0 0 216 151]{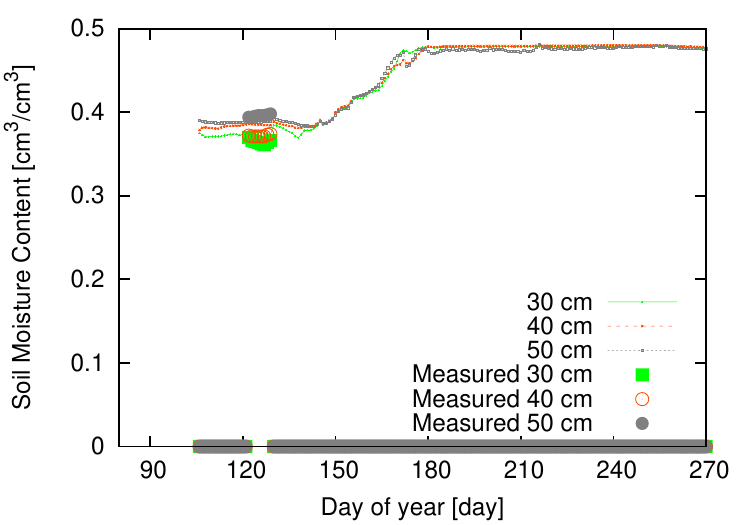}}
								\end{minipage} \\
			\hline 30th May  & \begin{minipage}{70mm}
									\centering
									\scalebox{0.29}{\includegraphics[scale=1.0, bb=0 0 216 151]{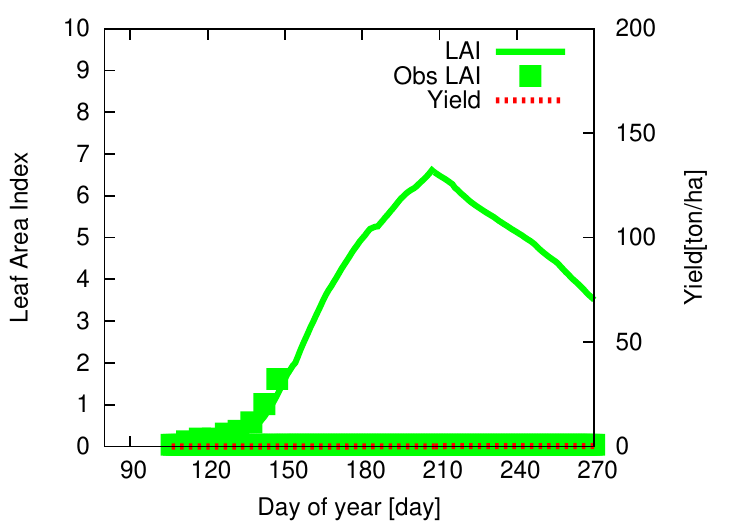}}
									\scalebox{0.29}{\includegraphics[scale=1.0, bb=0 0 216 151]{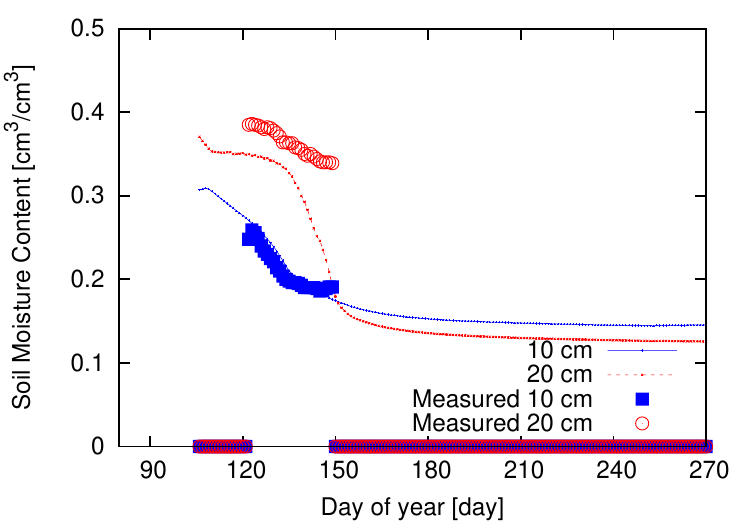}}
									\scalebox{0.29}{\includegraphics[scale=1.0, bb=0 0 216 151]{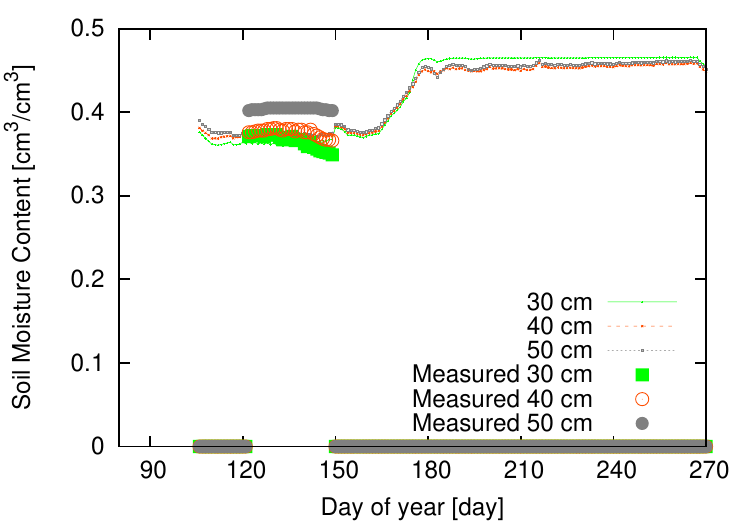}}
								\end{minipage} &
								\begin{minipage}{70mm}
									\centering
									\scalebox{0.29}{\includegraphics[scale=1.0, bb=0 0 216 151]{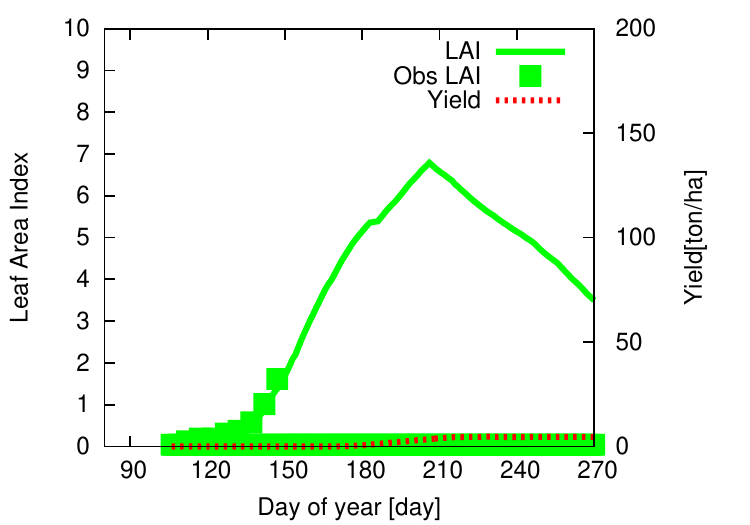}}
									\scalebox{0.29}{\includegraphics[scale=1.0, bb=0 0 216 151]{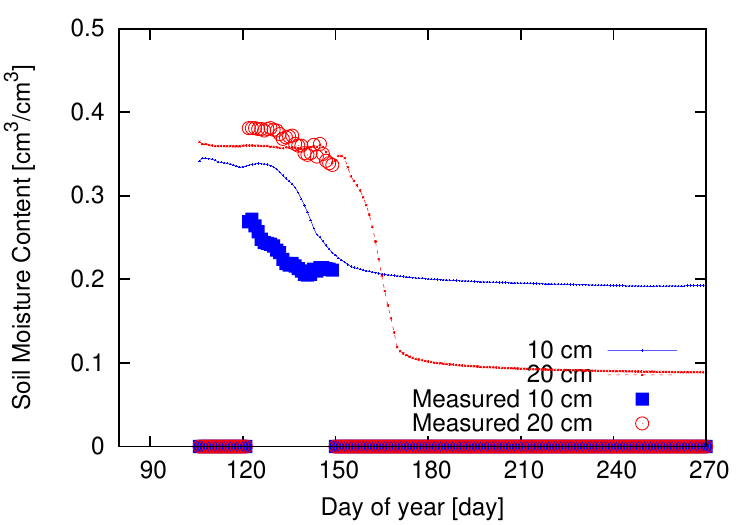}}
									\scalebox{0.29}{\includegraphics[scale=1.0, bb=0 0 216 151]{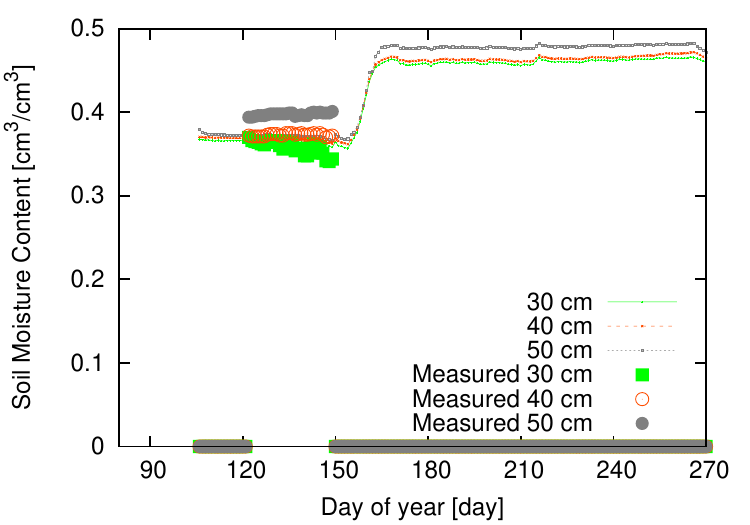}}
								\end{minipage} \\
			\hline 19th June & \begin{minipage}{70mm}
									\centering
									\scalebox{0.29}{\includegraphics[scale=1.0, bb=0 0 216 151]{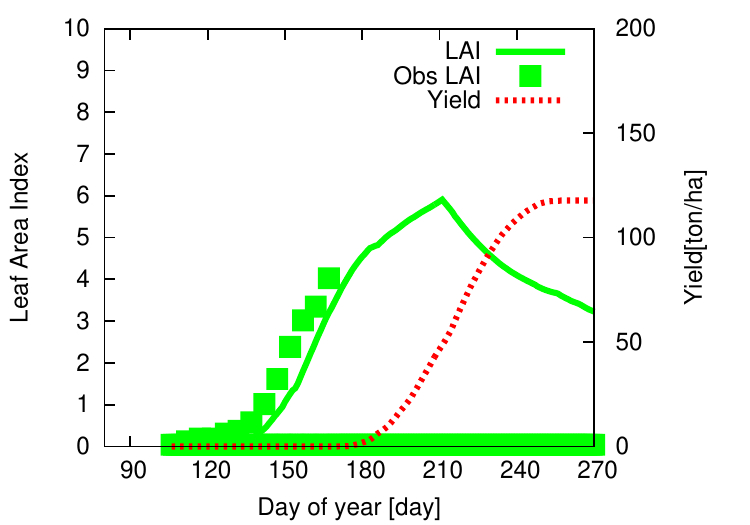}}
									\scalebox{0.29}{\includegraphics[scale=1.0, bb=0 0 216 151]{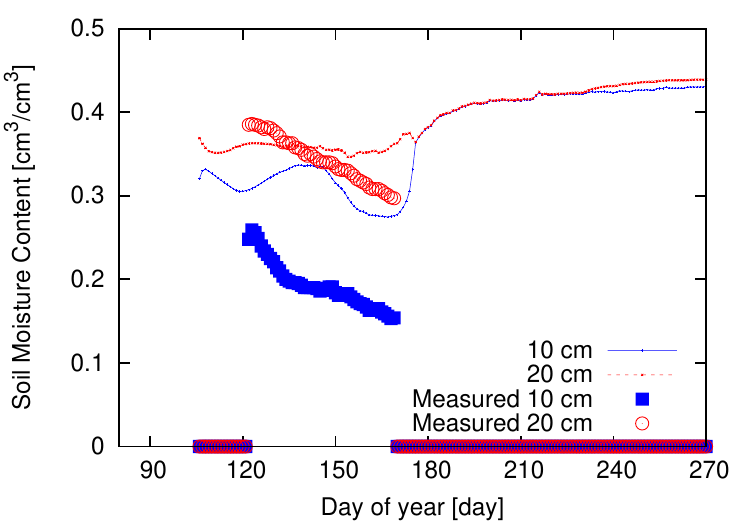}}
									\scalebox{0.29}{\includegraphics[scale=1.0, bb=0 0 216 151]{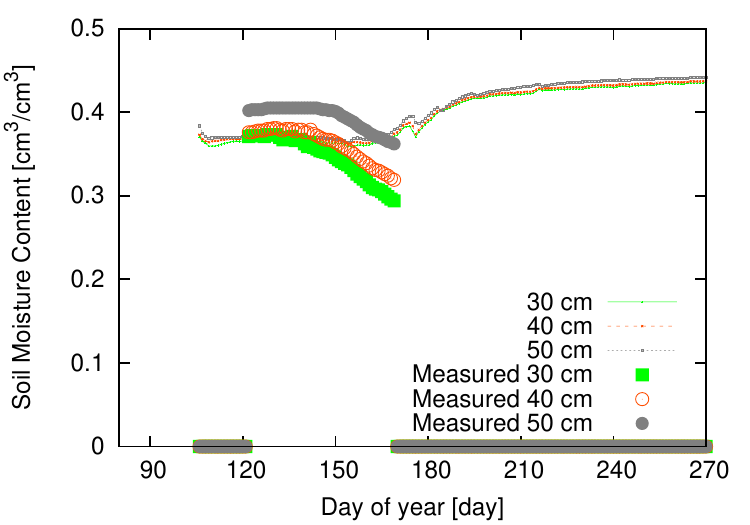}}
								\end{minipage} &
								\begin{minipage}{70mm}
									\centering
									\scalebox{0.29}{\includegraphics[scale=1.0, bb=0 0 216 151]{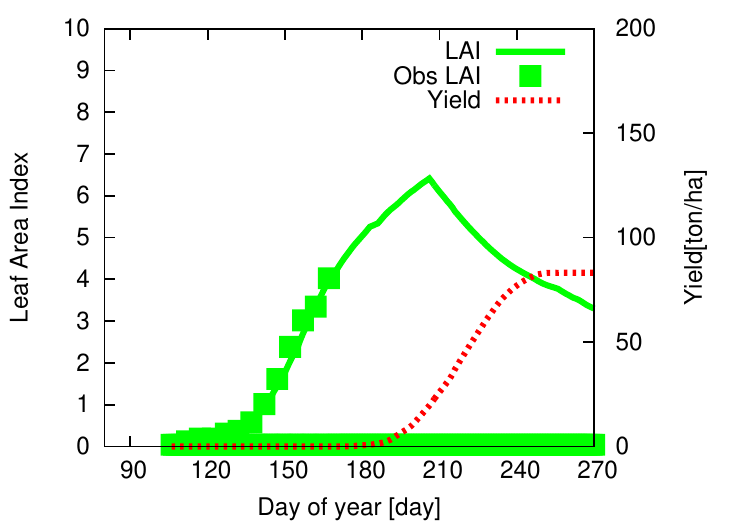}}
									\scalebox{0.29}{\includegraphics[scale=1.0, bb=0 0 216 151]{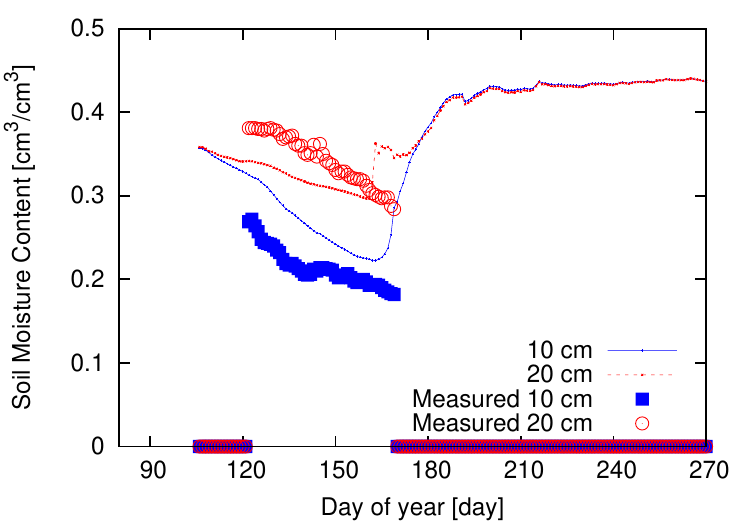}}
									\scalebox{0.29}{\includegraphics[scale=1.0, bb=0 0 216 151]{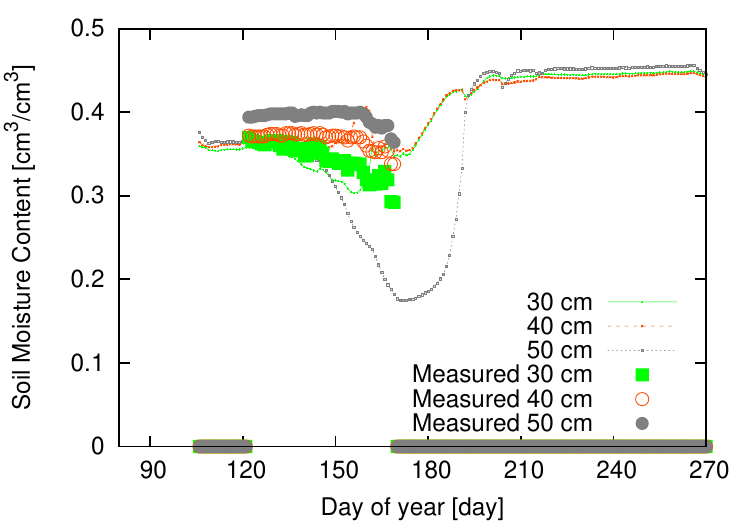}}
								\end{minipage} \\ \hline
		\end{tabular}
	\end{table}
	
	\begin{table}[H]
		\caption{Field B}
		\begin{tabular}{ccc}
			\hline & S1 & S2 \\
			\hline 10th May  & \begin{minipage}{70mm}
									\centering
									\scalebox{0.29}{\includegraphics[scale=1.0, bb=0 0 216 151]{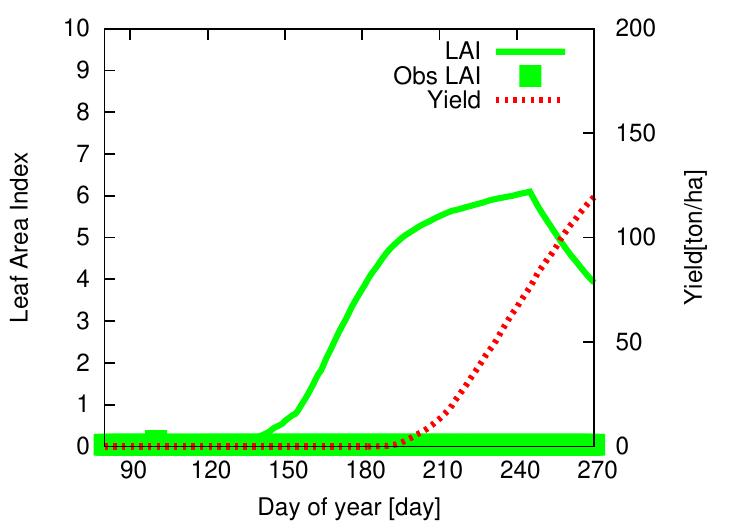}}
									\scalebox{0.29}{\includegraphics[scale=1.0, bb=0 0 216 151]{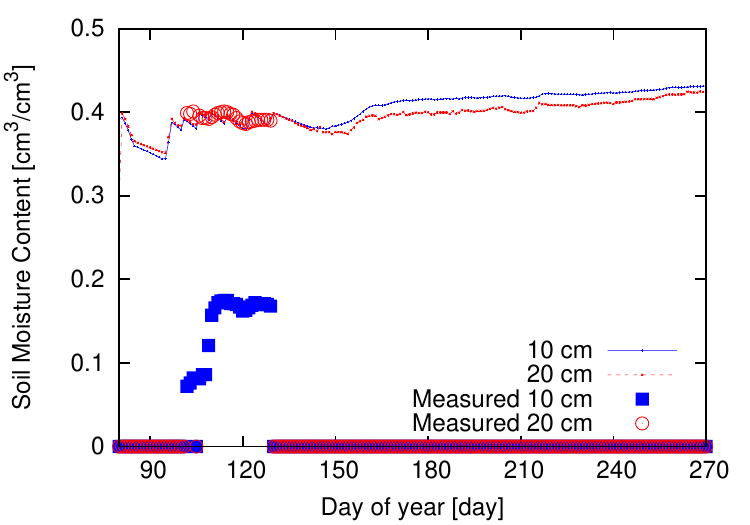}}
									\scalebox{0.29}{\includegraphics[scale=1.0, bb=0 0 216 151]{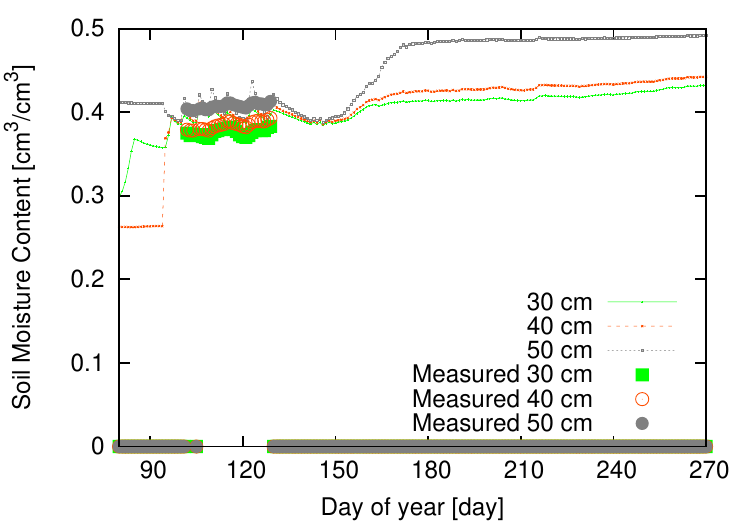}}
								\end{minipage} &
								\begin{minipage}{70mm}
									\centering
									\scalebox{0.29}{\includegraphics[scale=1.0, bb=0 0 216 151]{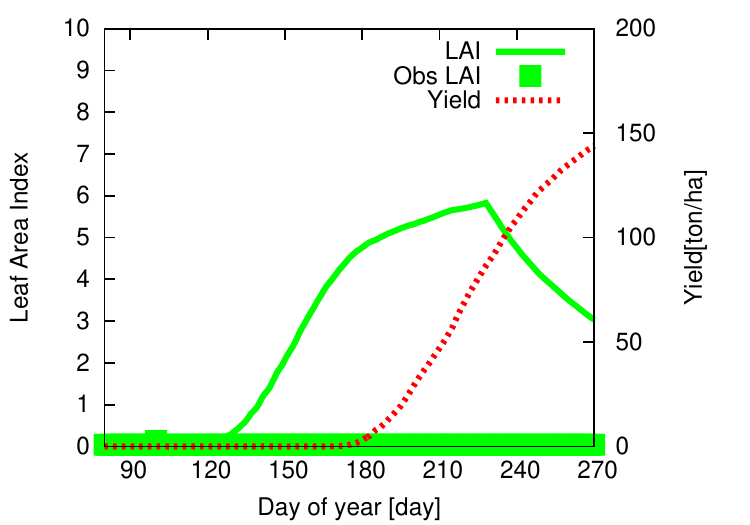}}
									\scalebox{0.29}{\includegraphics[scale=1.0, bb=0 0 216 151]{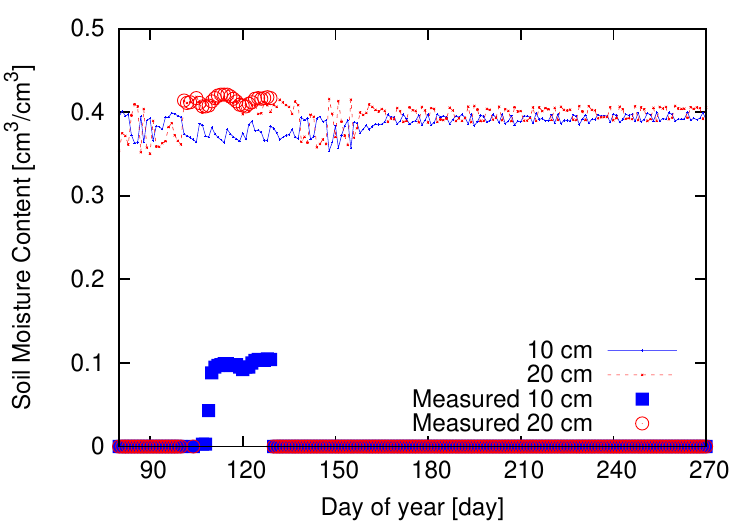}}
									\scalebox{0.29}{\includegraphics[scale=1.0, bb=0 0 216 151]{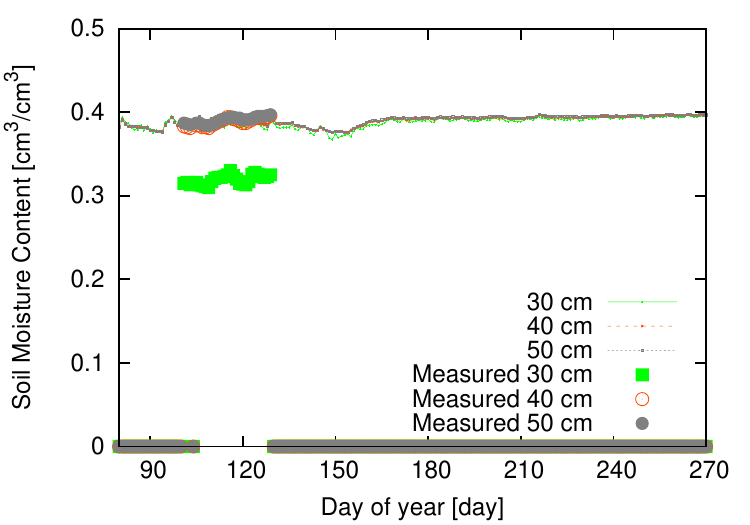}}
								\end{minipage} \\
			\hline 30th May  & \begin{minipage}{70mm}
									\centering
									\scalebox{0.29}{\includegraphics[scale=1.0, bb=0 0 216 151]{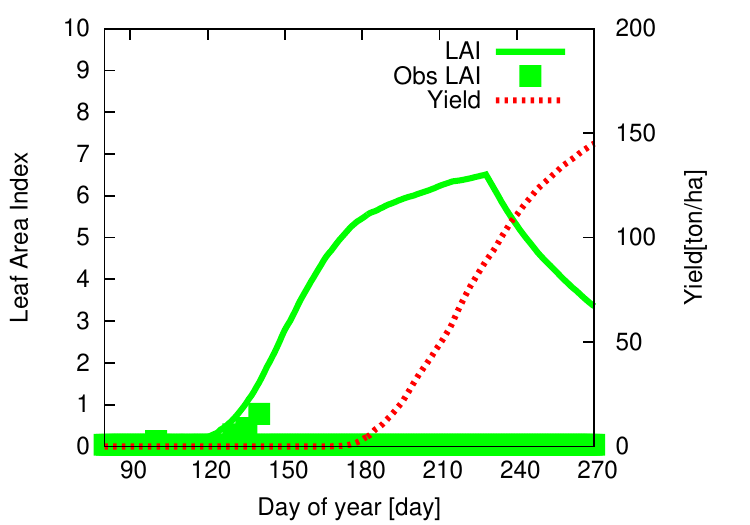}}
									\scalebox{0.29}{\includegraphics[scale=1.0, bb=0 0 216 151]{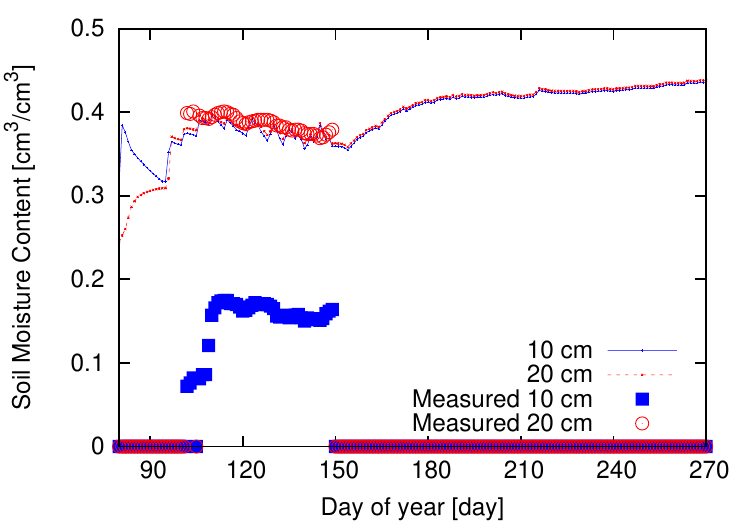}}
									\scalebox{0.29}{\includegraphics[scale=1.0, bb=0 0 216 151]{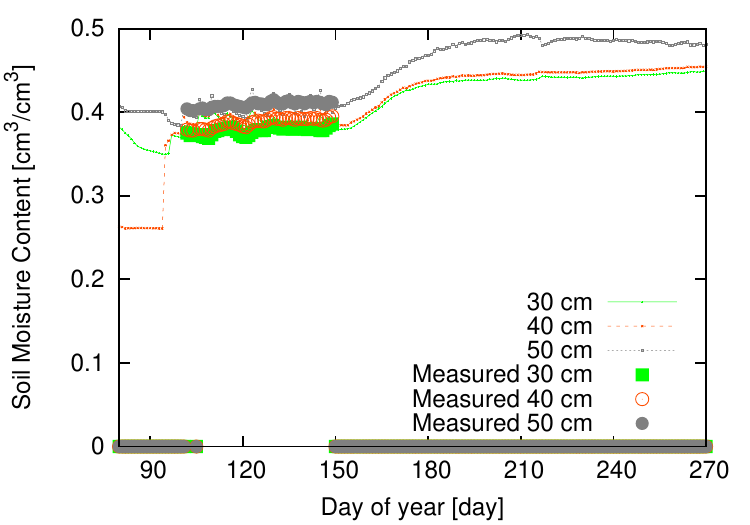}}
								\end{minipage} &
								\begin{minipage}{70mm}
									\centering
									\scalebox{0.29}{\includegraphics[scale=1.0, bb=0 0 216 151]{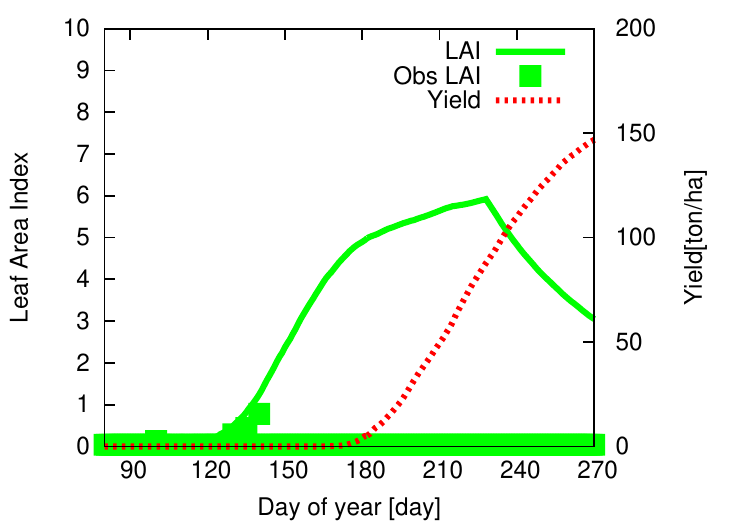}}
									\scalebox{0.29}{\includegraphics[scale=1.0, bb=0 0 216 151]{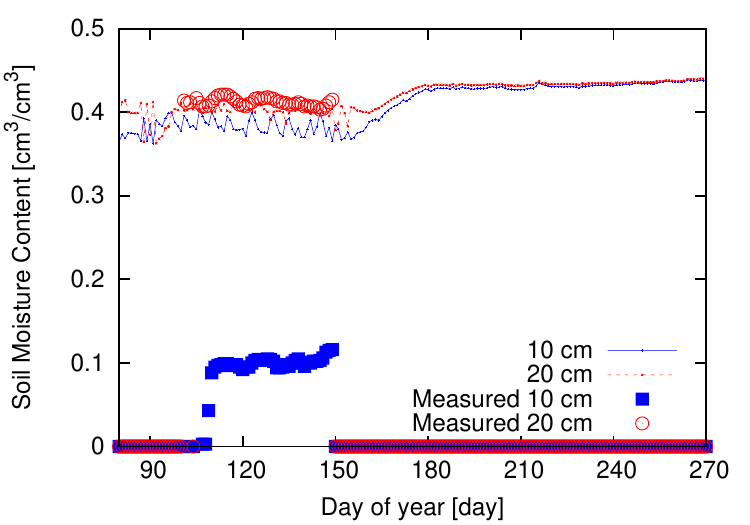}}
									\scalebox{0.29}{\includegraphics[scale=1.0, bb=0 0 216 151]{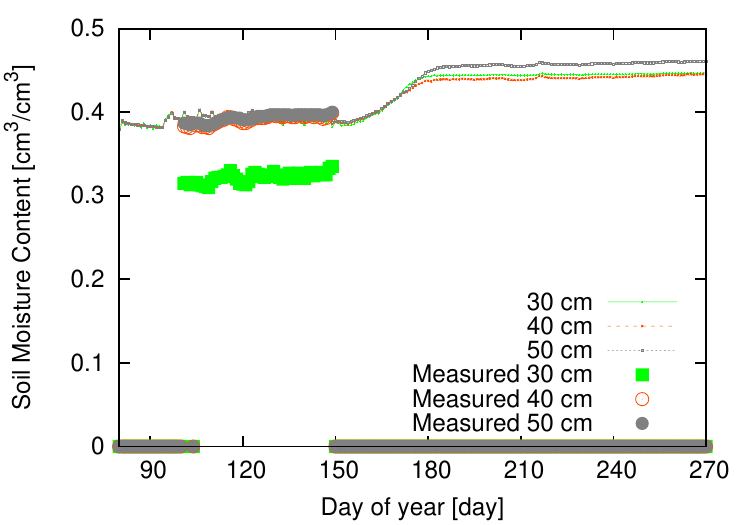}}
								\end{minipage} \\
			\hline 19th June & \begin{minipage}{70mm}
									\centering
									\scalebox{0.29}{\includegraphics[scale=1.0, bb=0 0 216 151]{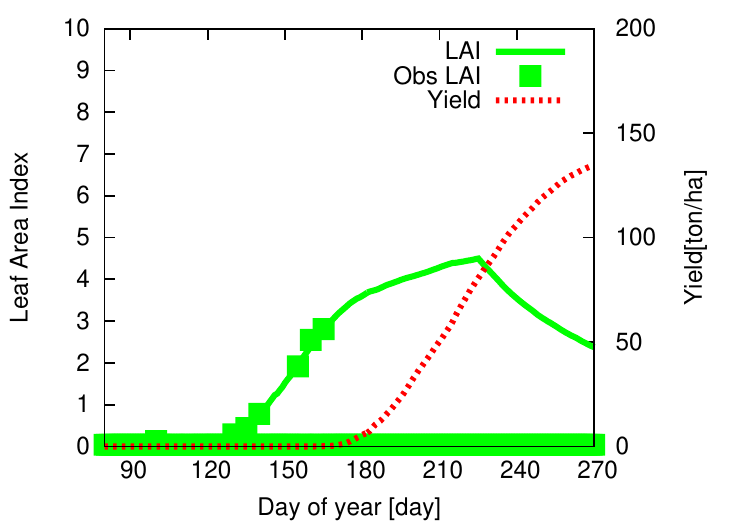}}
									\scalebox{0.29}{\includegraphics[scale=1.0, bb=0 0 216 151]{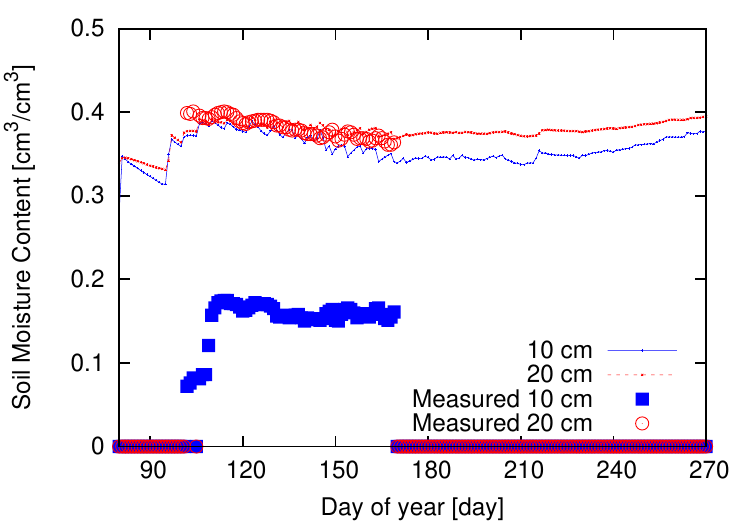}}
									\scalebox{0.29}{\includegraphics[scale=1.0, bb=0 0 216 151]{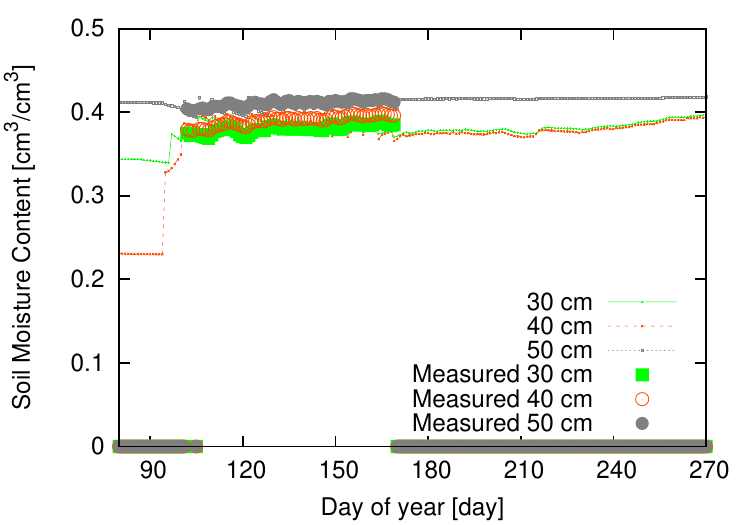}}
								\end{minipage} &
								\begin{minipage}{70mm}
									\centering
									\scalebox{0.29}{\includegraphics[scale=1.0, bb=0 0 216 151]{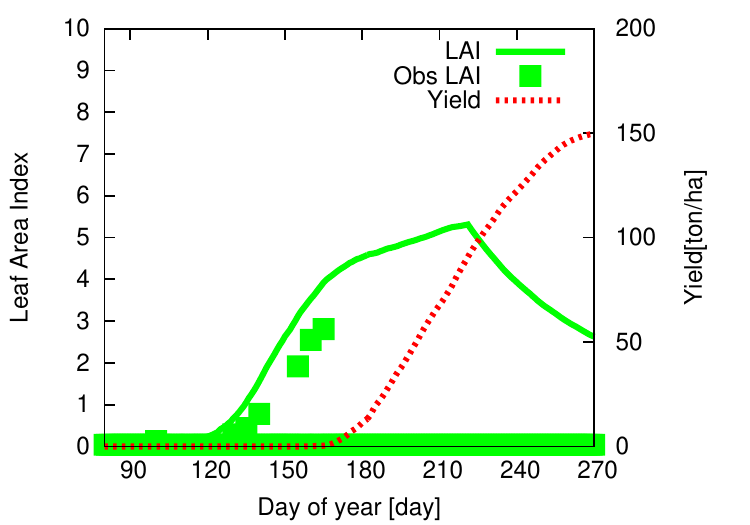}}
									\scalebox{0.29}{\includegraphics[scale=1.0, bb=0 0 216 151]{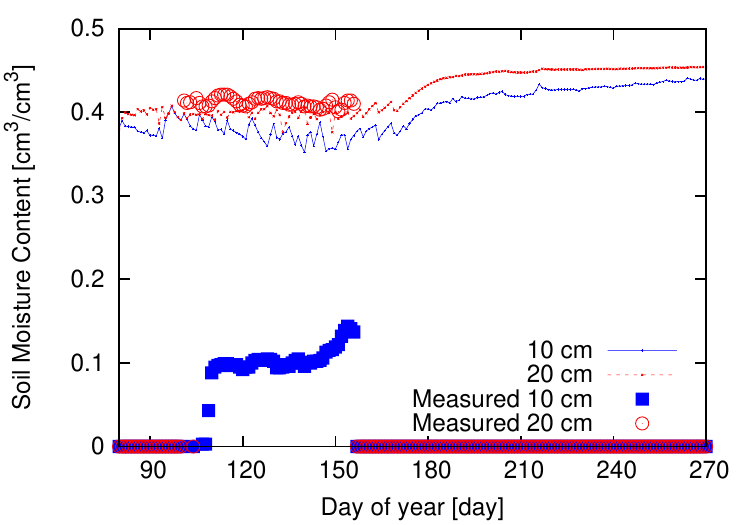}}
									\scalebox{0.29}{\includegraphics[scale=1.0, bb=0 0 216 151]{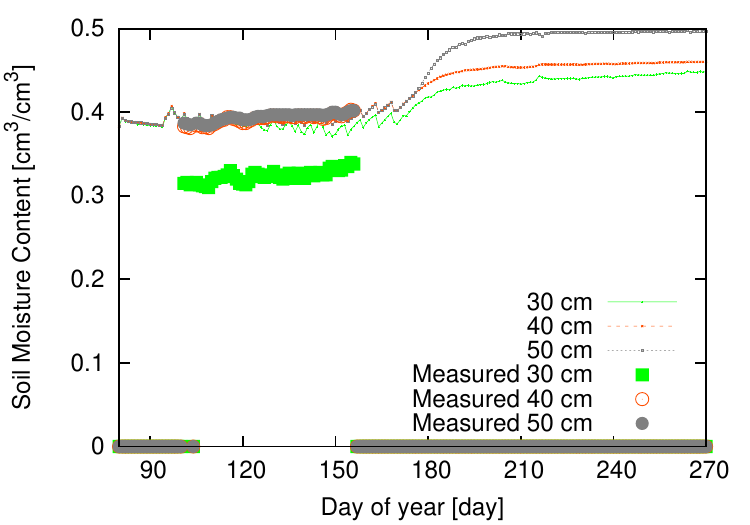}}
								\end{minipage} \\ \hline
		\end{tabular}
	\end{table}
	
	\begin{table}[H]
		\caption{Field C}
		\begin{tabular}{ccc}
			\hline & S1 & S2 \\
			\hline 10th May  & \begin{minipage}{70mm}
									\centering
									\scalebox{0.29}{\includegraphics[scale=1.0, bb=0 0 216 151]{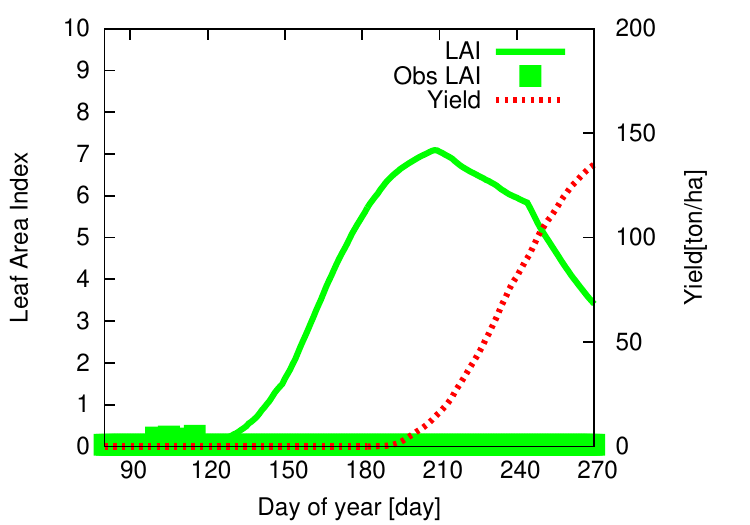}}
									\scalebox{0.29}{\includegraphics[scale=1.0, bb=0 0 216 151]{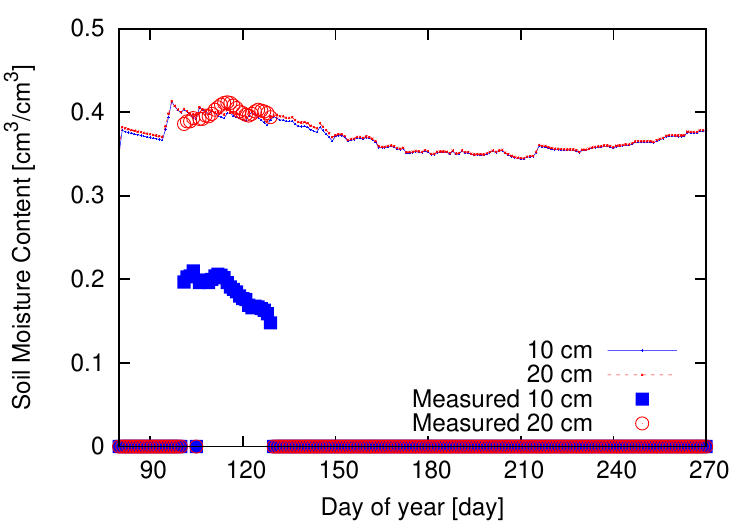}}
									\scalebox{0.29}{\includegraphics[scale=1.0, bb=0 0 216 151]{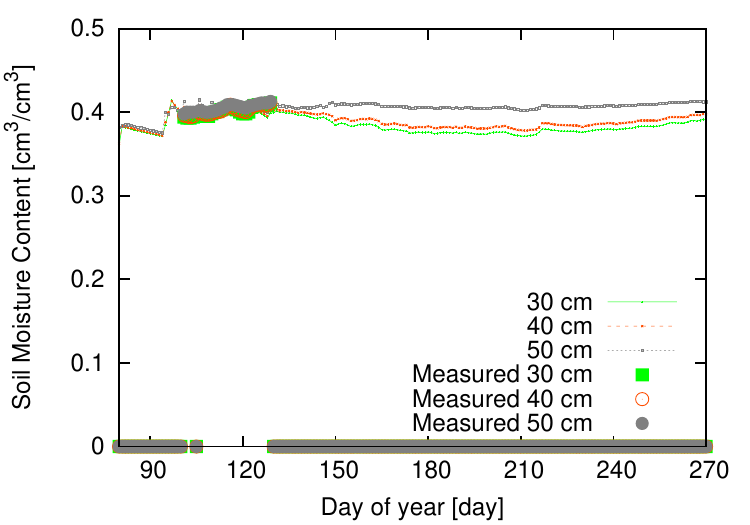}}
								\end{minipage} &
								\begin{minipage}{70mm}
									\centering
									\scalebox{0.29}{\includegraphics[scale=1.0, bb=0 0 216 151]{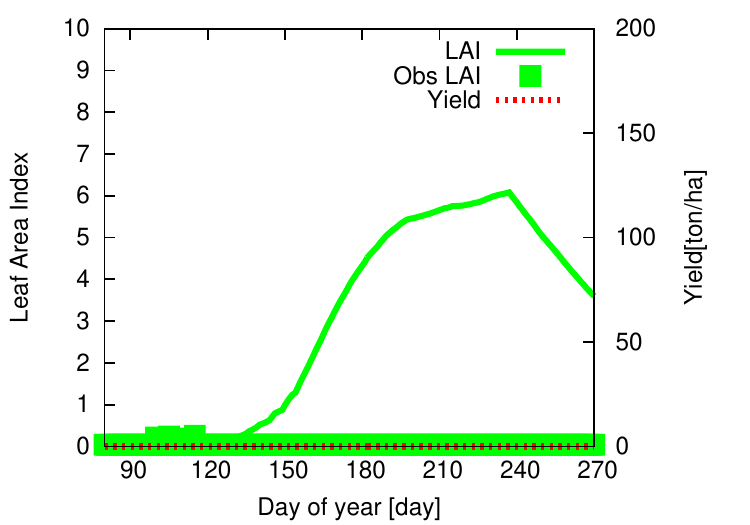}}
									\scalebox{0.29}{\includegraphics[scale=1.0, bb=0 0 216 151]{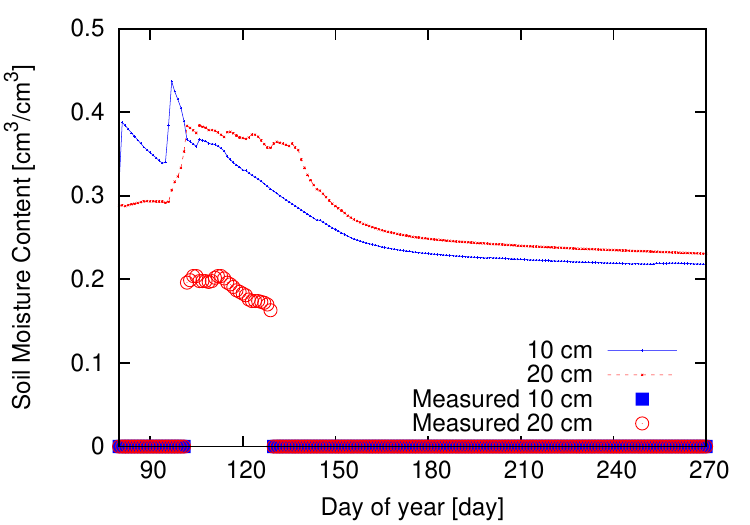}}
									\scalebox{0.29}{\includegraphics[scale=1.0, bb=0 0 216 151]{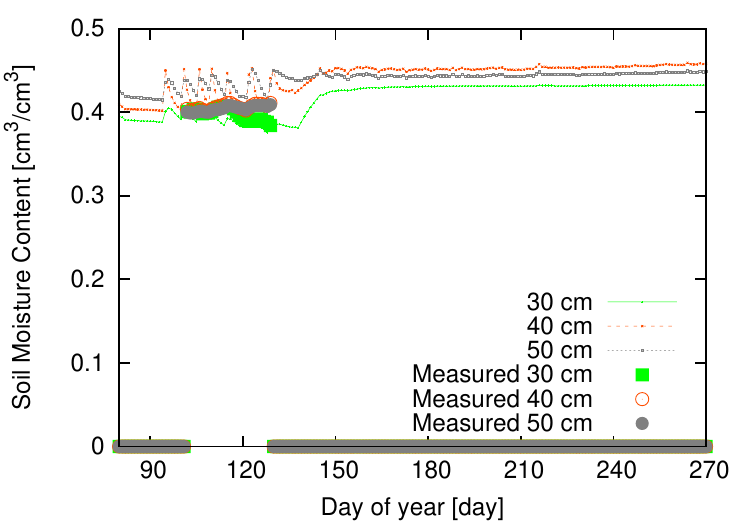}}
								\end{minipage} \\
			\hline 30th May  & \begin{minipage}{70mm}
									\centering
									\scalebox{0.29}{\includegraphics[scale=1.0, bb=0 0 216 151]{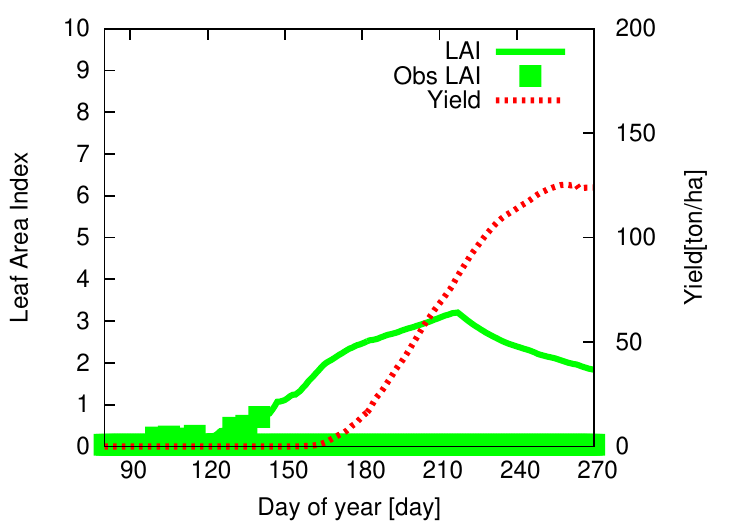}}
									\scalebox{0.29}{\includegraphics[scale=1.0, bb=0 0 216 151]{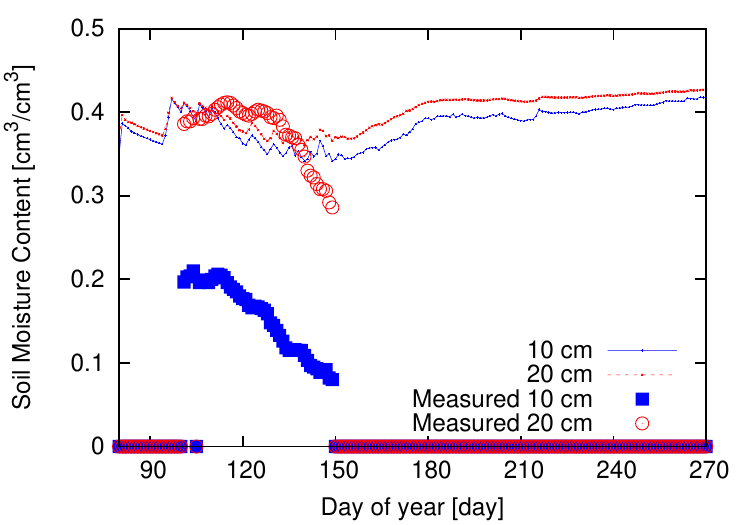}}
									\scalebox{0.29}{\includegraphics[scale=1.0, bb=0 0 216 151]{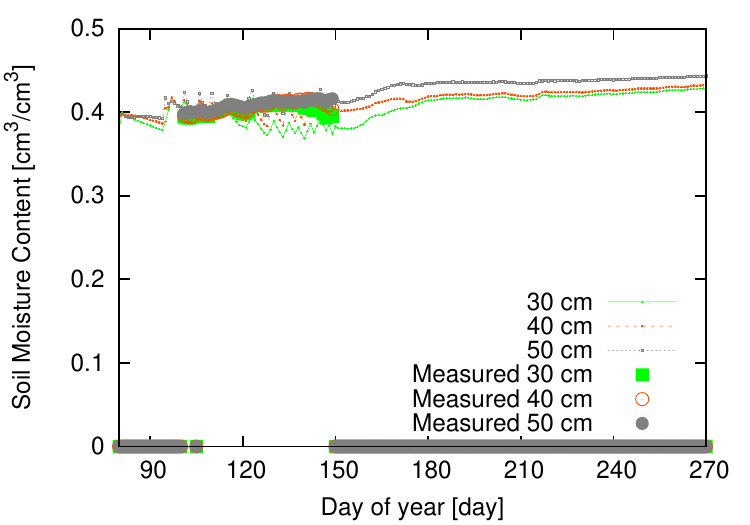}}
								\end{minipage} &
								\begin{minipage}{70mm}
									\centering
									\scalebox{0.29}{\includegraphics[scale=1.0, bb=0 0 216 151]{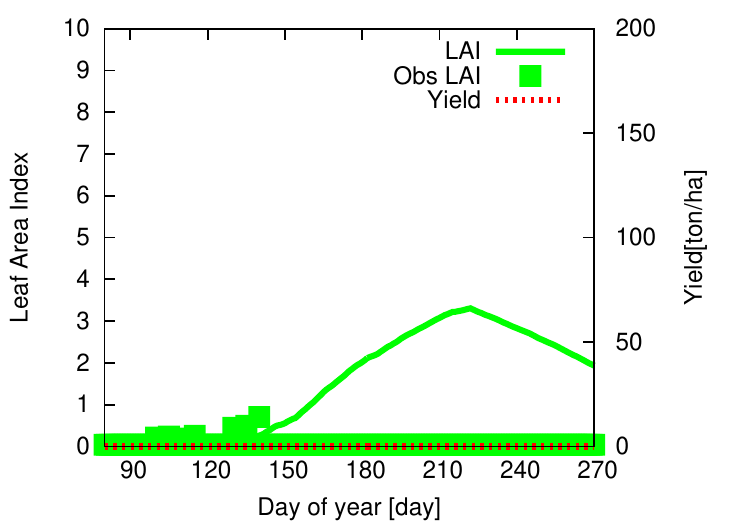}}
									\scalebox{0.29}{\includegraphics[scale=1.0, bb=0 0 216 151]{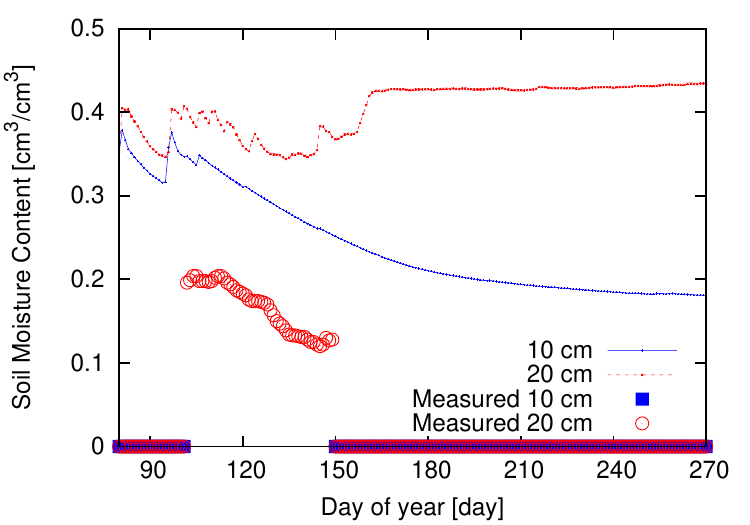}}
									\scalebox{0.29}{\includegraphics[scale=1.0, bb=0 0 216 151]{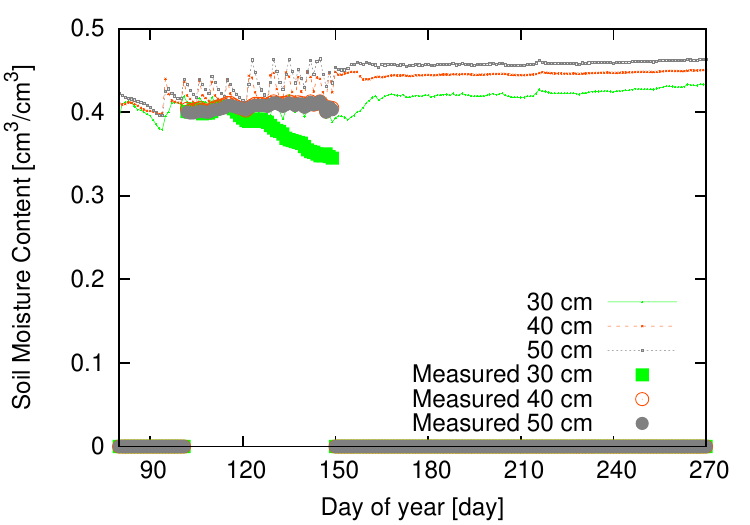}}
								\end{minipage} \\
			\hline 19th June & \begin{minipage}{70mm}
									\centering
									\scalebox{0.29}{\includegraphics[scale=1.0, bb=0 0 216 151]{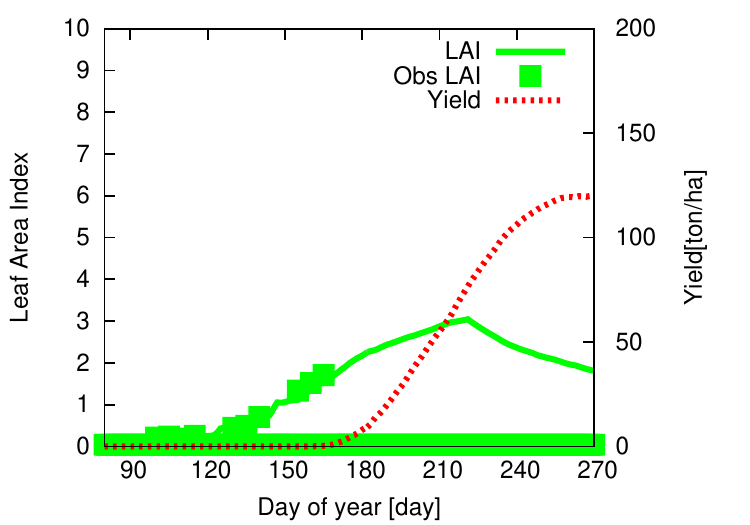}}
									\scalebox{0.29}{\includegraphics[scale=1.0, bb=0 0 216 151]{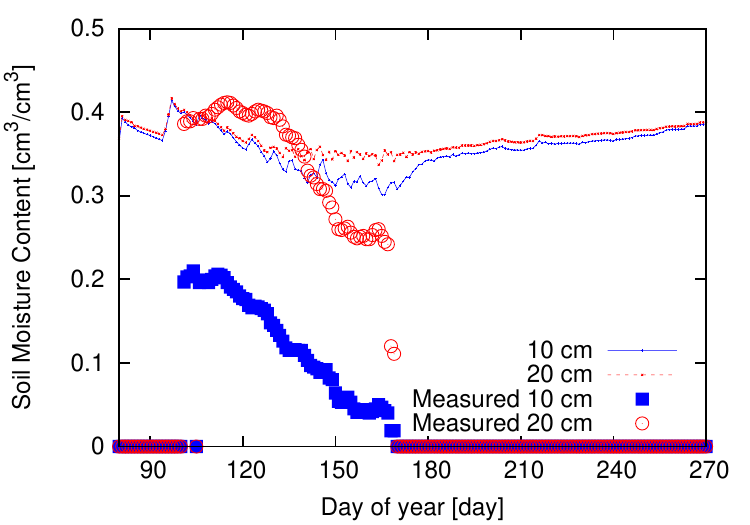}}
									\scalebox{0.29}{\includegraphics[scale=1.0, bb=0 0 216 151]{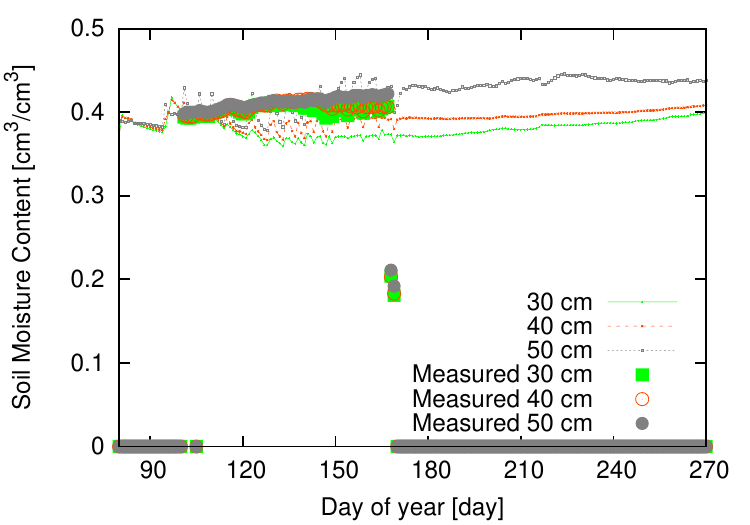}}
								\end{minipage} &
								\begin{minipage}{70mm}
									\centering
									\scalebox{0.29}{\includegraphics[scale=1.0, bb=0 0 216 151]{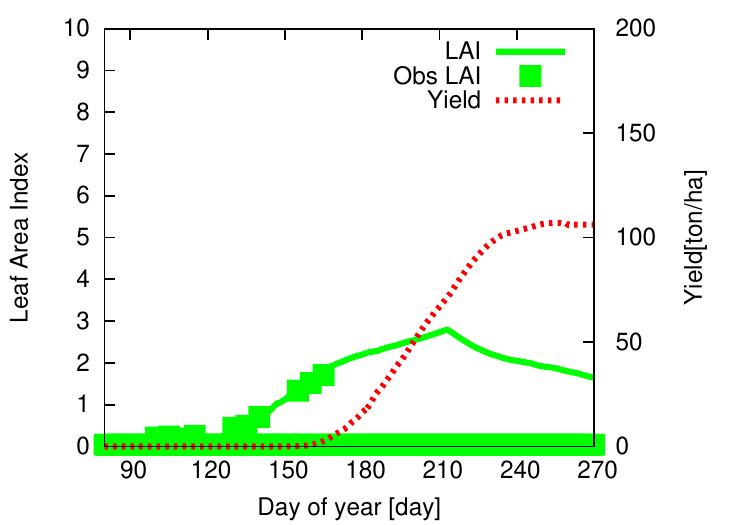}}
									\scalebox{0.29}{\includegraphics[scale=1.0, bb=0 0 216 151]{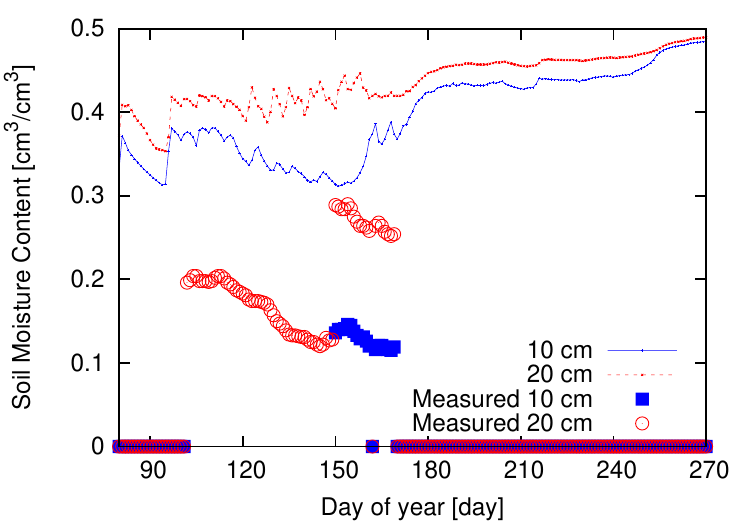}}
									\scalebox{0.29}{\includegraphics[scale=1.0, bb=0 0 216 151]{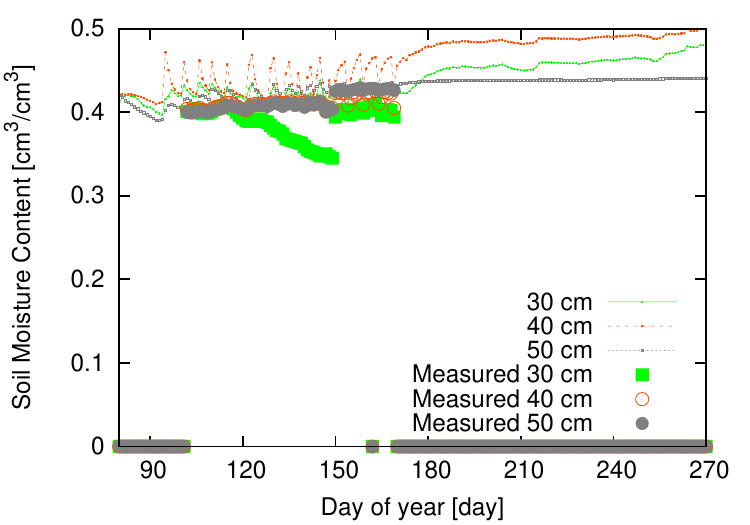}}
								\end{minipage} \\ \hline
		\end{tabular}
	\end{table}
	
	\begin{table}[H]
		\caption{Field D}
		\begin{tabular}{ccc}
			\hline & S1 & S2 \\
			\hline 10th May  & \begin{minipage}{70mm}
									\centering
									\scalebox{0.29}{\includegraphics[scale=1.0, bb=0 0 216 151]{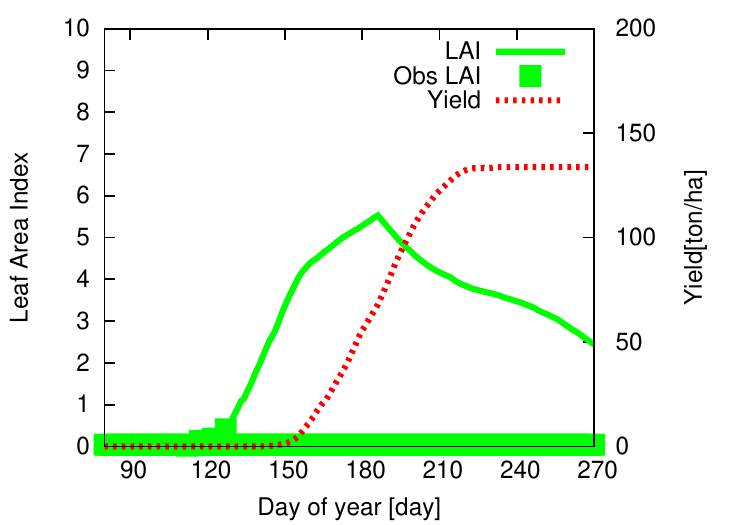}}
									\scalebox{0.29}{\includegraphics[scale=1.0, bb=0 0 216 151]{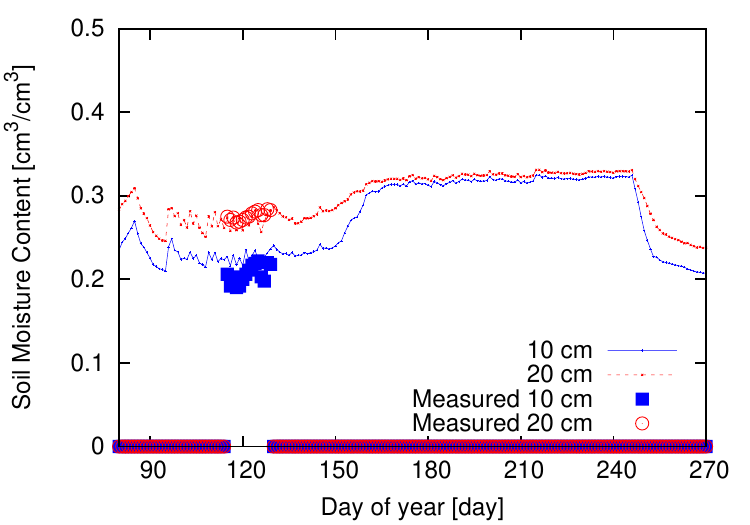}}
									\scalebox{0.29}{\includegraphics[scale=1.0, bb=0 0 216 151]{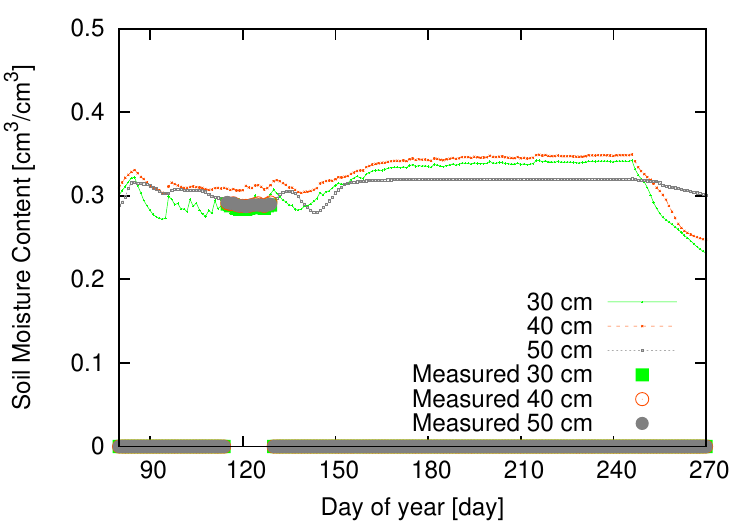}}
								\end{minipage} &
								\begin{minipage}{70mm}
									\centering
									\scalebox{0.29}{\includegraphics[scale=1.0, bb=0 0 216 151]{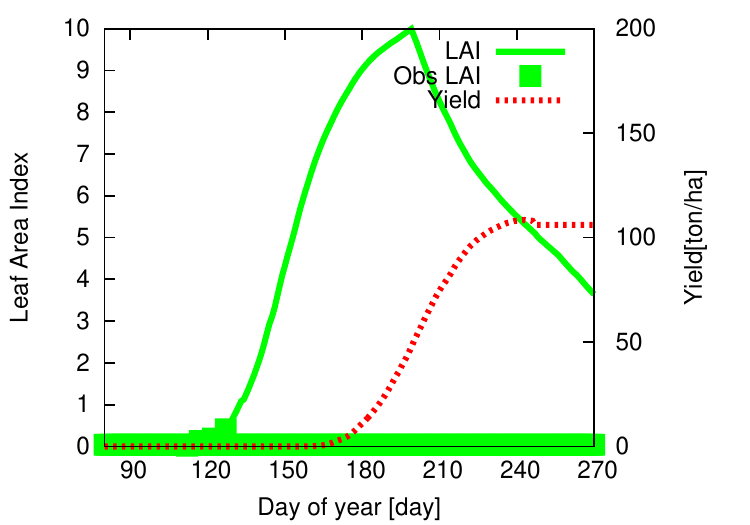}}
									\scalebox{0.29}{\includegraphics[scale=1.0, bb=0 0 216 151]{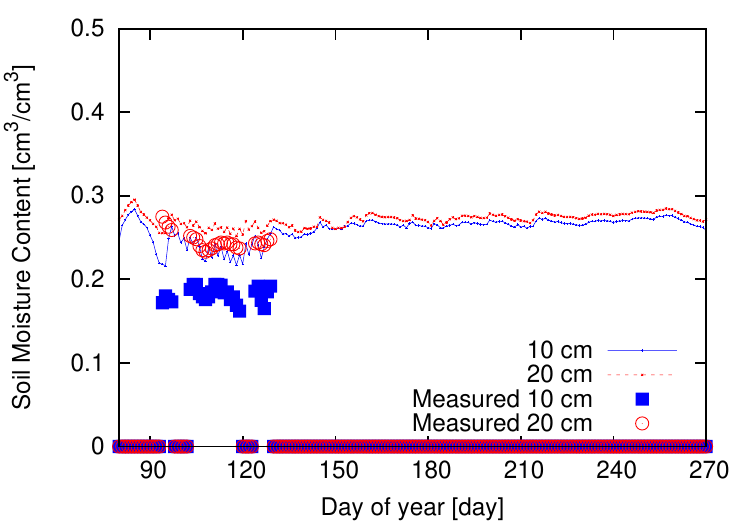}}
									\scalebox{0.29}{\includegraphics[scale=1.0, bb=0 0 216 151]{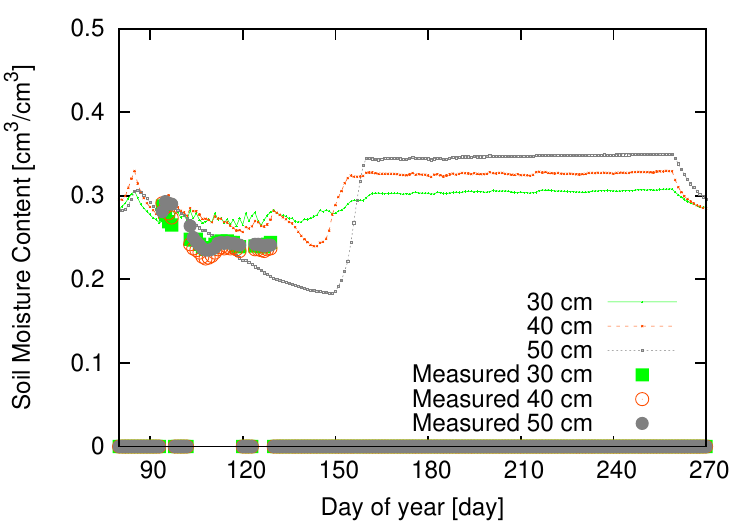}}
								\end{minipage} \\
			\hline 30th May  & \begin{minipage}{70mm}
									\centering
									\scalebox{0.29}{\includegraphics[scale=1.0, bb=0 0 216 151]{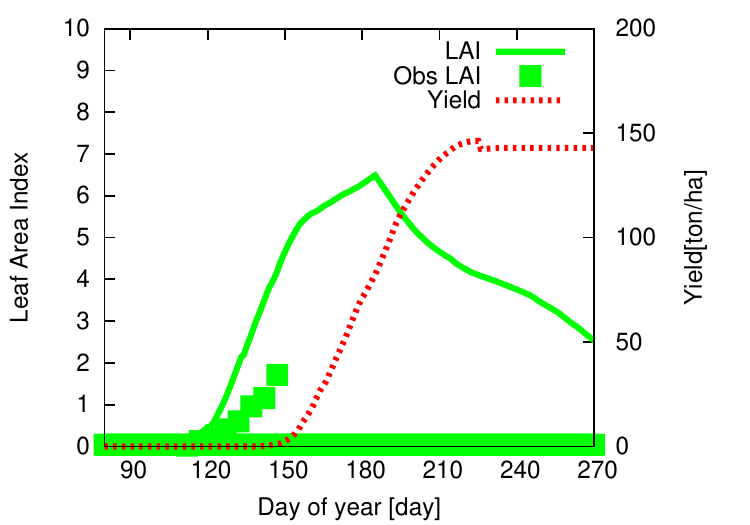}}
									\scalebox{0.29}{\includegraphics[scale=1.0, bb=0 0 216 151]{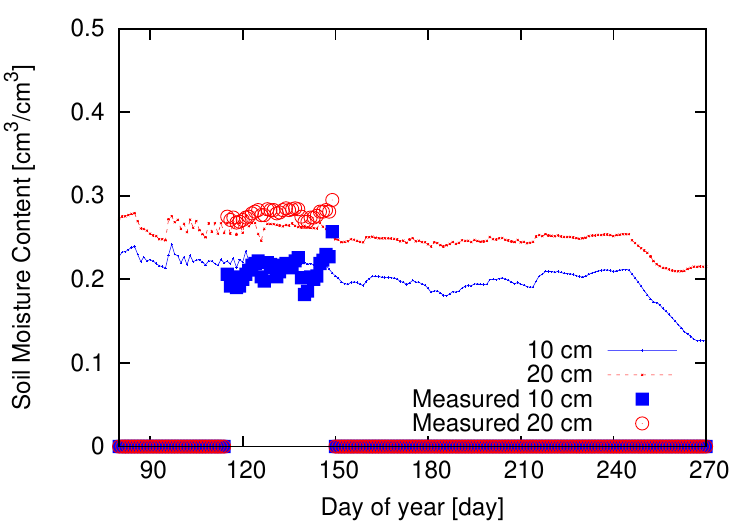}}
									\scalebox{0.29}{\includegraphics[scale=1.0, bb=0 0 216 151]{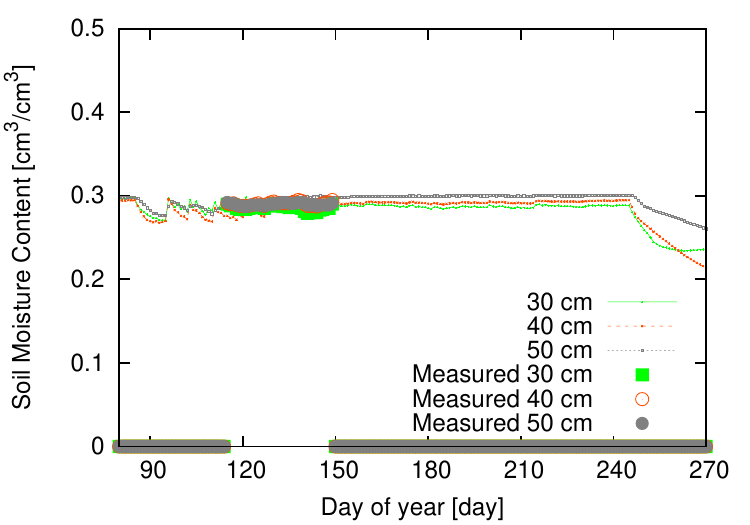}}
								\end{minipage} &
								\begin{minipage}{70mm}
									\centering
									\scalebox{0.29}{\includegraphics[scale=1.0, bb=0 0 216 151]{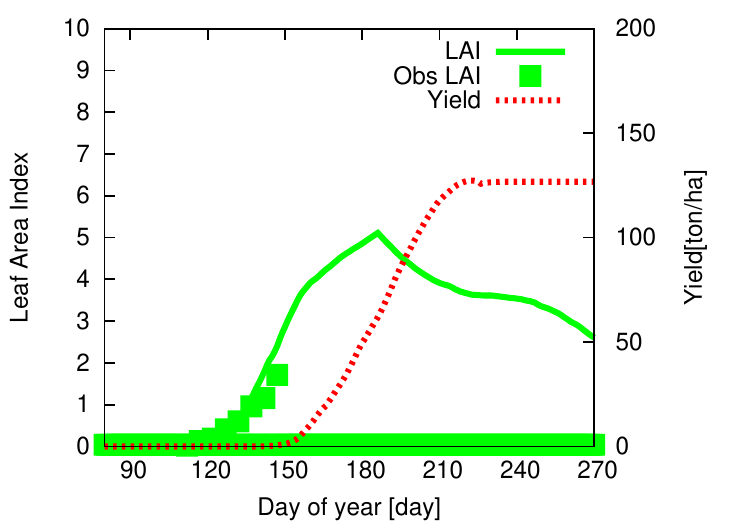}}
									\scalebox{0.29}{\includegraphics[scale=1.0, bb=0 0 216 151]{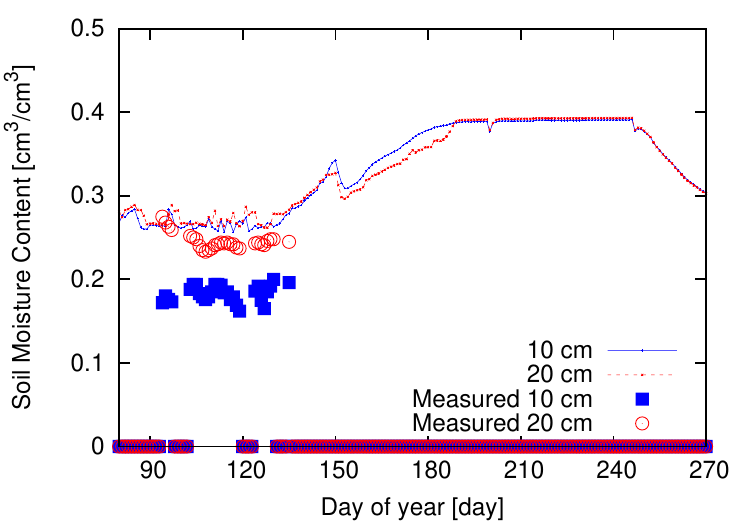}}
									\scalebox{0.29}{\includegraphics[scale=1.0, bb=0 0 216 151]{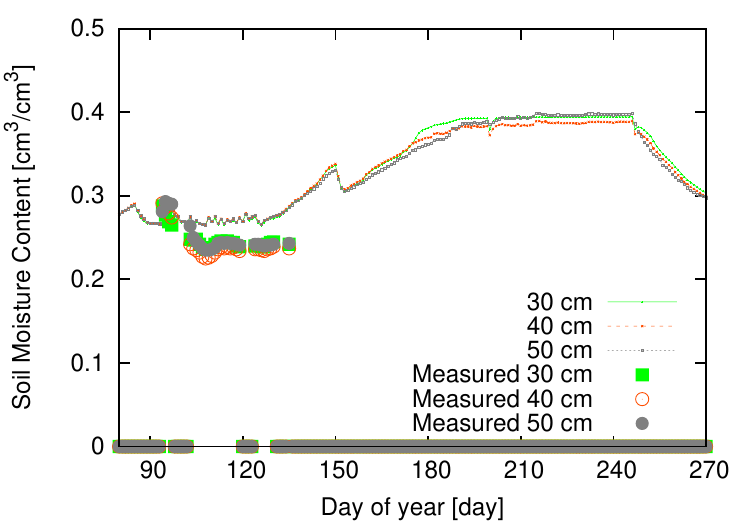}}
								\end{minipage} \\
			\hline 19th June & \begin{minipage}{70mm}
									\centering
									\scalebox{0.29}{\includegraphics[scale=1.0, bb=0 0 216 151]{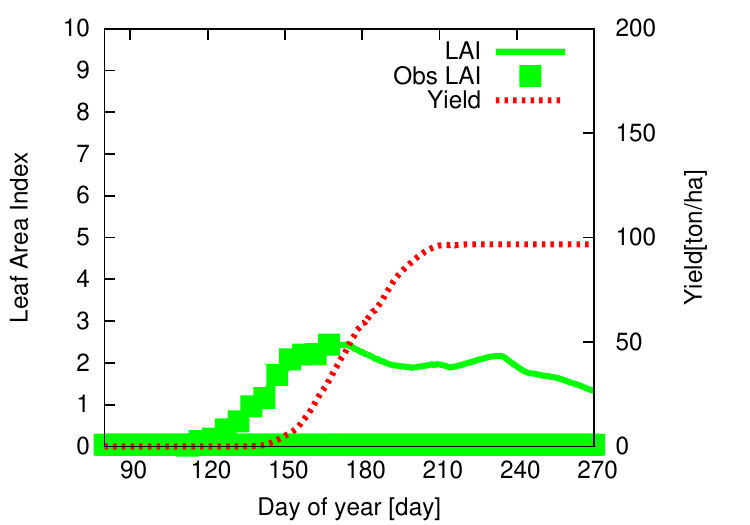}}
									\scalebox{0.29}{\includegraphics[scale=1.0, bb=0 0 216 151]{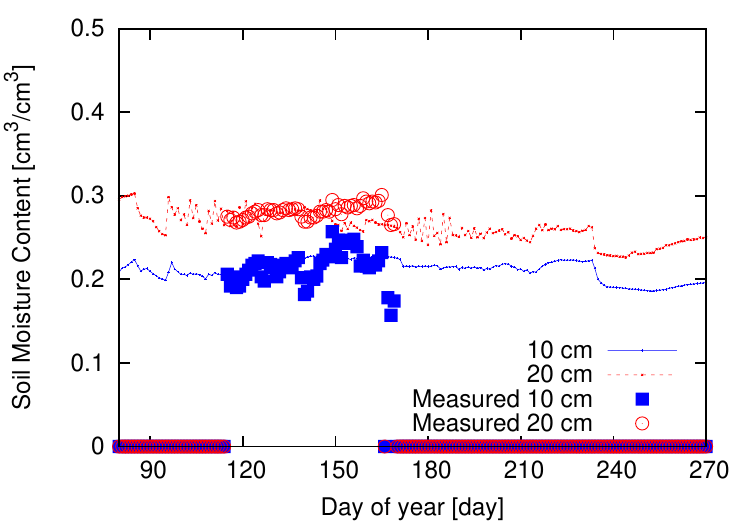}}
									\scalebox{0.29}{\includegraphics[scale=1.0, bb=0 0 216 151]{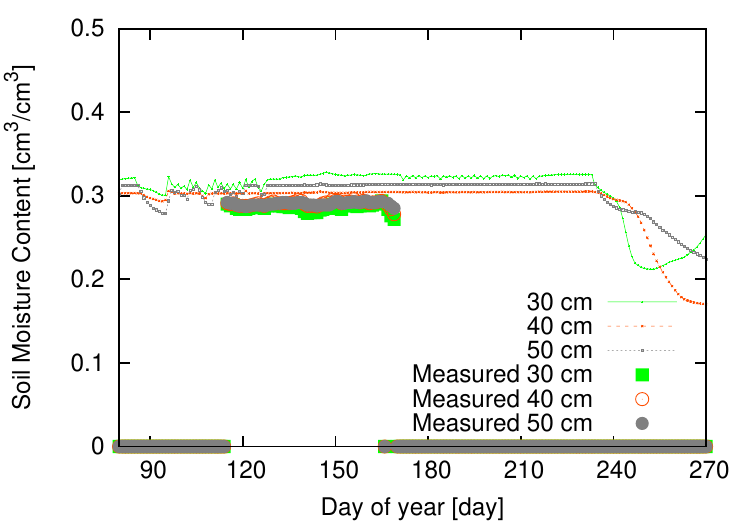}}
								\end{minipage} &
								\begin{minipage}{70mm}
									\centering
									\scalebox{0.29}{\includegraphics[scale=1.0, bb=0 0 216 151]{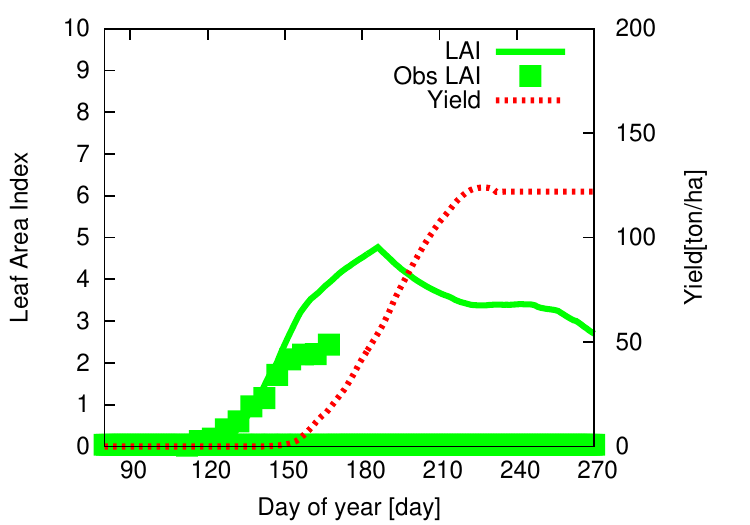}}
									\scalebox{0.29}{\includegraphics[scale=1.0, bb=0 0 216 151]{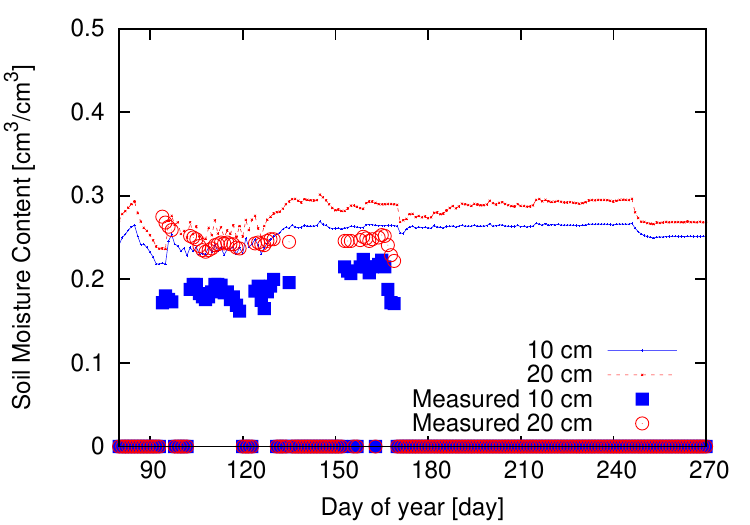}}
									\scalebox{0.29}{\includegraphics[scale=1.0, bb=0 0 216 151]{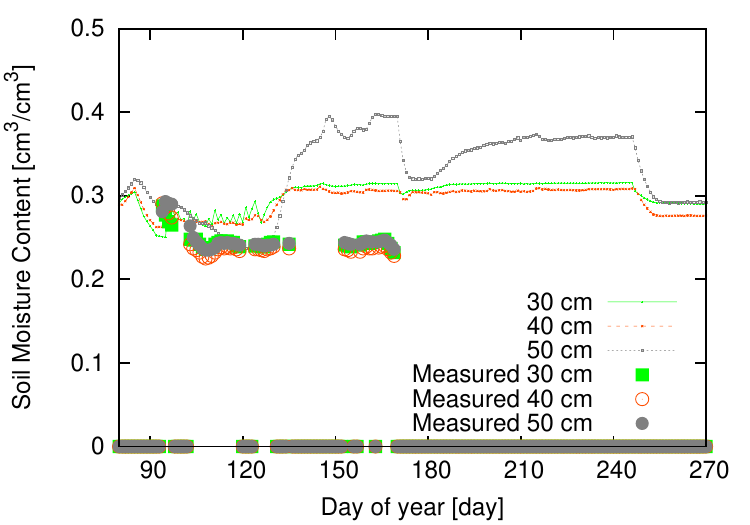}}
								\end{minipage} \\ \hline
		\end{tabular}
	\end{table}
	
	\begin{table}[H]
		\caption{Field E}
		\begin{tabular}{ccc}
			\hline & S1 & S2 \\
			\hline 10th May  & \begin{minipage}{70mm}
									\centering
									\scalebox{0.29}{\includegraphics[scale=1.0, bb=0 0 216 151]{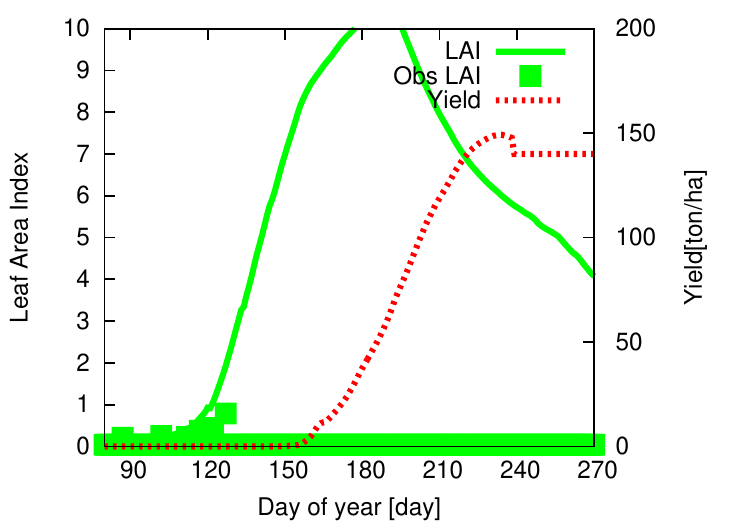}}
									\scalebox{0.29}{\includegraphics[scale=1.0, bb=0 0 216 151]{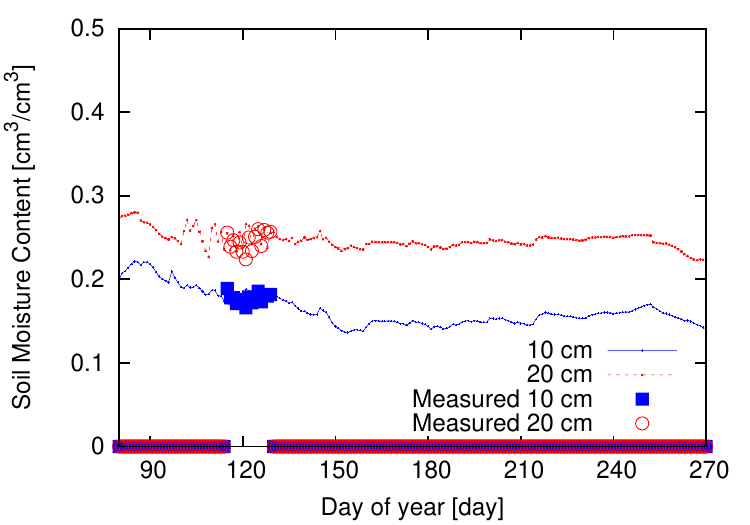}}
									\scalebox{0.29}{\includegraphics[scale=1.0, bb=0 0 216 151]{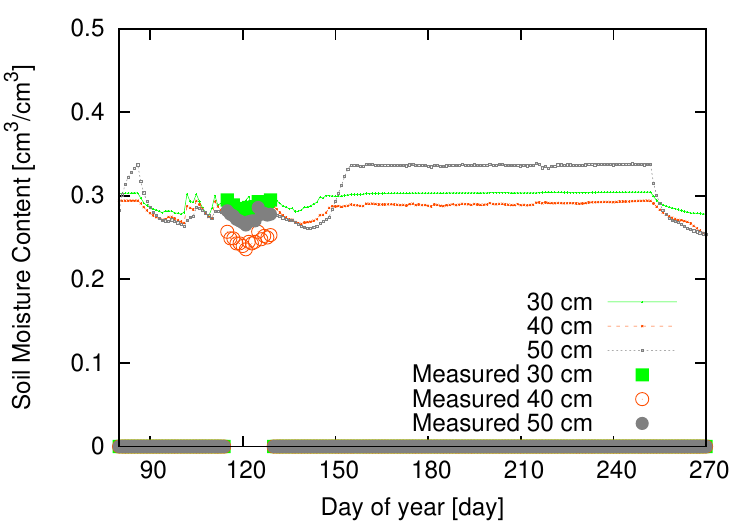}}
								\end{minipage} &
								\begin{minipage}{70mm}
									\centering
									\scalebox{0.29}{\includegraphics[scale=1.0, bb=0 0 216 151]{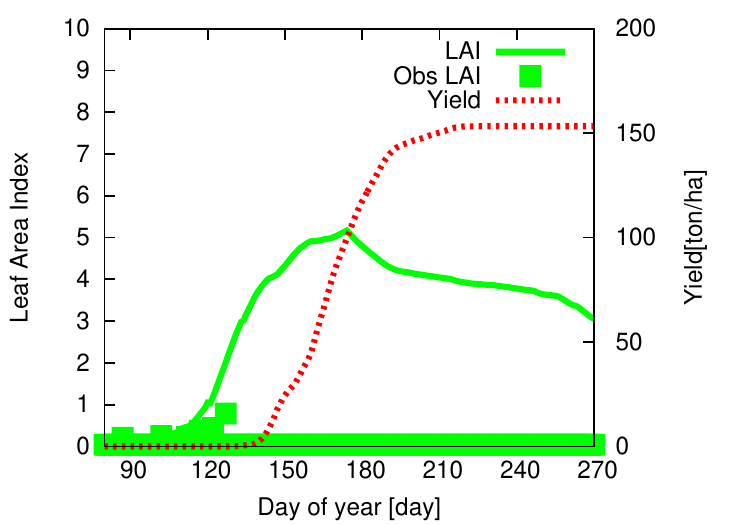}}
									\scalebox{0.29}{\includegraphics[scale=1.0, bb=0 0 216 151]{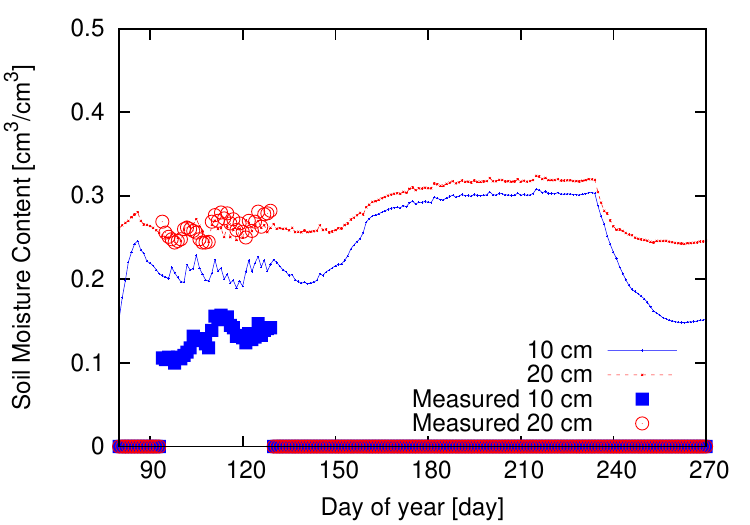}}
									\scalebox{0.29}{\includegraphics[scale=1.0, bb=0 0 216 151]{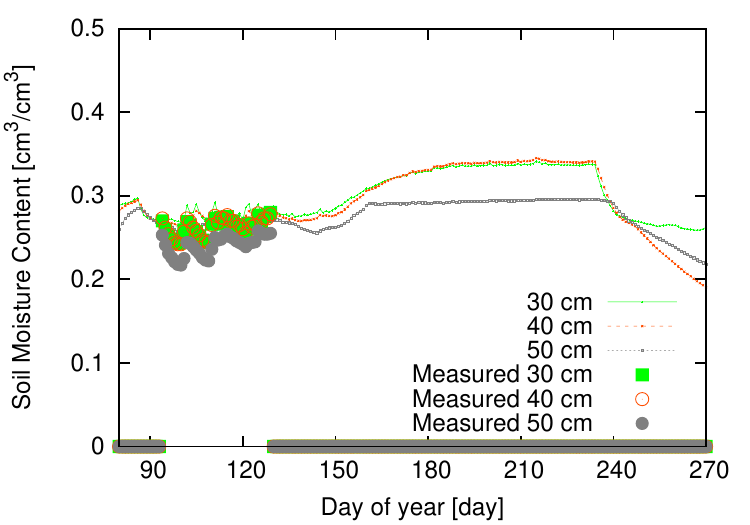}}
								\end{minipage} \\
			\hline 30th May  & \begin{minipage}{70mm}
									\centering
									\scalebox{0.29}{\includegraphics[scale=1.0, bb=0 0 216 151]{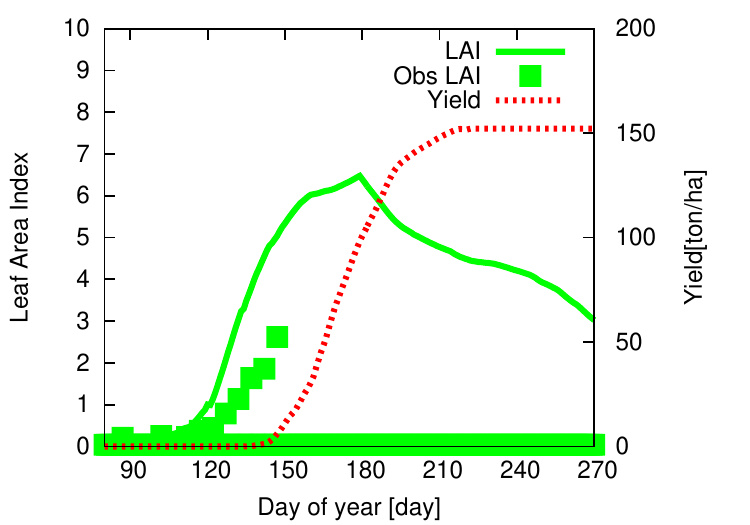}}
									\scalebox{0.29}{\includegraphics[scale=1.0, bb=0 0 216 151]{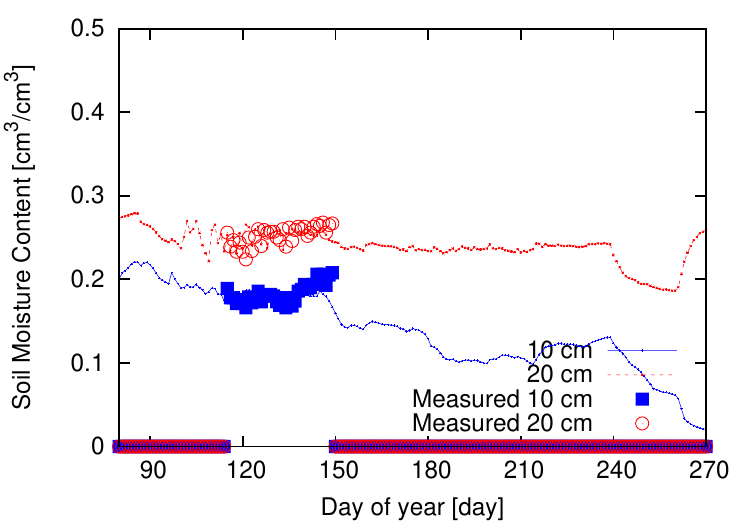}}
									\scalebox{0.29}{\includegraphics[scale=1.0, bb=0 0 216 151]{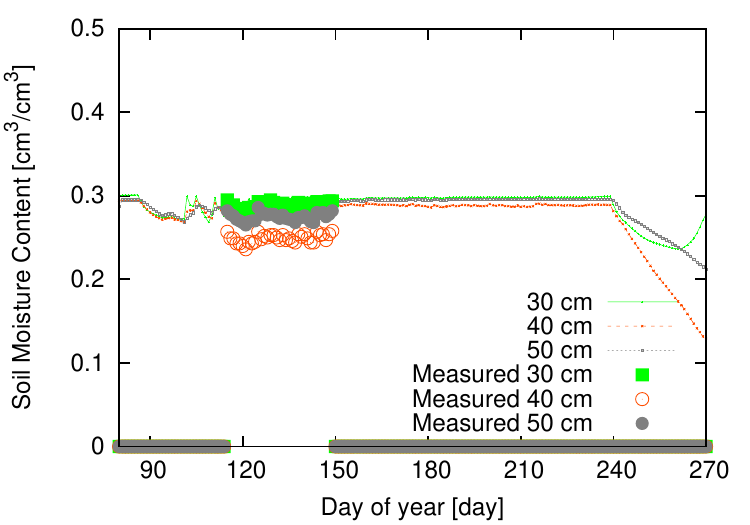}}
								\end{minipage} &
								\begin{minipage}{70mm}
									\centering
									\scalebox{0.29}{\includegraphics[scale=1.0, bb=0 0 216 151]{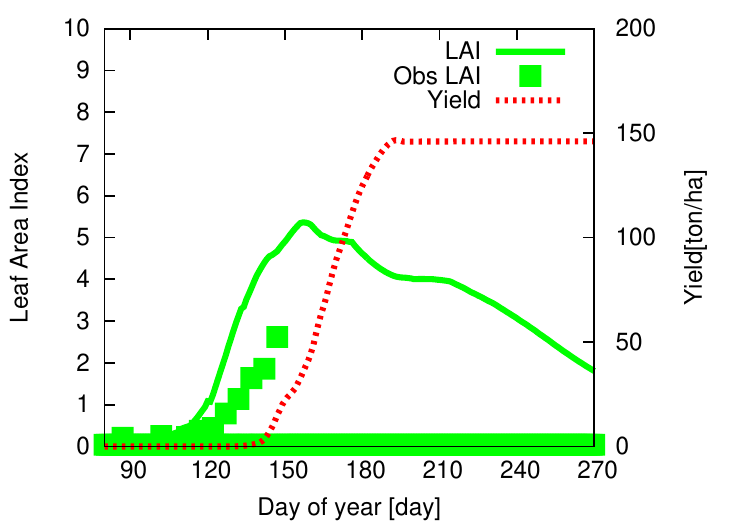}}
									\scalebox{0.29}{\includegraphics[scale=1.0, bb=0 0 216 151]{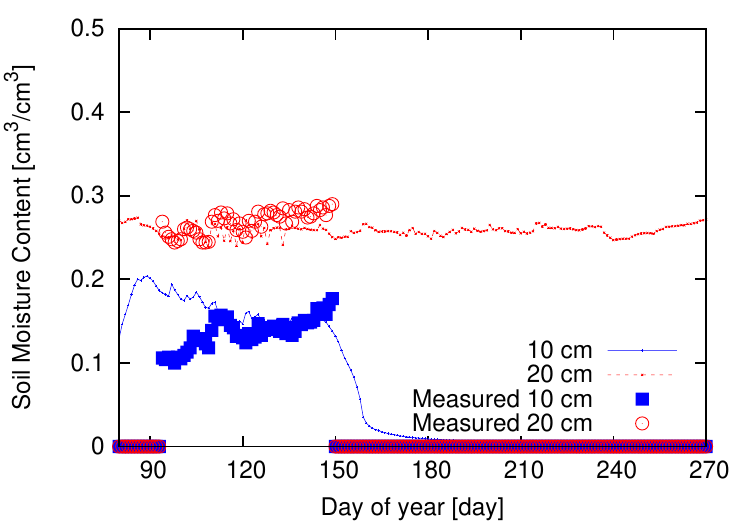}}
									\scalebox{0.29}{\includegraphics[scale=1.0, bb=0 0 216 151]{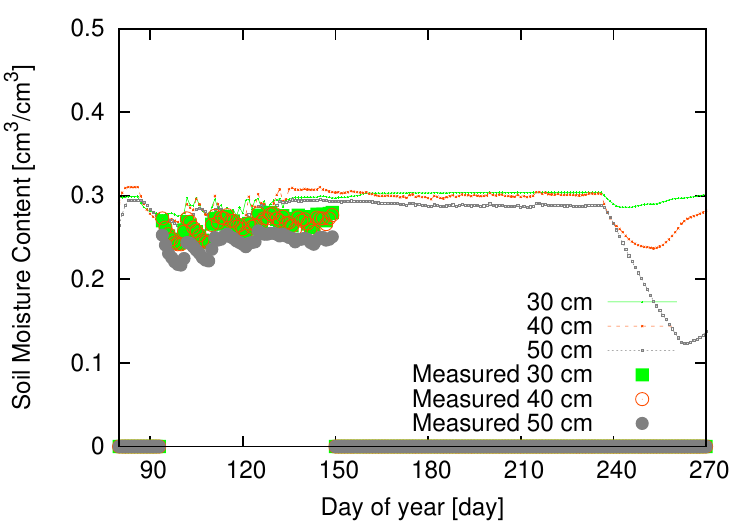}}
								\end{minipage} \\
			\hline 19th June & \begin{minipage}{70mm}
									\centering
									\scalebox{0.29}{\includegraphics[scale=1.0, bb=0 0 216 151]{./M_p_g_LAI_Y_us_FieldE_0619_190306001_S3.pdf}}
									\scalebox{0.29}{\includegraphics[scale=1.0, bb=0 0 216 151]{./M_p_g_SW_shallow_us_FieldE_0619_190306001_S3.pdf}}
									\scalebox{0.29}{\includegraphics[scale=1.0, bb=0 0 216 151]{./M_p_g_SW_deep_us_FieldE_0619_190306001_S3.pdf}}
								\end{minipage} &
								\begin{minipage}{70mm}
									\centering
									\scalebox{0.29}{\includegraphics[scale=1.0, bb=0 0 216 151]{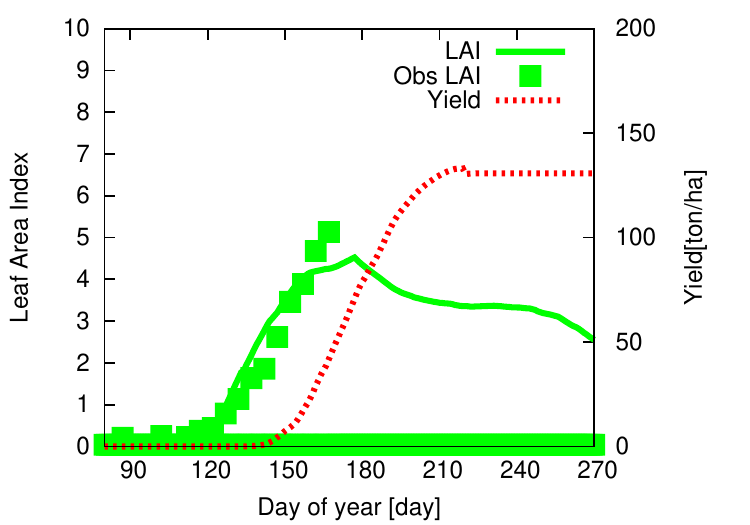}}
									\scalebox{0.29}{\includegraphics[scale=1.0, bb=0 0 216 151]{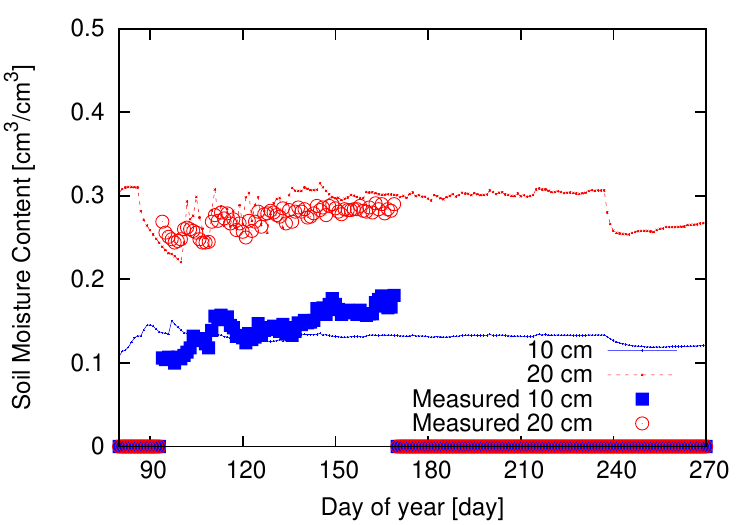}}
									\scalebox{0.29}{\includegraphics[scale=1.0, bb=0 0 216 151]{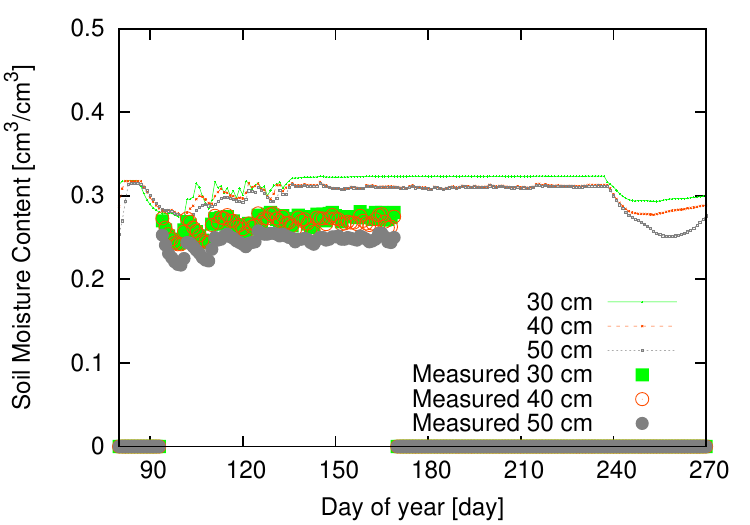}}
								\end{minipage} \\ \hline
		\end{tabular}
	\end{table}
	
\section{Difference between the soil moisture values during observation and analysis}
	
	\begin{table}[H]
	\caption{$\epsilon_{m, f, S, l, A_d}$: residue of soil moisture.
			Average soil moisture difference between observed data and simulation results for each soil layer.}
	\begin{tabular}{ccccccccccccc} \hline
		& & & \multicolumn{5}{p{5.5cm}}{Average difference between obs and sim by original DSSAT [m$^3$/m$^3$]} & \multicolumn{5}{p{5.5cm}}{Average difference between obs and sim by modified DSSAT [m$^3$/m$^3$]} \\ \cline{4-13}
		& & & \multicolumn{5}{p{5cm}}{Layer [cm]} & \multicolumn{5}{p{5cm}}{Layer [cm]} \\ \cline{4-13}
		Field & Sensor & Analysis day & 10 & 20 & 30 & 40 & 50 &  10 & 20 & 30 & 40 & 50 \\ \hline
		A & S1 & May 10  & 0.028 & 0.049 & 0.043 & 0.007 & 0.014 & 0.009 & 0.024 & 0.002 & 0.010 & 0.017 \\ \cline{2-13}
		  & S2 & May 10  & 0.015 & 0.049 & 0.028 & 0.021 & 0.008 & 0.018 & 0.017 & 0.011 & 0.013 & 0.007 \\ \cline{2-13}
		  & S1 & May 20  & 0.025 & 0.136 & 0.103 & 0.053 & 0.034 & 0.013 & 0.068 & 0.009 & 0.007 & 0.030 \\ \cline{2-13}
		  & S2 & May 20  & 0.022 & 0.114 & 0.083 & 0.047 & 0.022 & 0.078 & 0.014 & 0.009 & 0.005 & 0.026 \\ \cline{2-13}
		  & S1 & June 19 & 0.055 & 0.112 & 0.122 & 0.071 & 0.026 & 0.120 & 0.028 & 0.031 & 0.020 & 0.028 \\ \cline{2-13}
		  & S2 & June 19 & 0.025 & 0.142 & 0.141 & 0.043 & 0.028 & 0.051 & 0.040 & 0.025 & 0.017 & 0.102 \\ \hline
		B & S1 & May 10  & 0.252 & 0.025 & 0.077 & 0.075 & 0.053 & 0.246 & 0.006 & 0.020 & 0.016 & 0.011 \\ \cline{2-13}
		  & S2 & May 10  & 0.317 & 0.038 & 0.066 & 0.044 & 0.050 & 0.306 & 0.008 & 0.067 & 0.006 & 0.006 \\ \cline{2-13}
		  & S1 & May 20  & 0.184 & 0.194 & 0.131 & 0.113 & 0.040 & 0.228 & 0.009 & 0.014 & 0.011 & 0.008 \\ \cline{2-13}
		  & S2 & May 20  & 0.282 & 0.074 & 0.049 & 0.054 & 0.044 & 0.303 & 0.011 & 0.071 & 0.008 & 0.007 \\ \cline{2-13}
		  & S1 & June 19 & 0.183 & 0.103 & 0.201 & 0.083 & 0.089 & 0.218 & 0.009 & 0.011 & 0.013 & 0.006 \\ \cline{2-13}
		  & S2 & June 19 & 0.257 & 0.108 & 0.056 & 0.079 & 0.049 & 0.288 & 0.017 & 0.062 & 0.007 & 0.007 \\ \hline
		C & S1 & May 10  & 0.205 & 0.032 & 0.057 & 0.065 & 0.045 & 0.210 & 0.008 & 0.007 & 0.006 & 0.007 \\ \cline{2-13}
		  & S2 & May 10  & 0.375 & 0.086 & 0.073 & 0.019 & 0.032 & 0.365 & 0.185 & 0.011 & 0.022 & 0.029 \\ \cline{2-13}
		  & S1 & May 20  & 0.155 & 0.085 & 0.094 & 0.048 & 0.063 & 0.217 & 0.032 & 0.021 & 0.014 & 0.010 \\ \cline{2-13}
		  & S2 & May 20  & 0.342 & 0.122 & 0.079 & 0.039 & 0.041 & 0.322 & 0.205 & 0.031 & 0.021 & 0.033 \\ \cline{2-13}
		  & S1 & June 19 & 0.212 & 0.042 & 0.093 & 0.049 & 0.054 & 0.225 & 0.068 & 0.044 & 0.041 & 0.040 \\ \cline{2-13}
		  & S2 & June 19 & 0.302 & 0.082 & 0.103 & 0.044 & 0.090 & 0.332 & 0.227 & 0.039 & 0.031 & 0.015 \\ \hline
		D & S1 & May 10  & 0.044 & 0.019 & 0.006 & 0.033 & 0.025 & 0.021 & 0.010 & 0.007 & 0.020 & 0.002 \\ \cline{2-13}
		  & S2 & May 10  & 0.049 & 0.094 & 0.075 & 0.073 & 0.048 & 0.055 & 0.019 & 0.029 & 0.034 & 0.019 \\ \cline{2-13}
		  & S1 & May 20  & 0.067 & 0.070 & 0.034 & 0.013 & 0.062 & 0.018 & 0.018 & 0.008 & 0.008 & 0.007 \\ \cline{2-13}
		  & S2 & May 20  & 0.038 & 0.091 & 0.098 & 0.058 & 0.065 & 0.084 & 0.027 & 0.027 & 0.035 & 0.028 \\ \cline{2-13}
		  & S1 & June 19 & 0.113 & 0.072 & 0.072 & 0.052 & 0.052 & 0.020 & 0.016 & 0.036 & 0.013 & 0.021 \\ \cline{2-13}
		  & S2 & June 19 & 0.108 & 0.051 & 0.104 & 0.090 & 0.090 & 0.057 & 0.032 & 0.050 & 0.049 & 0.083 \\ \hline
		E & S1 & May 10  & 0.078 & 0.052 & 0.021 & 0.139 & 0.093 & 0.008 & 0.014 & 0.008 & 0.032 & 0.005 \\ \cline{2-13}
		  & S2 & May 10  & 0.051 & 0.078 & 0.036 & 0.075 & 0.096 & 0.081 & 0.012 & 0.018 & 0.014 & 0.025 \\ \cline{2-13}
		  & S1 & May 20  & 0.027 & 0.124 & 0.088 & 0.058 & 0.040 & 0.014 & 0.012 & 0.010 & 0.038 & 0.015 \\ \cline{2-13}
		  & S2 & May 20  & 0.063 & 0.077 & 0.063 & 0.065 & 0.096 & 0.038 & 0.018 & 0.025 & 0.028 & 0.040 \\ \cline{2-13}
		  & S1 & June 19 & 0.061 & 0.093 & 0.087 & 0.053 & 0.047 & 0.021 & 0.015 & 0.007 & 0.036 & 0.012 \\ \cline{2-13}
		  & S2 & June 19 & 0.062 & 0.053 & 0.099 & 0.064 & 0.104 & 0.025 & 0.021 & 0.044 & 0.039 & 0.056 \\ \hline
		
	\end{tabular}
	\label{residue}
	\end{table}
	
\section{Discretization errors depending on $\Delta t$}\label{dt_trend}
	\par
\begin{figure}[H]
		\begin{center}
		\includegraphics[scale=1.0, bb=0 0 461 346]{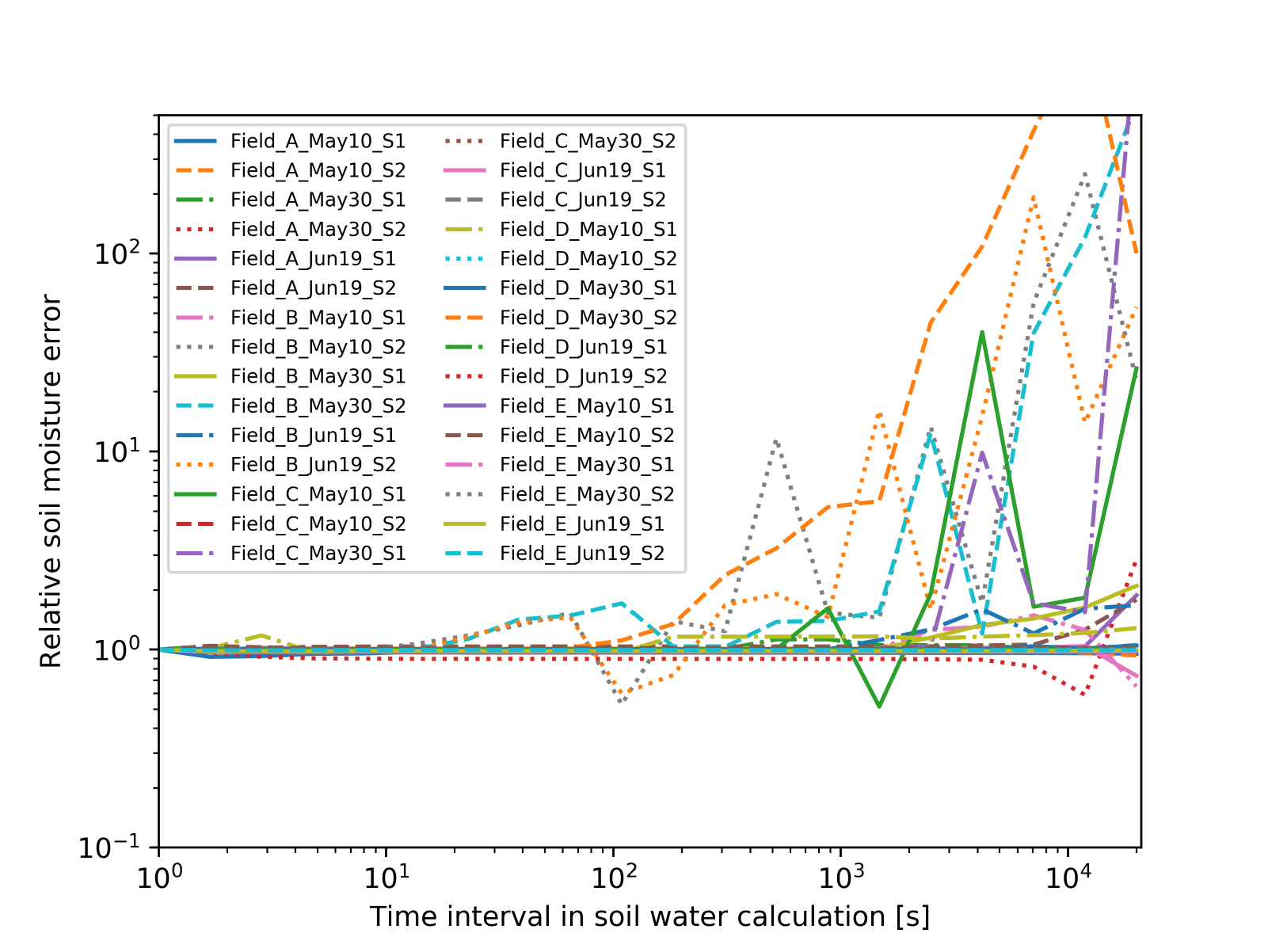}
		\caption{Relative soil moisture error based on simulation error using $\Delta t = 1\,s$. 
				The different colors represent the different field names, analysis days and sensor numbers.
				If large $\Delta t$ (e.g., 1,000 and 10,000\,s) is used, the soil moisture simulation error can rise much higher than the $\Delta t = 1\,{\rm s}$ case. If $\Delta t$ is 10\,s, the discretization error is as small as the case of $\Delta t = 1\,{\rm s}$.}
		\label{dt_fig}
		\end{center}
	\end{figure}

 To check whether the discretization errors caused by $\Delta t$ are small enough for soil moisture simulation, test calculations are performed, wherein we change the value of $\Delta t$ between 1 and 20,000\,s, and perform simulations under parameter optimized environments for all three analysis days of both sensors in all five fields.
	We calculate the relative soil moisture error $\gamma$ based on the simulation error using $\Delta t = 1\,s$.
	Thus, $\gamma$ can be expressed as follows:
	\begin{eqnarray}
		\gamma = \frac{\delta(\Delta t)}{\delta(\Delta t=1\,{\rm s})},
	\end{eqnarray}
	where $\delta(\Delta t)$ is
	\begin{eqnarray}
		\delta(\Delta t) = \Sigma_l\frac{1}{N_l}\sqrt{\frac{\Sigma_d\left( O_{d,l} - S_{d,l}(\Delta t) \right)^2}{N_d}}.
	\end{eqnarray}
	Here $l$ and $d$ are indexes for the soil layer and the day, respectively; $N_l$ and $N_d$ are the total number of layers and days, respectively; $O_{d,l}$ and $S_{d,l}(\Delta t)$ are the observed and simulated soil moisture values, respectively, wherein $S_{d,l}(\Delta t)$ depends on $\Delta t$ caused by discretization error.
	Figure \ref{dt_fig} shows the trend of the calculated $\gamma$.
	As can be seen therein, if $\Delta t$ is 10\,s, the error of all simulations is the same as that for $\Delta t = 1\,{\rm s}$, implying that the discretization error of the $\Delta t = 10\,{\rm s}$ case is small enough for the simulation.

\section*{Acknowledgment}
	\par The authors are grateful to T. Tchouboukjian for collecting the farming data and supporting the collection of the observation data.
	Further, the authors thank M. Satoh for useful discussions.
	The authors would like to take this opportunity to thank S. Matsumoto for leading our agriculture project in NEC and Enago (www.enago.jp) for the English language review.
	
\section*{Funding}
	\par This research did not receive any specific grant from funding agencies in the public, commercial, or not-for-profit sectors.
	
\section*{Declarations of interest}
	\par None

\end{document}